\documentclass[aps,pre,reprint,longbibliography,superscriptaddress,nobalancelastpage]{revtex4-2}

\usepackage{amsmath,amssymb}
\usepackage{graphicx}
\usepackage{xcolor}
\usepackage{mathrsfs}
\usepackage{newtxtext}
\usepackage[slantedGreek,cmintegrals]{newtxmath}
\usepackage{stmaryrd}
\usepackage{hyperref}
\usepackage{enumitem}
\usepackage{accents}
\usepackage{array}  
\usepackage{multirow}

\newcolumntype{C}{>{$}c<{$}}
\definecolor{linkcolor}{HTML}{0000ff}
\hypersetup{colorlinks,linkcolor={linkcolor},citecolor={linkcolor},urlcolor={linkcolor}}

\renewcommand{\vec}[1]{\boldsymbol{#1}}

\renewcommand{\pi}{\uppi}

\DeclareMathAlphabet{\mathcal}{OMS}{cmsy}{m}{n}
\DeclareMathAlphabet{\mathcalbf}{OMS}{cmsy}{b}{n}
\newcommand{\mat}[1]{\mathsf{#1}}

\newcommand{\figref}[2]{[Fig.~\hyperref[#1]{\ref*{#1}(#2)}]}
\newcommand{\figrefp}[2]{\hyperref[#1]{\ref*{#1}(#2)}}
\newcommand{\figrefi}[2]{[Fig.~\hyperref[#1]{\ref*{#1}(#2)}, inset]}
\newcommand{\textfigref}[2]{Fig.~\hyperref[#1]{\ref*{#1}(#2)}}
\newcommand{\wholefigref}[1]{(Fig.~\ref{#1})}
\newcommand{\textwholefigref}[1]{Fig.~\ref{#1}}
\newcommand{\citeref}[1]{Ref.~\cite{#1}}

\renewcommand{\leq}{\leqslant}
\renewcommand{\geq}{\geqslant}

\begin{document}

\title{Stabilization of Microbial Communities by Responsive Phenotypic Switching}
\author{Pierre A. Haas}
\email{haas@pks.mpg.de}
\affiliation{Max Planck Institute for the Physics of Complex Systems, N\"othnitzer Stra\ss e 38, 01187 Dresden, Germany}
\affiliation{\smash{Max Planck Institute of Molecular Cell Biology and Genetics, Pfotenhauerstra\ss e 108, 01307 Dresden, Germany}}
\affiliation{Center for Systems Biology Dresden, Pfotenhauerstra\ss e 108, 01307 Dresden, Germany}
\author{Maria A. Gutierrez}
\affiliation{Department of Applied Mathematics and Theoretical Physics, Centre for Mathematical Sciences, \\ University of Cambridge, 
Wilberforce Road, Cambridge CB3 0WA, United Kingdom}
\author{Nuno M. Oliveira}
\email{n.m.oliveira@damtp.cam.ac.uk}
\affiliation{Department of Applied Mathematics and Theoretical Physics, Centre for Mathematical Sciences, \\ University of Cambridge, 
Wilberforce Road, Cambridge CB3 0WA, United Kingdom}
\affiliation{\smash{Department of Veterinary Medicine, University of Cambridge, Madingley Road, Cambridge CB3 0ES, United Kingdom}}
\author{Raymond E. Goldstein}
\email{r.e.goldstein@damtp.cam.ac.uk}
\affiliation{Department of Applied Mathematics and Theoretical Physics, Centre for Mathematical Sciences, \\ University of Cambridge, 
Wilberforce Road, Cambridge CB3 0WA, United Kingdom}
\date{\today}%
\begin{abstract}
Clonal microbes can switch between different phenotypes and recent theoretical work has shown that stochastic switching between these subpopulations can stabilize microbial communities. This phenotypic switching need not be stochastic, however, but could also be in response to environmental factors, both biotic and abiotic. Here, motivated by the bacterial persistence phenotype, we explore the ecological effects of such responsive switching by analyzing phenotypic switching in response to competing species. We show that the stability of microbial communities with responsive switching differs generically from that of communities with stochastic switching only. To understand the mechanisms by which responsive switching stabilizes coexistence, we go on to analyze simple two-species models. Combining exact results and numerical simulations, we extend the classical stability results for the competition of two species without phenotypic variation to the case in which one species switches, stochastically and responsively, between two phenotypes. In particular, we show that responsive switching can stabilize coexistence even when stochastic switching on its own does not affect the stability of the community.
\end{abstract}

\maketitle
\section{Introduction}
One of the classical results of theoretical ecology is that large random ecological communities are very likely to be unstable~\cite{may72}. This is a statistical result, based on the analysis of random matrices representing the Jacobians of otherwise unspecified population dynamics~\cite{may72}. It may appear to contradict Nature's large biodiversity, but actual biological interactions are not random: Rather, they are the product of a long history of evolution, during which population dynamics pruned a possibly much larger set of species~\cite{servan18}. Nevertheless, the mathematical constraints from random matrix theory that restrict the evolution of this biodiversity can be revealed by this statistical take on population stability, because it enables some analysis of generic large ecological communities and their huge parameter space, and therefore complements the exact results for small systems that are not available for these larger systems. The power of this statistical approach has been demonstrated in a large body of work which, for example, analyzed the effect of the interaction type and structure on the stability of the community~\cite{allesina12,mougi12,coyte15,grilli16}, revealed the stabilizing effect of higher-order interactions and explicit resource dynamics~\cite{grilli17,butler18}, or explored yet other related problems~\cite{roberts74,grilli17b,gibbs18,stone18,barron20}.

In this context, recent theoretical work has revealed that subpopulation structure such as phenotypic variation can stabilize microbial communities~\cite{maynard19,haas20}. In particular, we have argued that abundant phenotypic variation is generically destabilizing~\cite{haas20}, essentially because introducing phenotypic variation increases the effective number of species in the system, which is known to be destabilizing~\cite{may72,allesina12}. More subtly however, stochastic switching to a rare phenotype such as the bacterial persister phenotype~\cite{maisonneuve14,harms16,radzikowski17} can stabilize communities~\cite{haas20}. 

Although such stochastic phenotypic switching is optimal in infrequently changing environments~\cite{kussell05}, frequent environmental cues, be they biotic or abiotic, favor responsive phenotypic switching associated to some kind of sensing mechanism~\cite{kussell05}. Indeed, recent experimental evidence suggests that such sensing is implicated in the bacterial stress response and, in particular, in switching to stress-resilient phenotypes akin to bacterial persisters~\cite{maisonneuve14,harms16,radzikowski17}. For example, formation of persisters under stress has recently been associated with production of the ``alarmone'' ppGpp, suggesting a stress-dependent persistence response~\cite{maisonneuve14,radzikowski17}. Data showing that sublethal antibiotic concentrations increase persister concentration~\cite{andersson14} are also consistent with responsive contributions to the persistence switching rates. (While early experiments exposing \emph{Escherichia coli} to antibiotics had suggested that direct sensing is absent from the persistence response~\cite{balaban04}, these could not, as already noted in Ref.~\cite{kussell05}, exclude effects such as a dependence of switching rates on antibiotic concentration.) The cues for this responsive switching may be toxins produced by other bacterial species, such as the colicinogens produced by certain strains of \emph{E. coli} that have garnered attention in the context of bacterial ``rock--paper--scissors'' games~\cite{kerr02,czaran02,kirkup04,reichenbach07}. Indeed, very recent work~\cite{oliveira22} provides experimental evidence linking such toxins to a stress response: a ``suicidal'' subpopulation of \emph{Pseudomonas aeruginosa} actively migrates up antibiotic gradients while upregulating the release of its own bacteriocins, suggesting an attack response against toxin-producing competitors~\cite{oliveira22}. Responsive mechanisms are thus starting to be appreciated as a feature of microbial populations, but their ecological role in microbial communities remains unclear.

Here, we address these ecological implications of responsive switching theoretically by analyzing its effects on the stability of competitive microbial communities with a rare, slowly growing and weakly competing, persister-like phenotype. In the first part of this paper, we extend the two-phenotype model of Ref.~\cite{haas20} to include responsive phenotypic switching, and show how the statistical stability properties of this model differ from those of a model with stochastic switching only, even if the second phenotype is rare. These statistical results emphasize the importance of the type of phenotypic switching for stability, but leave unanswered the question: Under which conditions does phenotypic switching, be it stochastic or responsive, stabilize or destabilize this community? We address this question in the second part of this paper: We extend the classical results for the stability of Lotka--Volterra systems~\cite{*[] [{, Chap. 3, pp. 79--118 and Appendices, pp. 501--511.}] murray} by analyzing a minimal model of two competing species in which one species switches, both stochastically and in response to the other species, between two phenotypes. Using numerical simulations and by deriving exact results for still simpler models, we show in particular that responsive switching can promote coexistence even in cases in which stochastic switching on its own does not affect the stability of the community.

\section{Statistical stability of responsive phenotypic switching}\label{sec:model}
In this Section, we introduce a model for the competition of $N$ species that switch, both responsively and stochastically, between two phenotypes. This model extends that of Ref.~\cite{haas20}. We show that its statistical stability properties are generically different from those of the corresponding model with stochastic switching only.
\subsection{Model}
\begin{figure}
\includegraphics{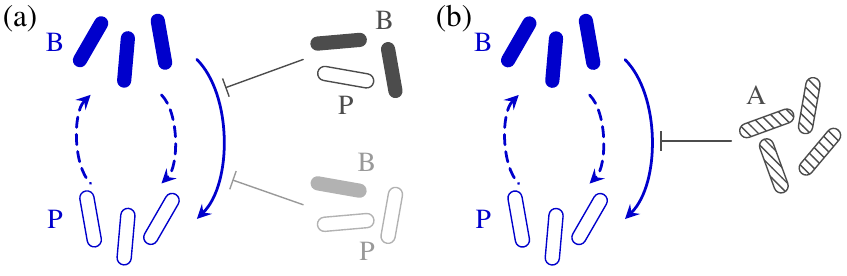}
\caption{Models of stochastic and responsive phenotypic switching. (a)~In the model of Sec.~\ref{sec:model}, each species has two phenotypes, B and P, and switches stochastically between them. Moreover, the B phenotype of each species responds to other species by switching to the P phenotype. (b)~In the minimal two-species model of Sec.~\ref{sec:2model}, the second species has a single competitor phenotype A, which causes B to switch to P. Dashed lines: stochastic switching. Solid lines: responsive switching.}\label{fig1} 
\end{figure}

We consider the competition of $N\geqslant 2$ well-mixed species that have two phenotypes, B and P, each, between which they switch stochastically. In addition, the B phenotypes of each species respond to the other species by switching to the corresponding P phenotype~\figref{fig1}{a}. We denote by $B_n$ and $P_n$ the respective abundances of the B and P phenotypes of species $n$, for $1\leq n\leq N$. With Lotka--Volterra competition terms~\cite{murray}, the dynamics of the vectors $\vec{B}=(B_1,B_2,\dots,B_N)$ and $\vec{P}=(P_1,P_2,\dots,P_N)$ are thus~\footnote{We imply, in Eqs.~\eqref{eq:fullmodel} and throughout Sec.~\ref{sec:model} and Appendix~\ref{appA}, elementwise multiplication of vectors and rows or columns of matrices by writing the corresponding symbols next to each other, and reserve dots to denote matrix multiplication.}
\begin{subequations}\label{eq:fullmodel}
\begin{align}
\vec{\dot{B}}&=\vec{B}\bigl(\vec{b}-\mat{C}\cdot\vec{B}-\varepsilon\mat{D}\cdot\vec{P}\bigr)-\varepsilon\vec{B}\bigl(\vec{k}+\mat{R}\cdot\vec{B}+\mat{S}\cdot\vec{P}\bigr)+\vec{\ell P},\\
\vec{\dot{P}}&=\varepsilon\vec{P}\bigl(\vec{p}-\mat{E}\cdot\vec{B}-\mat{F}\cdot\vec{P}\bigr)+\varepsilon\vec{B}\bigl(\vec{k}+\mat{R}\cdot\vec{B}+\mat{S}\cdot\vec{P}\bigr)-\vec{\ell P},
\end{align}
\end{subequations}
where $\vec{b},\vec{p}$ are growth rates, the nonnegative entries of the matrices $\mat{C},\mat{D},\mat{E},\mat{F}$ are competition strengths, $\vec{k},\vec{\ell}$ are nonnegative rates of stochastic switching, and $\mat{R},\mat{S}$ are nonnegative rates of responsive switching. The diagonal entries of $\mat{R},\mat{S}$ vanish, so that microbes do not switch phenotype in response to the presence of other microbes of their own species. We stress that Eqs.~\eqref{eq:fullmodel} are deterministic, as are all subsequent models in this paper: in particular, they do not resolve the individual, stochastic switching events, but only their mean behavior expressed by the deterministic rates of stochastic switching.

The functional form of the responsive switching rates in Eqs.~\eqref{eq:fullmodel}, with a direct dependence on the competitor abundances, implies a neglect of the dynamics of the chemical cues of the responsive switching. This is justified since our model aims to elucidate the effect of this responsive switching on stability, rather than the effect of these chemical dynamics that was analyzed, e.g., in Ref.~\cite{kelsic15}. The same argument justifies, for instance, neglecting resource dynamics in spite of their known effect on stability~\cite{butler18}. The linear dependence of responsive switching on the abundances of other species is, however, a simplifying assumption, just as the logistic Lotka--Volterra interaction terms~\cite{murray} in Eqs.~\eqref{eq:fullmodel} are the simplest choice of interaction terms. While the full dynamics of the system (and in particular, the existence of non-steady-state attractors like limit cycles) are expected to depend on the details of the functional forms of the interaction terms and switching rates, any such functional forms linearize to the logistic interaction terms and linear switching rates in Eqs.~\eqref{eq:fullmodel} close to an equilibrium. These details do not therefore affect the stability properties of equilibria. Moreover, as shown in Appendix~\ref{appA}, the logistic nonlinearities in Eqs.~\eqref{eq:fullmodel} are sufficient for their dynamics to be bounded. For these reasons, we believe these simplifying assumptions to be appropriate for the qualitative analysis of phenotypic dynamics and in particular of the stability of the equilibria of Eqs.~\eqref{eq:fullmodel} in this paper. 

The parameter $\varepsilon$ in Eqs.~\eqref{eq:fullmodel} scales the growth and competition of the P phenotypes, and the stochastic and responsive switching rates into the P phenotypes. In the limit $\varepsilon\ll 1$, the P~phenotype is therefore a slowly growing and weakly competing phenotype such as bacterial persisters~\cite{maisonneuve14,harms16,radzikowski17}. 

\subsection{Reduced and Averaged Models}
How does the stability of a microbial community with responsive phenotypic switching differ from that of a community with stochastic switching only? To answer this biological question and thus understand the ecological effects of the type of phenotypic switching, it is tempting to compare the mathematical stability of equilibria of Eqs.~\eqref{eq:fullmodel} to that of equilibria of a corresponding \emph{reduced model} with stochastic switching only, in which $\mat{R}=\mat{S}=\mat{O}$. Such a comparison does not however answer our biological question, because the populations at equilibrium of Eqs.~\eqref{eq:fullmodel} and this reduced model are in general different. This direct comparison cannot therefore decide whether stability differences result from this difference of populations or from the differences in phenotypic switching.

To understand the two types of phenotypic switching, we do not therefore compare Eqs.~\eqref{eq:fullmodel} to this reduced model, but instead to an \emph{averaged model} that has the same population at equilibrium as Eqs.~\eqref{eq:fullmodel}. While the reduced model takes values of interaction parameters or switching rates from Eqs.~\eqref{eq:fullmodel}, the averaged model replaces these values with effective values that are determined by the condition of equality of populations at equilibrium. We introduced such averaging of models in Ref.~\cite{haas20}. There, we pointed out that equality of populations at equilibrium does not lead to simple relations between the eigenvalues of the corresponding Jacobians, so there is no reason to expect their stability properties to be the same. The present discussion develops these ideas. 

While stability differences between Eqs.~\eqref{eq:fullmodel} and the corresponding averaged model thus reveal the ecological roles of responsive and stochastic phenotypic switching, stability differences between the reduced and averaged models stem from the differences of the reduced and the averaged, effective model parameters. In an actual biological community, these parameters can evolve independently from the evolution of a phenotypic substructure, and so stability differences need not result from this phenotypic substructure. This raises interesting questions akin to those asked in Ref.~\cite{kussell05}: For example, does evolving a more complex phenotypic substructure require more evolutionary adaptations than evolving the effective parameters directly? Answering such questions requires, however, coupling Eqs.~\eqref{eq:fullmodel} to an evolutionary model. This is beyond the scope of this paper, in which we will therefore focus on the stability differences between Eqs.~\eqref{eq:fullmodel} and the corresponding averaged model, which can be imputed directly to responsive phenotypic switching.

After this rather abstract discussion of reduced and averaged models, we now write down the averaged model with stochastic switching only:
\begin{subequations}\label{eq:fullmodels}
\begin{align}
\vec{\dot{B}}&=\vec{B}\bigl(\vec{b}-\mat{C}\cdot\vec{B}-\varepsilon\mat{D}\cdot\vec{P}\bigr)-\varepsilon\vec{k'}\vec{B}+\vec{\ell P},\\
\vec{\dot{P}}&=\varepsilon\vec{P}\bigl(\vec{p}-\mat{E}\cdot\vec{B}-\mat{F}\cdot\vec{P}\bigr)+\varepsilon\vec{k'}\vec{B}-\vec{\ell P}.
\end{align}
\end{subequations}
This is the model that we have analyzed in Ref.~\cite{haas20}. To establish the correspondence between Eqs.~\eqref{eq:fullmodel} and Eqs.~\eqref{eq:fullmodels}, we notice that an equilibrium $\mathcalbf{E}=(\vec{B_\ast},\vec{P_\ast})$ of Eqs.~\eqref{eq:fullmodel} is also an equilibrium of Eqs.~\eqref{eq:fullmodels} if $\vec{k'}=\vec{k}+\mat{R}\cdot\vec{B_\ast}+\mat{S}\cdot\vec{P_\ast}$ and all other growth rates, competition strengths, and switching rates are unchanged. At equilibrium, responsive switching thus modifies only the effective rate of stochastic switching. It is not therefore possible to distinguish, at equilibrium, between purely stochastic switching and a dependence of switching rates on the presence of other species. This aspect, we have noted in the Introduction, has previously been discussed from an experimental point of view in the context of \emph{E. coli} persistence~\cite{kussell05,balaban04}. Somewhat conversely, in a reduced model, the rate $\vec{k}$ of stochastic switching would remain unchanged from Eqs.~\eqref{eq:fullmodel}; stability differences between Eqs.~\eqref{eq:fullmodels} and this reduced model could therefore also result, independently from responsive switching, from evolution of the rates of stochastic switching.

Despite the correspondence at $\mathcalbf{E}$ of Eqs.~\eqref{eq:fullmodel} and Eqs.~\eqref{eq:fullmodels}, it is clear that their dynamics away from $\mathcalbf{E}$ are in general different. It is for this reason that correspondence at equilibrium does not, as we have already noted above, translate to corresponding stability properties. In what follows, we therefore analyze these stability properties to understand the ecological effect of responsive switching.

\subsection{Results}
In the spirit of the random matrix approach to ecological stability, we compare the stability of Eqs.~\eqref{eq:fullmodel} and Eqs.~\eqref{eq:fullmodels} by sampling their parameters randomly and computing stability statistics. Since the coexistence equilibria of Eqs.~\eqref{eq:fullmodel} and Eqs.~\eqref{eq:fullmodels} cannot be found in closed form, we cannot sample the model parameters directly; rather, as discussed in more detail in Appendix~\ref{appA}, we sample the coexistence state itself and some model parameters directly, leaving linear equations to be solved for the remaining parameters to ensure that the chosen coexistence state is steady~\cite{haas20}. 

\begin{figure}[t]
\includegraphics{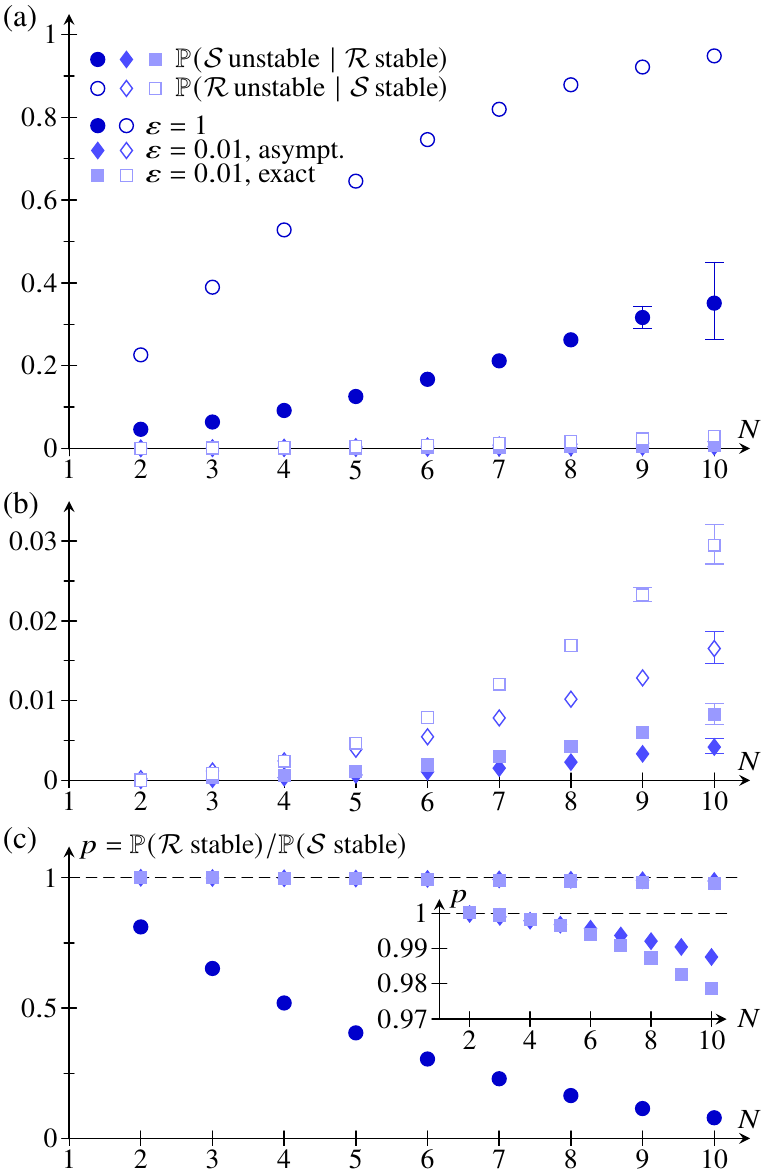}
\caption{Stability of random microbial communities with responsive phenotypic switching. (a)~Probability of a random equilibrium that is stable in the model with responsive switching $\mathcal{R}$ [Eqs.~\eqref{eq:fullmodel}] or in the model with stochastic switching only $\mathcal{S}$ [Eqs.~\eqref{eq:fullmodels}] being unstable in the other model, as a function of the number $N$ of species in the system. (b)~Same plot, but focused on low probabilities. (c) Ratio of the probabilities of random equilibria of $\mathcal{R}$ and $\mathcal{S}$ being stable. Inset: same plot, focused on small probability differences. Probabilities were estimated from up to $5\cdot10^8$ random systems each. Parameter values: $\varepsilon=1$ and $\varepsilon=0.01$~\cite{[{Relative persister abundances are typically very small, {say} $\smash{\varepsilon=10^{-5}}$ for \emph{E. coli}~\protect{\cite{balaban04}}, but can vary widely, as reported by }][{. }]hofsteenge13,*[{The value $\varepsilon=0.01$ used in Figs.~\ref{fig2}--\ref{fig4} is towards the upper end of the experimental range for \emph{E. coli} in culture (\emph{vide ibid.}), but typical for \emph{P. aeruginosa} biofilms [}][{]. Anyway, the mechanism underlying the differences between $\varepsilon=0$ and $\varepsilon\ll1$ that we have identified in Ref.~\protect{\cite{haas20}} is generic, which justifies choosing $\varepsilon=0.01$ for numerical convenience.}]lewis08}; for the latter value, both exact and asymptotic equilibria were computed. Error bars are $95\%$ confidence intervals~\cite{*[{The confidence intervals computed for Figs.~\figrefp{fig2}{a),(b}, \ref{fig3}, \ref{fig4} are Wilson intervals [see, e.g., }] [{]; those computed for~\textfigref{fig2}{c} are derived from the Fieller-type statistic introduced by }] brown01,*nam02} larger than the plot markers.}\label{fig2}
\end{figure}

For these random systems, we find that, as the number $N$ of species increases, stable coexistence states of Eqs.~\eqref{eq:fullmodel} are increasingly unlikely to be stable with the dynamics of Eqs.~\eqref{eq:fullmodels}, and vice versa: Stable coexistence with responsive switching is increasingly unlikely to be stable with stochastic switching only, and vice versa~\figref{fig2}{a}. This trend persists for $\varepsilon\ll 1$, although the probabilities are reduced in magnitude~\figref{fig2}{b}. Responsive switching can thus stabilize and destabilize equilibria of Eqs.~\eqref{eq:fullmodel}. However, coexistence is less likely to be stable with responsive switching than with stochastic switching only~\figref{fig2}{c}, although the probabilities are nearly equal for small $\varepsilon$~\figrefi{fig2}{c}.

In the limit $\varepsilon\ll 1$, coexistence states can be determined in closed form by asymptotic expansion in the small parameter~$\varepsilon$. Writing \mbox{$\vec{B_\ast}=\vec{B_0}+\varepsilon\vec{B_1}+O\bigl(\varepsilon^2\bigr)$}, $\vec{P_\ast}=\vec{P_0}+\varepsilon\vec{P_1}+O\bigl(\varepsilon^2\bigr)$, we find $\smash{\vec{B_0}=\mat{C}^{-1}\cdot\vec{b}}$, $\vec{P_1}=(\vec{k}+\mat{R}\cdot\vec{B_0})\vec{B_0}/\vec{\ell}$, but \mbox{$\vec{B_1}=\vec{P_0}=\vec{0}$}, as derived in more detail in Appendix~\ref{appA}. We thence obtain asymptotic expressions for the Jacobians (Appendix~\ref{appA}), similarly to calculations in Ref.~\cite{haas20}. Sampling asymptotic coexistence equilibria using this solution, we confirm our findings above, although results differ at a quantitative level because of the different sampling methods~\figref{fig2}{b}.

The models with and without responsive switching thus generically lead to different stability results. The Jacobians at $\mathcalbf{E}$, $\mat{J_\ast}$ with responsive switching and $\mat{K_\ast}$ with stochastic switching only, are related by
\begin{align}
\mat{J_\ast}=\mat{K_\ast}+\varepsilon\left(\begin{array}{c|c}
-\vec{B_\ast}\mat{R}&-\vec{B_\ast}\mat{S}\\
\hline
\vec{B_\ast}\mat{R}&\vec{B_\ast}\mat{S}\\
\end{array}\right),\label{eq:JK}
\end{align}
which follows from the calculations in Appendix~\ref{appA}. We stress that this is an exact result and not an asymptotic approximation. Even in view of this linear relation, the fact that the two models give different stability results is not fundamentally surprising~\cite{haas20}: this simply reflects the fact that linear relations between matrices do not imply linear relations between their determinants. What is perhaps more unexpected is that, although $\mat{J_\ast}=\mat{K_\ast}+O(\varepsilon)$, the stability results differ even in the limit $\varepsilon\ll1$. We have previously related behavior of this ilk to the possibility of eigenvalues with small real parts being stabilized or destabilized by higher-order terms in the expansion~\cite{haas20}. 

We had to restrict to small values of $N$ when computing the numerical results in \textwholefigref{fig2}, because it becomes increasingly difficult to sample increasingly rare~\cite{may72,allesina12} stable systems as $N$ increases. The strength of the classical random matrix approach~\cite{may72,allesina12} to stability is that it can often circumvent this difficulty by analyzing random Jacobians without further specification of the underlying population dynamics and hence of their equilibria. This simplification is not however available for the questions addressed here, because we need to relate ``full'' and averaged models (in this case, with and without responsive switching). Nonetheless, the trends in \textwholefigref{fig2} suggest that the stability differences found numerically there are amplified for larger values of $N$.

There is, however, a more important limitation of this stability analysis: one may argue that instability of an equilibrium is not biologically or ecologically significant, because it does not imply extinction of a species, since the dynamics, perturbed away from the unstable equilibrium, may converge to a different equilibrium, a limit cycle, or a more complex attractor, and so the species may still coexist. A more relevant question is therefore: does stability of one model (with responsive switching or with stochastic switching only) not only fail to predict stability, but also fail to predict coexistence in the other model? Unlike predicting stability, predicting coexistence requires the full dynamics of the system and its non-steady-state attractors, and so the predictions depend, as noted above, on the details of the interactions and switching rates. Analyzing the effect of these details on coexistence is beyond the scope  of this paper. Here, we restrict to the logistic interactions and linear switching rates in Eqs.~\eqref{eq:fullmodel}, evolve unstable equilibria numerically and, as described in Appendix~\ref{appA}, determine whether the species coexist permanently~\cite{[{We will use ``permanent coexistence'' rather loosely to mean the alternative to extinction of one or more species, but note that different mathematical definitions, some of which go (rather unhelpfully in the biological context of this paper) by the name of ``persistence'' rather than ``permanence'', are associated with this concept in the dynamical systems literature: See, e.g., the definitions in Sec. 3 of }][{ and the related discussion in Sec. 2 of }]butler86,*hutson83}.

\begin{figure}[b]
\includegraphics{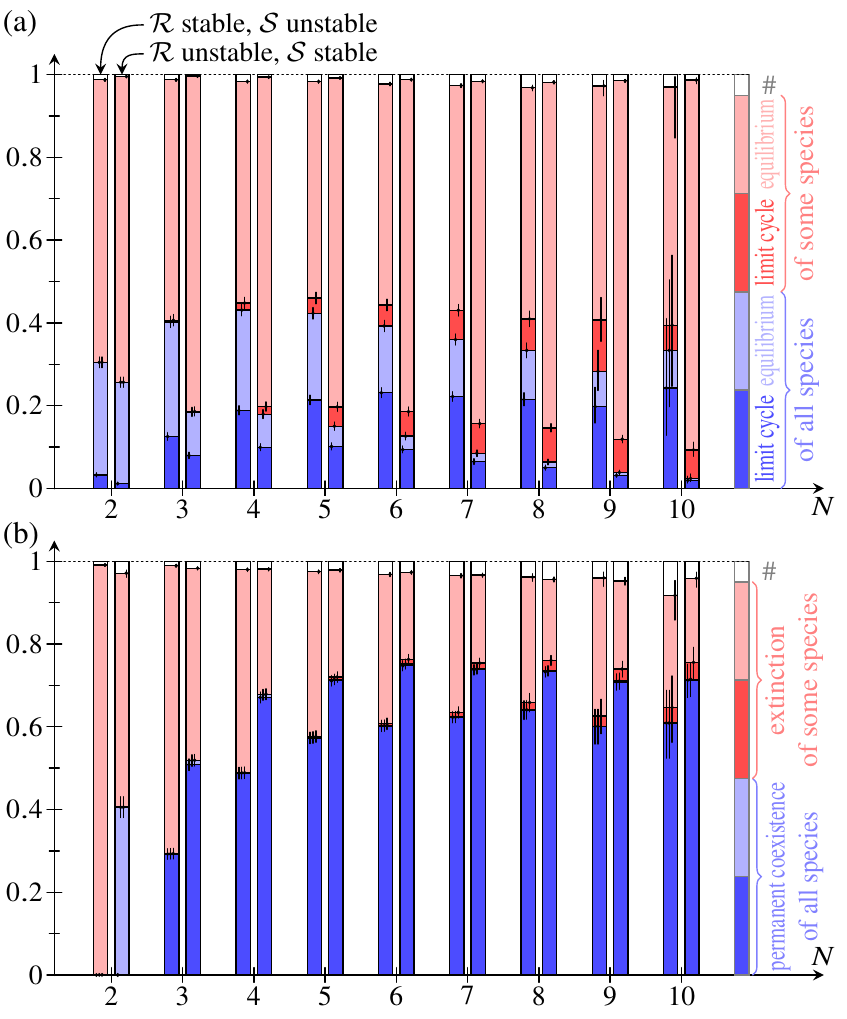} 
\caption{Distributions, in the models with responsive switching $\mathcal{R}$ [Eqs.~\eqref{eq:fullmodel}] or with stochastic switching only $\mathcal{S}$ [Eqs.~\eqref{eq:fullmodels}], of the long-time dynamics of unstable equilibria that are stable in the other model. Distributions are shown for exact equilibria with (a) $\varepsilon=1$, (b)~${\varepsilon=0.01}$~\cite{hofsteenge13}, and different numbers of species $N$. Each distribution was estimated by numerical integration of up to $5\cdot 10^3$ unstable systems. For a small proportion of the systems ($\#$), the numerical solution did not converge. Vertical bars represent $95\%$ confidence intervals~\cite{brown01}.}
\label{fig3}
\end{figure}

Figure~\ref{fig3} shows the distributions of the long-time dynamics of equilibria that are unstable with responsive switching or with stochastic switching only, respectively, but are stable in the other model. The possible long-time dynamics are permanent coexistence or extinction of some species; in both cases, we distinguish between convergence of the remaining species to an equilibrium or to a limit cycle. The small proportion of systems for which the numerical solution does not converge~\wholefigref{fig3} may include systems in which more complex attractors arise. We note in particular that extinction of some species is not a rare outcome~\wholefigref{fig3}: it is actually the most likely outcome if both phenotypes have similar abundances [$\varepsilon=1$, \textfigref{fig3}{a}], while convergence to a limit cycle of all species is more likely if one phenotype is rare [$\varepsilon\ll 1$, \textfigref{fig3}{b}]. 

\begin{figure}[t]
\includegraphics{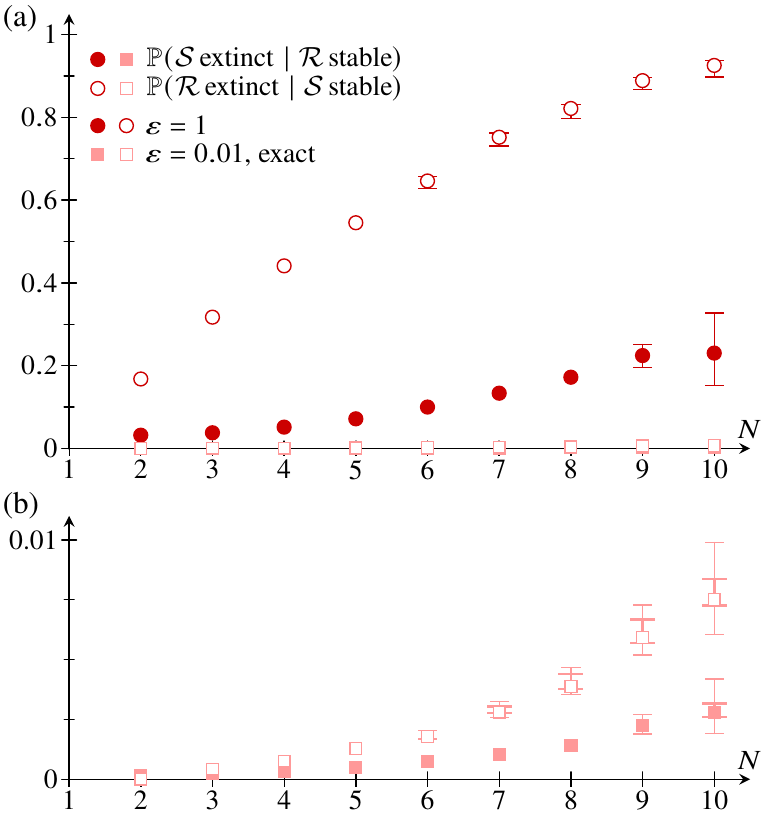} 
\caption{Permanent coexistence in random microbial communities with responsive phenotypic switching. (a)~Probability, in the models with responsive switching $\mathcal{R}$ [Eqs.~\eqref{eq:fullmodel}] and with stochastic switching only $\mathcal{S}$ [Eqs.~\eqref{eq:fullmodels}], of extinction of some species of a random system perturbed away from an equilibrium that is stable in the other model, as a function of the number of species in the system, $N$. (b)~Same plot, but focused on low probabilities. Exact equilibria were obtained for $\varepsilon=1$ and $\varepsilon=0.01$~\cite{hofsteenge13}. Each probability was computed from up to $5\cdot 10^8$ random systems and up to $5\cdot 10^3$ random systems having an equilibrium with different stability in models $\mathcal{R}$ and $\mathcal{S}$. Thick error bars correct the estimated probabilities for the systems in which the numerical solution of the long-time behavior did not converge~\wholefigref{fig3}; thin error bars add $95\%$ confidence intervals~\cite{brown01}. Only those error bars larger than the plot markers are shown.}
\label{fig4}
\end{figure}

From these distributions, we estimate, for both models, the probabilities of extinction of some species for systems perturbed away from an equilibrium that is stable in the other model~\wholefigref{fig4}. The increase of these probabilities with the number of species $N$ qualitatively matches the increase of the probabilities of an equilibrium that is stable in one model being unstable in the other model~\wholefigref{fig2}. We can therefore extend our previous conclusion: stability of an equilibrium in one model does not even in general imply permanent coexistence in the other model.

Finally, we could similarly compare the stability of Eqs.~\eqref{eq:fullmodel} to that of an averaged model without phenotypic switching (Appendix~\ref{appA}). However, in view of the correspondence of Eqs.~\eqref{eq:fullmodel} and Eqs.~\eqref{eq:fullmodels}, the qualitative conclusions of such a comparison must parallel those of comparing models with stochastic switching only and averaged models without phenotypic variation. These two models we compared in Ref.~\cite{haas20}, where we concluded that stochastic switching to an abundant phenotype is destabilizing, but stochastic switching to a rare phenotype, corresponding to $\varepsilon\ll 1$, is stabilizing. Here, we can therefore conclude similarly that responsive switching to an abundant phenotype generically destabilizes the community (compared to the case in which there is no phenotypic variation), but that responsive switching to a rare phenotype has a stabilizing effect.

However, all of these results are fundamentally statistical in nature. While they show how responsive phenotypic switching affects stability on average, they do not yield any insight into the conditions under which phenotypic switching stabilizes or destabilizes the community. To understand the mechanisms underlying these stability differences, we therefore complement this statistical analysis with an analysis of reduced two-species models in the next Section.

\section{Stabilization of two-species coexistence by responsive switching}\label{sec:2model}
In this Section, we reduce the $N$-species model~\eqref{eq:fullmodel} to a minimal two-species model of responsive phenotypic switching. In the context of this model, we analyze the mechanisms by which stochastic and responsive switching affect stability numerically. We confirm some of these numerical results by deriving exact results for simplified models in Appendix~\ref{appB}. These exact results extend the classical results~\cite{murray}, which we rederive in Appendix~\ref{appC}, for the stability of the two-species Lotka--Volterra competition model. 

\subsection{Minimal two-species model}
The simplest setting for responsive phenotypic switching is the competition of two well-mixed species~\figref{fig1}{b}. The first species has a single phenotype, A, while the second one has two phenotypes, B and P, between which it switches stochastically. Moreover, phenotype B responds to the competitor phenotype A by switching to P~\figref{fig1}{b}. To interpret this model in terms of actual biological systems, it is useful to think of P as a phenotype resilient to stress conditions, akin to the bacterial persister phenotype~\cite{maisonneuve14,harms16,radzikowski17}, and to which the common, ``normal'' phenotype B switches in response to toxins produced by A. With this interpretation and for simplicity, we shall sometimes refer to B, P, A and their respective abundances $B$, $P$, $A$ as bacteria, persisters, and competitors~\footnote{The name ``bacteria'', which we use to have a simple way of referring to phenotype B, is of course not strictly biologically accurate, because the persisters (phenotype P) and competitors (phenotype A) are, biologically, bacteria, too.}, and we shall sometimes assume the model parameters to scale accordingly.

Thus, again using Lotka--Volterra competition terms~\cite{murray}, the nondimensionalized dynamics of the system are
\begin{subequations}\label{eq:model}
\begin{align}
\dot{B}&=B(1-\alpha A-B-\kappa P)-\beta AB-\gamma B+\delta P,\label{eq:B}\\
\dot{P}&= P(\mu-\xi A-\varpi B-\varsigma P)+\beta A B+\gamma B-\delta P,\label{eq:P}\\
\dot{A}&=A(\zeta-\eta A-\vartheta B-\iota P),
\end{align}
\end{subequations}
wherein $\alpha,\beta,\gamma,\delta,\zeta,\eta,\vartheta,\iota,\kappa,\mu,\xi,\varpi,\varsigma\geq 0$ are dimensionless nonnegative parameters. To obtain this form of the equations, we have scaled time and the population sizes so as to remove the parameters that would otherwise appear in the logistic growth term $B(1-B)$ of the bacteria in the absence of persisters and competitors~\footnote{The variables $B$ and $P$ in Eqs.~\eqref{eq:model} must be nondimensionalized using the same dimensional scalings lest the switching terms in Eqs.~\eqref{eq:B} and \eqref{eq:P} become unbalanced. We could have chosen a different scaling for $A$ in Eqs.~\eqref{eq:model} from that of $B$ and $P$ to set e.g. $\alpha=1$, but equal scalings make comparing competition strengths easier. In the same spirit, we have reused the dimensional scalings from Eqs.~\eqref{eq:model} to nondimensionalize $B'$, $A'$ in the averaged model described by Eqs.~\eqref{eq:lv} to avoid having to unravel different dimensional scalings when writing down the consistency conditions expressed by Eqs.~\eqref{eq:cca}.}. The interpretation of the model parameters in Eqs.~\eqref{eq:model} is given in Table~\ref{tab0}.

\begin{table}[t]
\caption{Parameters of the two-species model~\eqref{eq:model}, representing the interactions in~\textfigref{fig1}{b}, and their interpretations. The parameters have been divided into three groups. Only one of the persister logistic parameters appears in each of the simplified models~\eqref{eq:model2}, \eqref{eq:model2b}, \eqref{eq:model2c}. } \label{tab0}
\begin{ruledtabular}
\begin{tabular}{cCl}
group&\text{parameter}\footnote{The equations have been nondimensionalized by rescaling time and population abundances to remove two (dimensional) parameters, viz. the growth rate and the strength of the within-species competition of B.}&interpretation\footnote{Phenotype abbreviations (B: bacteria; P: persisters; A: competitors) used in this column are as in~\textfigref{fig1}{b}.}\\
\hline
\multirow{4}{*}{\rotatebox{90}{\parbox{15mm}{competitor logistic parameters}}}&\alpha&magnitude of effect of B--A interaction on B\\
&\zeta&growth rate of A\\
&\eta&strength of within-species competition of A\\
&\vartheta&magnitude of effect of B--A interaction on A\\
\hline
\multirow{3}{*}{\rotatebox{90}{\parbox{12.5mm}{phenotypic\\switching\\rates}}}&\beta&rate of responsive switching from B to P\rule{0pt}{5mm}\\
&\gamma&rate of stochastic switching from B to P\\
&\delta&rate of stochastic switching from P to B\rule[-3mm]{0pt}{0pt}\\
\hline
\multirow{6}{*}{\rotatebox{90}{\parbox{23mm}{persister logistic parameters}}}&\iota&magnitude of effect of P--A interaction on A\\
&\kappa&magnitude of effect of B--P interaction on B\\
&\mu&growth rate of P\\
&\xi&magnitude of effect of P--A interaction on P\\
&\varpi&magnitude of effect of P--B interaction on P\\
&\varsigma&strength of within-species competition of P
\end{tabular}
\end{ruledtabular}
\end{table}

We have not explicitly introduced a positive parameter $\varepsilon\lesssim 1$ scaling the competition dynamics of and switching rates to persisters, but we expect 
\begin{align}
&\alpha,\delta,\zeta,\eta,\vartheta=O(1),&&\beta,\gamma,\iota,\kappa,\mu,\xi,\varpi,\varsigma=O(\varepsilon) \label{eq:scalings}
\end{align}
from the interpretation of the model in terms of bacteria, persisters, and competitors and by comparison of Eqs.~\eqref{eq:fullmodel} and~\eqref{eq:model}. We will not assume $\varepsilon\ll 1$, but we will sometimes invoke $\varepsilon\lesssim1$ below and in the calculations in Appendices~\ref{appB}, \ref{appC}, \ref{appD} to restrict parameter ranges or impose inequalities between parameters.

\subsection{Results}
In this subsection, we analyze model~\eqref{eq:model} numerically and ask: under which conditions and to what extent does responsive switching stabilize or destabilize coexistence?

\subsubsection{Numerical setup}
Although the full model \eqref{eq:model} is too complicated for meaningful analytical progress to be made, its equilibria can be found numerically and efficiently by precomputing, using \textsc{Mathematica}, the exact polynomial equations satisfied by the equilibria from Gr\"obner bases~\cite{groebner}. For each computed equilibrium, we check its accuracy using the values of the right-hand sides of Eqs.~\eqref{eq:model} evaluated there, and verify that all solutions have been found using a test based on Sturm's theorem~\cite{algebra}. Finally, we determine the stability of the computed equilibria using the Routh--Hurwitz conditions~\cite{murray}. Similarly, we determine the stability of equilibria of the averaged models (with stochastic switching only and without phenotypic variation) corresponding to Eqs.~\eqref{eq:model}. Unstable equilibria are integrated numerically, similarly to the calculations in Sec.~\ref{sec:model}, to determine whether the species coexist permanently notwithstanding a particular coexistence equilibrium being unstable. We reduce the number of systems that need to be integrated in this way by showing, in Appendix~\ref{appD}, that coexistence is permanent if all trivial steady states are unstable provided that the persister scalings~\eqref{eq:scalings} are satisfied.

\begin{figure}[b]
\includegraphics{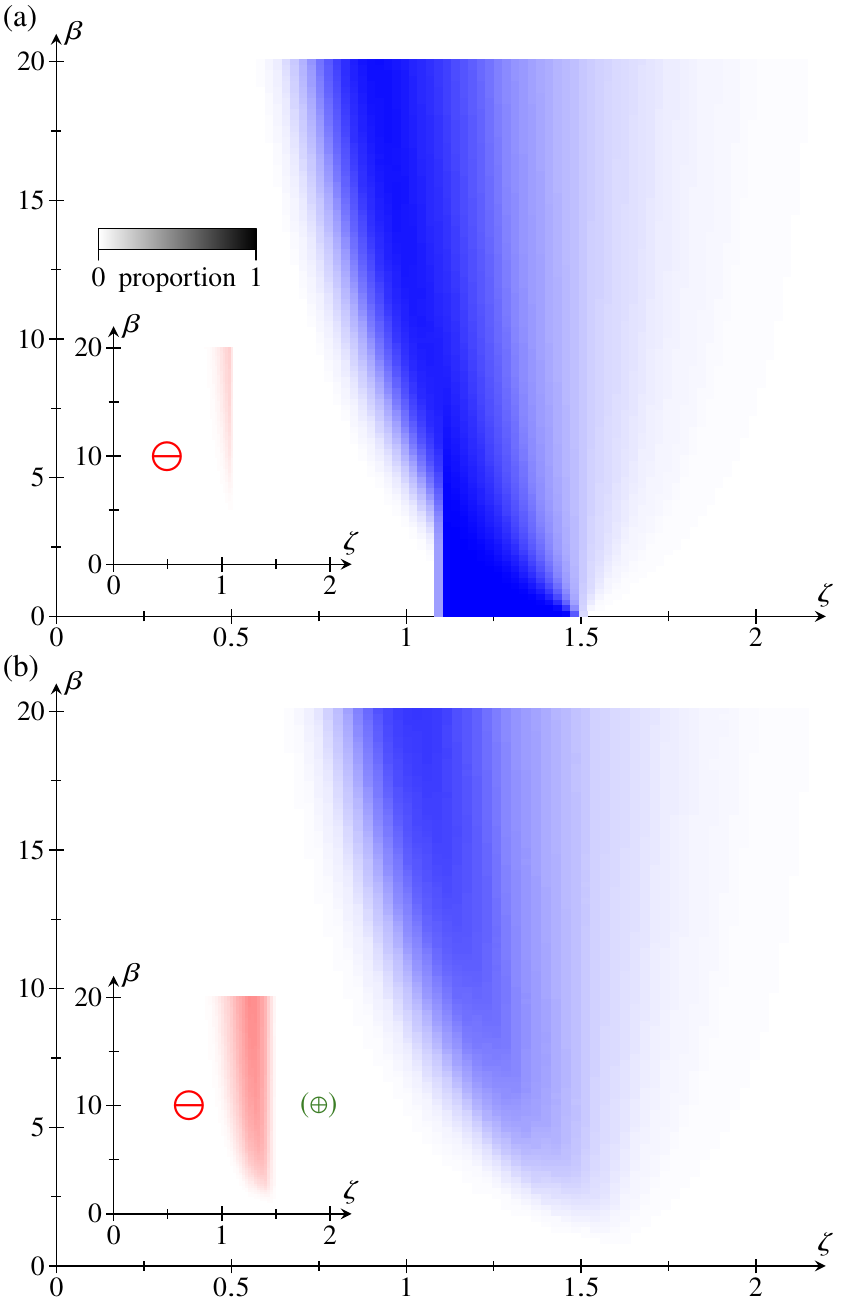}
\caption{Numerical stability diagrams of Eqs.~\eqref{eq:model} in the $(\zeta,\beta)$ diagram in the cases (a) $\eta/\alpha>\vartheta$ and (b)~${\eta/\alpha<\vartheta}$. The color of each point in the stability diagrams represents the proportion of $N=1000$ random systems for which coexistence is stable or permanent at that point. The insets plot the proportion of systems for which coexistence is destabilized, $\ominus$, compared to the averaged model with stochastic switching only. The symbol $\oplus$ in parentheses [panel (b), inset] indicates that a nonzero proportion of systems (too small to be visualizable by the color scheme) is stabilized (or becomes permanent) compared to the averaged model. Parameter values: $\alpha=0.8$, $\eta=1.2$, $\vartheta=1.1$ [panel (a)] or $\vartheta=1.9$ \mbox{[panel (b)]}, $\gamma,\iota,\kappa,\mu,\xi,\varpi,\varsigma\sim\mathcal{U}[\varepsilon,2\varepsilon]$, with $\varepsilon=0.1$, and $\delta\sim\mathcal{U}[0.8,1.6]$.}\label{fig5} 
\end{figure}

\begin{figure*}
\includegraphics{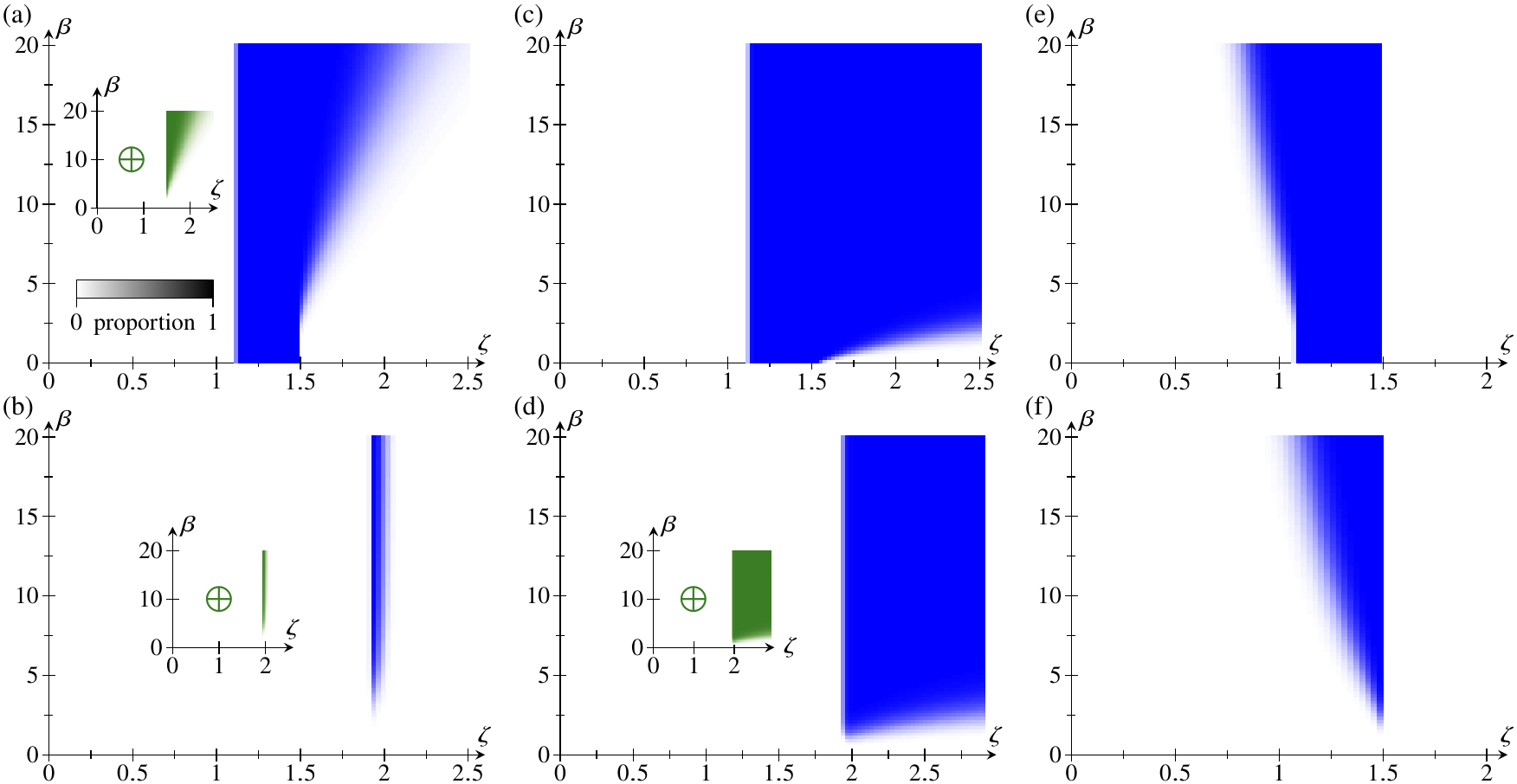}
\caption{Effect of the parameters $\iota,\mu,\kappa$ on the stability and permanence of coexistence. Numerical stability and permanence results for Eqs.~\eqref{eq:model} if only one of the persister parameters $\iota,\kappa,\mu,\xi,\varpi,\varsigma$ is nonzero: $\iota>0$ [panels (a), (b)], $\mu>0$ [panels (c), (d)], $\kappa>0$ [panels (e), (f)]. The analytical results in Appendix~\ref{appB} confirm some of these results. The cases $\eta/\alpha>\vartheta$ [panels (a), (c), (e)] and $\eta/\alpha<\vartheta$ [panels (b), (d), (f)] lead to qualitatively different diagrams. The color of each point in the stability diagrams represents the proportion of $N=1000$ random systems for which coexistence is stable or permanent at that point. The insets similarly plot the proportion of systems for which coexistence is stabilized (or becomes permanent), $\oplus$, or destabilized, $\ominus$, compared to the averaged model with stochastic switching only; there is no (de)stabilization where there is no inset. Parameter values: $\alpha=0.8$, $\eta=1.2$, and $\vartheta=1.1$ [panels (a), (c), (e)] or $\vartheta=1.9$ \mbox{[panels (b), (d), (f)]}. The remaining parameters (if not set to zero) were sampled uniformly and independently, constrained by the persister scalings~\eqref{eq:scalings}: $\gamma,\iota,\kappa,\mu,\xi,\varpi,\varsigma\sim\mathcal{U}[\varepsilon,2\varepsilon]$, with $\varepsilon=0.1$, and $\delta\sim\mathcal{U}[0.8,1.6]$.}\label{fig6} 
\end{figure*}

\begin{figure*}
\includegraphics{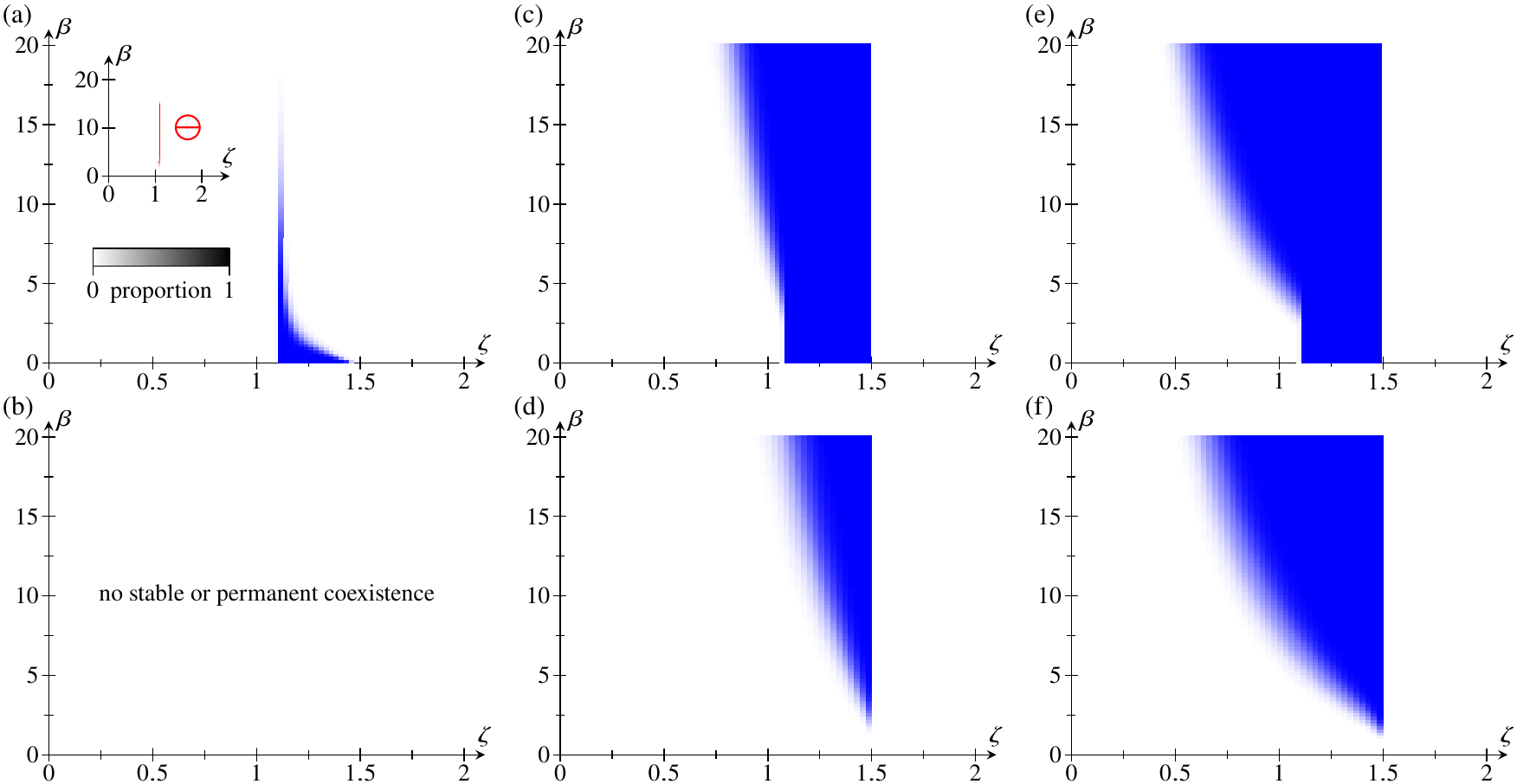}
\caption{Effect of the parameters $\xi,\varpi,\varsigma$ on the stability and permanence of coexistence. Numerical stability and permanence results for Eqs.~\eqref{eq:model} if only one of the persister parameters $\iota,\kappa,\mu,\xi,\varpi,\varsigma$ is nonzero: $\xi>0$ [panels (a), (b)], $\varpi>0$ \mbox{[panels (c), (d)]}, $\varsigma>0$ [panels (e), (f)]. Panels (a), (c), (e) have $\eta/\alpha>\vartheta$, while panels (b), (d), (f) have $\eta/\alpha<\vartheta$. Plots are analogous to and parameter values are equal to those in \textwholefigref{fig6}.}\label{fig7} 
\end{figure*}

The setup for our numerical calculations is as follows: we fix a population of bacteria and competitors by fixing the competition parameters $\alpha,\eta,\vartheta$, and compute stability diagrams in the $(\zeta,\beta)$ plane for random choices of the remaining model parameters $\gamma,\delta,\iota,\kappa,\mu,\xi,\varpi,\varsigma$, which we constrain to satisfy the persister scalings~\eqref{eq:scalings}. 
We report these numerical results by plotting, for each point in the $(\zeta,\beta)$ diagram, stability statistics (such as the proportion of random systems in which coexistence is stable or permanent at this point), but it is important to note that these statistics, while providing a convenient way of visualizing the generic properties of the stability diagrams, have no real biological meaning because the averaged models vary between these random instantiations of the full model, since the bacteria-persister competition parameters are held fixed while the switching parameters and the remaining competition parameters are varied.

\subsubsection{Discussion}
Figure~\ref{fig5} shows numerical stability diagrams of Eqs.~\eqref{eq:model}. The classical results for two-species Lotka--Volterra model (Appendix~\ref{appC}) suggest distinguishing the cases $\eta/\alpha>\vartheta$ and $\eta/\alpha<\vartheta$, which indeed give rise to qualitatively different stability diagrams~\wholefigref{fig5}. 

In both cases however, sufficient levels of responsive switching destabilize coexistence compared to the averaged model with stochastic switching only. This effect is less pronounced for $\vartheta<\eta/\alpha$~\figref{fig5}{a} than for $\vartheta>\eta/\alpha$~\figref{fig5}{b}. This destabilization is that which we already noted when discussing the statistical stability of responsive switching in the previous section: Figure~\figrefp{fig2}{c} has already shown that coexistence is more likely to be stable in models with stochastic switching only than in those with responsive switching.

In the context of the minimal two-species model~\eqref{eq:model}, we can however address the mechanisms underlying this average destabilization, and hence even identify the conditions under which responsive switching can be stabilizing. To this end, we analyze the effect of the P phenotype, i.e. the effect of the parameters $\iota,\kappa,\mu,\xi,\varpi,\varsigma$ that describe its interactions with the A and B phenotypes (Table~\ref{tab0}), on the stability and permanence of coexistence: We plot, in Figs.~\ref{fig6} and \ref{fig7}, stability diagrams of Eqs.~\eqref{eq:model} for the cases in which one and only one among $\iota,\kappa,\mu,\xi,\varpi,\varsigma$ is nonzero. We emphasize that the these parameters are of the same order, as expressed by the persister scalings~\eqref{eq:scalings}, so neglecting the effect represented by one (say, persister growth) compared to another (say, persister-competitor interactions) is not in general ecologically consistent. For this reason, these ``one-parameter'' models are not ``more minimal'' than Eqs.~\eqref{eq:model}, but the ``one-parameter'' diagrams in Figs.~\ref{fig6} and~\ref{fig7} can reveal the individual mathematical effects of these parameters, and our results will show how they can help unravel mechanisms of (de)stabilization.

The full stability diagram of Eqs.~\eqref{eq:model} is not of course a trivial superposition of the stability diagrams in Figs.~\ref{fig6} and~\ref{fig7}. Nonetheless, we recognize features of these diagrams in the stability diagrams of Eqs.~\eqref{eq:model} plotted in Fig.~\ref{fig5}. It is therefore all the more striking that, for the same parameter values for which \textwholefigref{fig5} shows destabilization of coexistence due to responsive switching, the ``one-parameter'' diagrams show the possibility of stabilization of coexistence: Persister-competitor interactions [$\iota>0$, Figs.~\figrefp{fig6}{a),(b}] and persister growth [$\mu>0$, Figs.~\figrefp{fig6}{c),(d}] are stabilizing. Moreover, Figs.~\figrefp{fig7}{a),(b} show a very slight destabilizing effect of competitor-persister interactions ($\xi>0$). Finally, the different types of competition between bacteria and persisters (Table~\ref{tab0}) correspond to the cases $\kappa>0$ [Figs.~\figrefp{fig6}{e),(f}], $\varpi>0$ [Figs.~\figrefp{fig7}{c),(d}], and $\varsigma>0$ [Figs.~\figrefp{fig7}{e),(f}]. They yield very similar stability diagrams and are neither stabilizing nor destabilizing for the generic parameter values used in Figs.~\ref{fig6} and~\ref{fig7}. Of the six persister parameters $\iota,\mu,\kappa,\xi,\varpi,\varsigma$, only the first two have thus, on their own, a strong effect on stability or permanence of coexistence when coupled to responsive switching.

\begin{figure}[b]
\includegraphics{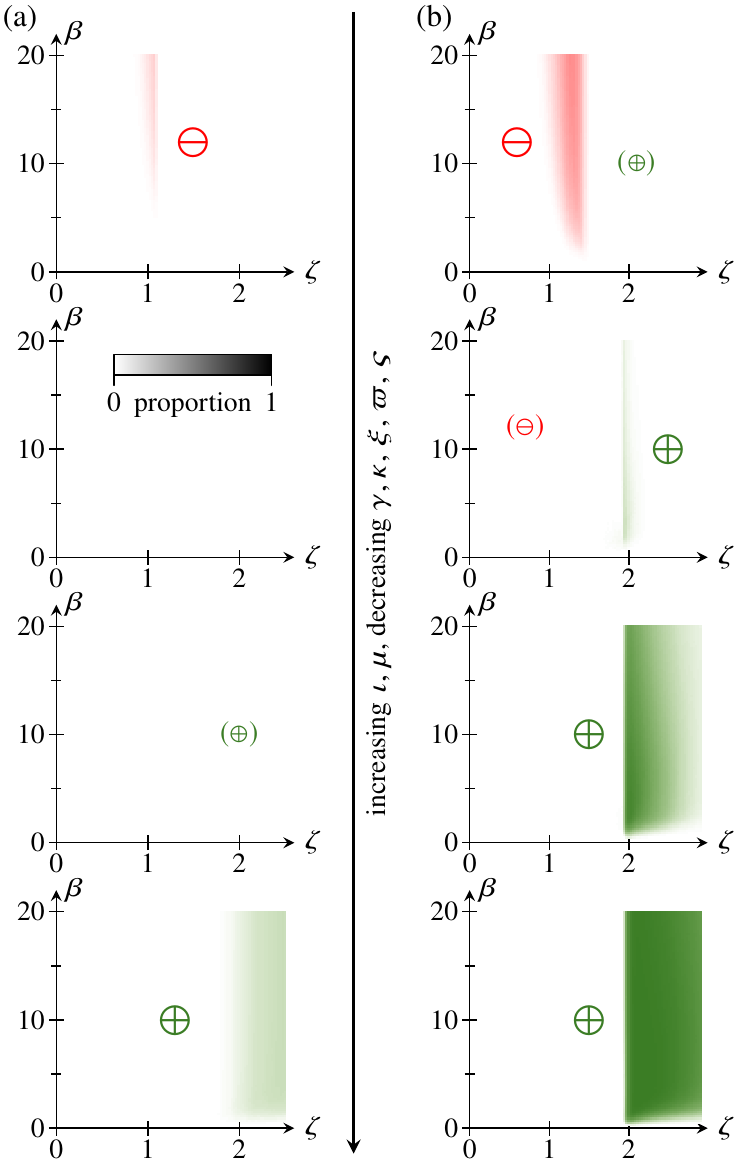}
\caption{Stabilization of coexistence by ``elevated'' persister growth~$\mu$ or persister-competitor interaction $\iota$. Plots, in $(\zeta,\beta)$ space and for (a)~$\eta/\alpha>\vartheta$ and (b) $\eta/\alpha<\vartheta$, of the proportion of $N=1000$ random instantiations of Eqs.~\eqref{eq:model} for which coexistence is stabilized (or becomes permanent), $\oplus$, or destabilized, $\ominus$, compared to the averaged model with stochastic switching only. Proportions are represented by the colors of the points in the diagrams, with symbols in parentheses indicating that the stability or permanence of a proportion of systems too small to be visualizable by the color scheme changes compared to the averaged model. The ``elevation'' of $\iota,\mu$ increases from top to bottom. Parameter values: $\alpha=0.8$, $\eta=1.2$, (a)~$\vartheta=1.1$ or (b) $\vartheta=1.9$, and $\delta\sim\mathcal{U}[0.8,1.6]$, as in Figs.~\ref{fig5}--\ref{fig7}; also, $\gamma,\mu,\xi,\varpi,\varsigma\sim\mathcal{U}[\varepsilon/f,2\varepsilon/f]$, $\mu,\iota\sim\mathcal{U}[\varepsilon f,2\varepsilon f]$, with $\varepsilon=0.1$ and where $f\in\{1\;\text{(top)},1.5,2,2.5\;\text{(bottom)}\}$ ``elevates'' the values of $\iota,\mu$.} \label{fig8}
\end{figure}

This suggests that coexistence can be stabilized by tuning the persister parameters so that the stabilizing parameters $\iota,\mu$ exceed the others, while still remaining ``small'', as required by the persister scalings~\eqref{eq:scalings}. This is confirmed by the numerical results in \textwholefigref{fig8}. The stabilizing effect of ``elevating'' $\iota,\mu$ is less pronounced for ${\vartheta<\eta/\alpha}$~\figref{fig8}{a} than for $\vartheta>\eta/\alpha$~\figref{fig8}{b}, mirroring the weaker destabilizing effect of ``unelevated'' $\iota,\mu$ in the former case.

How do we interpret this stabilization by elevated $\iota,\mu$ in the context of our analysis based on a weakly-competing, persister-like phenotype P~\figref{fig1}{b}? The ``elevated'' growth rate $\mu$ that stabilizes coexistence with responsive switching might correspond to a phenotype that relies on a different food source (and therefore competes weakly with both the normal phenotype B and the competitors A); the growth rate $\mu$ may still be small because the alternative food source may be scarce, or more difficult to metabolize. Similarly, an elevated, but small persister-competitor interaction rate $\iota$ may result from toxins produced by the persisters and acting specifically on the competitors. That the ecological fitness of the phenotypes we have just described should be linked to responsive switching in particular is not surprising. 

The elevated growth and and interaction with competitors of this phenotype are consistent with a persister phenotype, because the corresponding parameters remain ``small'' in the sense of the persister scalings~\eqref{eq:scalings}. While the bacterial persister phenotype~\cite{maisonneuve14,harms16,radzikowski17} is sometimes considered to be a dormant phenotype~\cite{[{Whether persisters are dormant is still a subject of debate in the microbiological literature: See, e.g., }] [{ and }] wood13,*zou22}, small levels of growth or interaction with competitors are mathematically different from absence thereof. Our theoretical results stress this point by suggesting that even such phenotypes can contribute crucially to stability and permanence of coexistence in a microbial community. 

\subsubsection{Analytical results}
The full model~\eqref{eq:model} does not allow meaningful analytical progress to be made. However, some of the effects of the parameters $\iota,\mu,\kappa$ and hence some of the features in their stability diagrams \wholefigref{fig6} and in the full stability diagram of Eqs.~\eqref{eq:model} in \textwholefigref{fig5}~ can be understood analytically. (We are not aware of similar analytical results for the other persister parameters $\xi,\varpi,\varsigma$.) In Appendix~\ref{appB}, we derive and discuss these analytical results, which extend the classical results for the two-species Lotka--Volterra model~\cite{murray} to the simplest mathematical models of responsive switching. Again, these ``one-parameter'' mathematical models are not ``more minimal'' ecological models than Eqs.~\eqref{eq:model}, but they are sufficiently simple to allow analytical understanding of the numerical results that have led us to identifying the stabilizing phenotype discussed above. We shall postpone the detailed discussion of these analytical results to Appendix~\ref{appB}, but emphasize three of their features here.

First, the exact calculations show that there are regions of parameter space in which all steady states of the ``one-parameter'' models involving $\iota,\mu,\kappa$ are unstable, but in which coexistence is still permanent. This emphasizes the importance of non-steady-state attractors for permanence of coexistence and stabilization of coexistence by responsive switching for these models and stresses how phenotypic switching increases the complexity of the dynamics: no limit cycles (and no more complex attractors) arise in classical two-species Lotka--Volterra competition model~\cite{murray}. Thus analysis of the linear stability of the steady states~(Ref.~\cite{murray} and Appendix~\ref{appC}) provides a complete biological picture of the community for the average model, but does not yield a similarly complete picture of a two-species community with phenotypic variation.

Second, these results and further calculations in Appendix~\ref{appC} show that stochastic switching on its own does not affect the stability of coexistence in these models. Responsive switching can thus be stabilizing even when stochastic switching has no effect on stability. This statement can be extended numerically to the other ``one-parameter'' models and permanence of coexistence. Indeed, the ``one-parameter'' (de)stabilization diagrams comparing the stability or permanence of coexistence in the full model~\eqref{eq:model} and in its average without phenotypic variation (not shown) are identical to those comparing Eqs.~\eqref{eq:model} and its average with stochastic switching only (Figs.~\ref{fig6} and~\ref{fig7}, insets), so stochastic switching on its own does not even affect the permanence of coexistence in these ``one-parameter'' models. However, these results do not carry over to the full model~\eqref{eq:model}: the (de)stabilization diagrams for the comparison of Eqs.~\eqref{eq:model} and its average without phenotypic variation (not shown) are different from those in Figs.~\ref{fig5} and \ref{fig8}. From these numerical results, we can in fact conclude that elevated $\iota,\mu$ as in~\textwholefigref{fig8} also stabilize the averaged model with stochastic switching only compared to that without phenotypic variation.

Finally, the analytical results stress that even small rates of stochastic switching, ${\beta=O(\varepsilon)}$, affect stability, but only if the B and A phenotypes are similar enough, as expressed by the scaling requirement $\eta-\alpha\vartheta\lesssim O\bigl(\varepsilon^2\bigr)$. Our choice of allowing, in our numerical calculations, ``large'' values $\beta=O(1)$, which are not consistent with the persister scalings~\eqref{eq:scalings}, is therefore one of numerical convenience that does not affect the biological validity of our analysis. The existence of such a requirement is not unexpected: an asymptotically rare phenotype P should not change the stability of communities in which one of the A and B phenotypes strongly dominates the other, i.e. $\eta-\alpha\vartheta=O(1)$. This case of strong dominance is perhaps less relevant to our analysis, because one expects such strong dominance to lead to one species simply outcompeting the other. By contrast, weak dominance of one species may build up the evolutionary pressure that could lead to the emergence of stabilizing features such as the responsive switching analyzed here. In this context, our earlier results in Sec.~\ref{sec:model} and in Ref.~\cite{haas20}, showing that the effect of a rare phenotype is amplified as the number of species increases, might be relatable to a combinatorial and statistical increase of species interactions lacking such strong dominance.

\section{Conclusion}
In this paper, we have analyzed the ecological implications of phenotypic variation and, in particular, responsive phenotypic switching in the context of microbial communities in which the species have a rare, slowly growing and weakly competing, persister-like phenotype. We have shown that the statistical properties of stability and permanence of coexistence are different in models with responsive phenotypic switching and in corresponding averaged models with stochastic phenotypic switching only, and we have emphasized the need to define these averaged models carefully. Although this statistical analysis showed that coexistence is less likely to be stable on average with responsive switching than with stochastic switching only, numerical results for a minimal two-species model revealed those parameters (and hence the ecological conditions) in combination with which responsive switching can stabilize two-species communities. Exact results for simplified mathematical models showed further that responsive switching can stabilize coexistence even when stochastic switching on its own does not affect the stability of the community. Additionally, our numerical results emphasized the importance of non-steady-state attractors for coexistence of all species, even for the simplified two-species models, but analytical understanding of these attractors is still lacking. 

All of our results thus hint at a complex relationship between ecological stability and phenotypic variation, of which this and previous studies~\cite{maynard19,haas20} have only scratched the surface. Focusing, as we did in this paper, on a minimal two-phenotype structure with the asymptotic separation afforded by a rare, persister-like phenotype has enabled more detailed and mechanistic understanding. However, extending these results to many-species systems and more general phenotypic structures remains an important challenge for future work. Reference~\cite{maynard19} has already begun to address these questions in the context of stochastic phenotypic switching, but to understand which phenotypic interactions stabilize or destabilize many-species systems, further analytical progress, for simple models of communities of more than two species, will be crucial to guide statistical exploration of many-species systems due to the quadratic increases of the number of interaction parameters with the number of species. Such analytical progress may benefit from the information-theoretic approaches used in evolutionary population dynamics~\cite{rivoire16,zhang20}. Moreover, now that a range of stability-affecting structures of microbial communities, from phenotypic variation (this work and Refs.~\cite{maynard19,haas20}) and species modularity~\cite{grilli16} to higher-order interactions~\cite{grilli17} or resource and signaling dynamics~\cite{butler18,kelsic15}, has been identified, further theoretical work will need to analyze the interplay of such effects and ask: Which of these effects (if any) dominates the overturning of May's stability paradigm~\cite{may72} in actual ecological communities?

The deterministic equations in this paper describe well-mixed populations, with the deterministic parameter that we call the stochastic switching rate representing a mean over individual, stochastic events of phenotypic switching. For a rare phenotype such as the persister phenotype, this approximation may break down because persister abundances may be small not only in relative terms~\cite{hofsteenge13}, but also in absolute terms~\cite{[{For example, }] [{ estimate that bacteria in soil communities interact with at most $\smash{O\bigl(10^3\bigr)}$ neighbors, so the low relative persister abundances~\cite{hofsteenge13} mean that, even if the community counts many fewer than the $\smash{O\bigl(10^4\bigr)}$ species typical for soil communities (\emph{vide ibid.}), interactions with persisters are rare, stochastic events in such communities.}] raynaud14}. Future work will need to address more generally such stochastic effects in simulations resolving individual rare stochastic events to disentangle their contributions to the stability of ecological communities and the contributions of their mean deterministic behavior, as studied in this paper.

Meanwhile, our work has predicted that the combination of responsive switching to a rare persister-like phenotype and its slow growth and weak interaction with competitors can favor coexistence in microbial communities. This ecological prediction remains to be tested experimentally and the weak growth, interaction with competitors, and responsive switching rates of persisters remain to be quantified in microbial communities. Some experimentally support for the importance of this parameter combination might already be given by the recent work~\cite{oliveira22} showing that a non-reproducing subpopulation of \emph{P. aeruginosa} releases bacteriocins while migrating up antibiotic gradients. More importantly, these experiments emphasize the role of spatial dynamics in the response to antibiotic stress. Future theoretical work will therefore also need to add such spatial dynamics into the models that we have analyzed here.

More generally, identifying experimental systems in which the role of responsive phenotypic switching can be addressed experimentally is another key challenge for future work. In particular, experimental systems for these questions must allow disentangling the effect of responsive phenotypic switching, say, and the effect of the change of the equilibrium population that results from ``turning off'' responsive switching. This is precisely the problem that we highlighted when discussing reduced models and solved, theoretically, by defining averaged models for stability comparison. By introducing these averaged models, we have been able to make complex, albeit mathematical ecological perturbations. It is this ability that explains the power of theoretical approaches for guiding experimental exploration of problems in ecology. For this reason, the importance of these theoretical approaches mirrors the outstanding importance of microbial communities~\cite{stubbendieck16}---and hence that of understanding the biological and mathematical effects that stabilize or destabilize them---for the world that surrounds us.

\begin{acknowledgments}
We thank B. G. Stokell for a conversation on confidence intervals. This work was supported in part by the Schlumberger Chair Fund. R.E.G. was supported by Established Career Fellowship No. EP/M017982/1 from the Engineering and Physical Sciences Research Council and Grant No. 7523 from the Marine Microbiology Initiative of the Gordon and Betty Moore Foundation. N.M.O. was supported by Discovery Fellowship No. BB/T009098/1 from the Biotechnology and Biological Sciences Research Council, a Wellcome Trust Interdisciplinary Fellowship, and, at earlier stages of this work, a Herchel Smith Postdoctoral Research Fellowship. P.A.H. was supported by the Max Planck Society, and, at earlier stages of this work, by a Nevile Research Fellowship from Magdalene College, Cambridge and a Hooke Research Fellowship at the Mathematical Institute, University of Oxford.
\end{acknowledgments}

\appendix
\section{\uppercase{Details of the statistical analysis of Eqs.}~(\ref{eq:fullmodel})}\label{appA}
This Appendix provides details of the statistical analysis of Eqs.~\eqref{eq:fullmodel}: It discusses the conditions under which the dynamics of Eqs.~\eqref{eq:fullmodel} are bounded, explains the random sampling of systems, provides the calculations of the Jacobians of the equilibria of Eqs.~\eqref{eq:fullmodel} and the asymptotic calculations in the limit $\varepsilon\ll 1$, and defines an averaged model without phenotypic variation for completeness.
\subsection{Bounded Dynamics of Eqs.~(\ref{eq:fullmodel})}
We begin by establishing sufficient conditions for the dynamics of Eqs.~\eqref{eq:fullmodel} to be bounded, and therefore to be realistic biologically. 

We assume that, for each $n\in\{1,2,\dots,N\}$, the B phenotype of species $n$ satisfies $C_{nn}\not=0$ or $b_n<\varepsilon k_n$, and that its P phenotype satisfies $F_{nn}\not=0$ or $p_n<\ell_n$. Then, from Eqs.~\eqref{eq:fullmodel},
\begin{align}
\dot{B}_n+\dot{P}_n&\leq B_n\left[\left(b_n-\varepsilon k_n\right)-C_{nn}B_n\right]\nonumber\\
&\qquad+\varepsilon P_n\left[(p_n-\ell_n)-F_{nn}C_n\right].
\end{align}
Consider first the generic case in which $C_{nn},F_{nn}\not=0$. Then $\dot{B}_n+\dot{P}_n<0$ if $B_n>b_n/C_{nn}$ and $P_n>p_n/F_{nn}$. Moreover, if $P_n<p_n/F_{nn}$ or $B_n<b_n/C_{nn}$, then from Eqs.~\eqref{eq:fullmodel},
\begin{subequations}
\begin{align}
\dot{B}_n&\leq B_n\left(b_n-\varepsilon k_n\right)-C_{nn}B_n^2+\ell_nP_n\nonumber\\
&<B_n\left(b_n-\varepsilon k_n\right)-C_{nn}B_n^2+\ell_np_n/F_{nn},\\
\dot{P}_n&\leq \varepsilon\left[P_n(p_n-\ell_n)-F_{nn}C_n^2+k_nB_n\right]\nonumber\\
&<\varepsilon\left[P_n(p_n-\ell_n)-F_{nn}C_n^2+k_nb_n/C_{nn}\right].
\end{align}
\end{subequations}
It follows that $\dot{B}_n<0$ if $P_n<p_n/F_{nn}$ and $B_n>B_n^{\min}$, for some $B_n^{\min}>0$. We will not need an explicit expression for $B_n^{\min}$, but can assume without loss of generality that ${B_n^{\min}>b_n/C_{nn}}$; similarly, $\dot{P}_n<0$ if $B_n<b_n/C_{nn}$ and $P_n>P_n^{\min}$, for some $P_n^{\min}>p_n/F_{nn}$. These bounds are independent of the other species, and thus show that, irrespective of the initial conditions, the dynamics of $(B_n,P_n)$ will enter the bounded region in \textwholefigref{fig9}, and remain in that region.

\begin{figure}[b]
\includegraphics{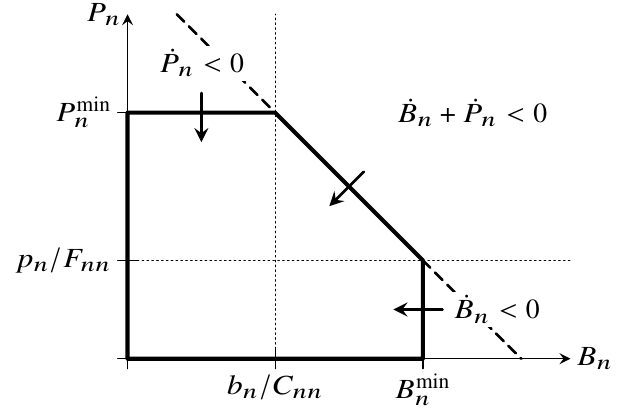}
\caption{Bounded dynamics of Eqs.~\eqref{eq:fullmodel}. For any nonnegative initial conditions, $(B_n,P_n)$ will eventually enter the pentagonal bounding region (thick lines), and remain in that region for all times. If $C_{nn}=0$ and $b_n<\varepsilon k_n$ or $F_{nn}=0$ and $p_n<\ell_n$, the sides of the bounding region are modified (dashed lines).}\label{fig9} 
\end{figure}

A similar argument shows that the dynamics are bounded if $C_{nn}=0$, but $b_n<\varepsilon k_n$ or $F_{nn}=0$, but $p_n<\ell_n$~\wholefigref{fig9}.

On identifying parameters appropriately, these conditions also provide sufficient conditions for the dynamics of the two-species model~\eqref{eq:model}, or the simplified models~\eqref{eq:model2}, \eqref{eq:model2b}, \eqref{eq:model2c}, to be bounded.

\subsection{Sampling of Random Systems}
We follow the approach that we introduced in Ref.~\cite{haas20} to sample random instantiations of Eqs.~\eqref{eq:fullmodel} and their coexistence equilibria. 

In more detail, we choose the competition strengths $\mat{C},\mat{D},\mat{E},\mat{F}$, the stochastic switching rates $\vec{k},\vec{\ell}$, the responsive switching rates $\mat{R},\mat{S}$, and the equilibria $\vec{B_\ast},\vec{P_\ast}$ independently from the uniform $\mathcal{U}[0,1]$ distribution. This leaves linear equations to be solved for the remaining parameters $\vec{b},\vec{p}$; to ensure that they are nonnegative, we choose a common random scaling for the switching rates $\vec{k},\vec{\ell}$ and $\mat{R},\mat{S}$.

Using the asymptotic solution derived below, the model parameters can be sampled directly, i.e. we can sample $\vec{b},\vec{p}$ randomly, and use that solution to compute $\vec{B_\ast},\vec{P_\ast}$. To avoid a breakdown of asymptoticity, we sample parameters in the interval $[\varepsilon^{1/4},1]$ rather than $[0,1]$. We also discard those sampled systems for which any component of the right-hand sides of Eqs.~\eqref{eq:fullmodel} is greater than $\varepsilon$. Moreover, we ensure feasibility of $\vec{B_\ast}$ by sampling $\vec{b}$ as a linear combination of the (normalized) columns of $\mat{C}$.

Finally, to sample exact equilibria (again indirectly) for ${\varepsilon\ll 1}$, we adapt our previous strategy by imposing ${\vec{P}=\varepsilon\vec{B}(\vec{k}+\mat{R}\cdot\vec{B})}$, up to a random $O\bigl(\varepsilon^2\bigr)$ correction, to ensure that $\vec{p}=O(1)$.
\subsection{Jacobian of Equilibria of Eqs.~(\ref{eq:fullmodel})}
The Jacobian of an equilibrium $\mathcalbf{E}=(\vec{B_\ast},\vec{P_\ast})$ of Eqs.~\eqref{eq:fullmodel} is
\begin{align}
\mat{J_\ast}=\left(\begin{array}{c|c}
\mat{J_1}&\mat{J_2}\\
\hline
\mat{J_3}&\mat{J_4}\\
\end{array}\right),\label{eq:Japp}
\end{align}
wherein
\begin{subequations}
\begin{align}
\mat{J_1}&=\bigl(\vec{b}-\mat{C}\cdot\vec{B_\ast}-\varepsilon\mat{D}\cdot\vec{P_\ast}\bigr)\mat{I}-\vec{B_\ast}\mat{C}\nonumber\\
&\hspace{18mm}-\varepsilon\bigl(\vec{k}+\mat{R}\cdot\vec{B_\ast}+\mat{S}\cdot\vec{P_\ast}\bigr)\mat{I}-\varepsilon\vec{B_\ast}\mat{R},\\
\mat{J_2}&=-\varepsilon\vec{B_\ast}\mat{D}+\vec{\ell}\mat{I}-\varepsilon\vec{B_\ast}\mat{S},\\
\mat{J_3}&=-\varepsilon\vec{P_\ast}\mat{E}+\varepsilon\bigl(\vec{k}+\mat{R}\cdot\vec{B_\ast}+\mat{S}\cdot\vec{P_\ast}\bigr)\mat{I}+\varepsilon\vec{B_\ast}\mat{R},\\
\mat{J_4}&=\varepsilon\bigl(\vec{c}-\mat{E}\cdot\vec{B_\ast}-\varepsilon\mat{F}\cdot\vec{P_\ast}\bigr)\mat{I}-\varepsilon\vec{P_\ast}\mat{F}-\vec{\ell}\mat{I}+\varepsilon\vec{B_\ast}\mat{S},
\end{align}
\end{subequations}
with $\mat{I}$ being the identity. Using $\vec{k'}=\vec{k}+\mat{R}\cdot\vec{B_\ast}+\mat{S}\cdot\vec{P_\ast}$, Eq.~\eqref{eq:JK} follows immediately.

\subsection{Asymptotic Coexistence Equilibria of Eqs.~(\ref{eq:fullmodel}) for $\boldsymbol{\varepsilon\ll 1}$}
As announced in the main text, we seek an expansion of the coexistence state $(\vec{B_\ast},\vec{P_\ast})$ in powers of $\varepsilon\ll 1$ by writing
\begin{align}
\vec{B_\ast}=\vec{B_0}+\varepsilon\vec{B_1}+O\bigl(\varepsilon^2\bigr), &&\vec{P_\ast}=\vec{P_0}+\varepsilon\vec{P_1}+O\bigl(\varepsilon^2\bigr).
\end{align}
On expanding Eqs.~\eqref{eq:fullmodel}, we find
\begin{subequations}
\begin{align}
\vec{0}&=\vec{B_0}\bigl(\vec{b}-\mat{C}\cdot\vec{B_0}\bigr)+\vec{\ell P_0}+\varepsilon\bigl[\vec{B_1}\bigl(\vec{b}-\mat{C}\cdot\vec{B_0}\bigr)-\vec{B_0}\mat{C}\cdot\vec{B_1}\nonumber\\
&\hspace{4mm}-\vec{B_0}\mat{D}\cdot\vec{P_0}+\vec{\ell P_1}-\vec{B_0}\bigl(\vec{k}+\mat{R}\cdot\vec{B_0}+\mat{S}\cdot\vec{P_0}\bigr)\bigr]+O\bigl(\varepsilon^2\bigr),\\
\vec{0}&=-\vec{\ell P_0}+\varepsilon\bigl[\vec{P_0}\bigl(\vec{p}-\mat{E}\cdot\vec{B_0}-\mat{F}\cdot\vec{P_0}\bigr)-\vec{\ell P_1}\nonumber\\
&\hspace{19mm}+\vec{B_0}\bigl(\vec{k}+\mat{R}\cdot\vec{B_0}+\mat{S}\cdot\vec{P_0}\bigr)\bigr]+O\bigl(\varepsilon^2\bigr).
\end{align}
\end{subequations}
Solving at order $O\bigl(\varepsilon^0\bigr)$, $\vec{P_0}=\vec{0}\Longrightarrow\vec{B_0}=\mat{C}^{-1}\cdot\vec{b}$, unless $\det{\mat{C}}=0$, which we assume not to be the case. Then, at order $O\bigl(\varepsilon^1\bigr)$, $\vec{P_1}=\bigl(\vec{k}+\mat{R}\cdot\vec{B_0}\bigr)\vec{B_0}/\vec{\ell}$, and hence $\vec{B_0}\mat{C}\cdot\vec{B_1}=\vec{0}$, which implies $\vec{B_1}=\vec{0}$, as claimed in the main text. On substituting these results into Eq.~\eqref{eq:Japp}, we find 
\begin{align}
\mat{J_\ast}&=\underbrace{\left(\begin{array}{c|r}
-\vec{B_0}\mat{C}&\vec{\ell}\mat{I}\\
\hline
\mat{O}&-\vec{\ell}\mat{I}\end{array}\right)+\varepsilon\left(\begin{array}{r|c}
-(\vec{k}+\mat{R}\cdot\vec{B_0})\mat{I}&-\vec{B_0}\mat{D}\\
\hline
(\vec{k}+\mat{R}\cdot\vec{B_0})\mat{I}&(\vec{p}-\mat{E}\cdot\vec{B_0})\mat{I}\\
\end{array}\right)}_{\mat{K_\ast}\,+\,O\left(\!\varepsilon^2\right)}\nonumber\\
&\hspace{8mm}+\varepsilon\left(\begin{array}{r|r}
-\vec{B_0}\mat{R}&-\vec{B_0}\mat{S}\\
\hline
\vec{B_0}\mat{R}&\vec{B_0}\mat{S}\\
\end{array}\right)+O\bigl(\varepsilon^2\bigr),\label{eq:Ja}
\end{align}
in which $\mat{O}$ is the zero matrix and $\mat{I}$ is again the identity. In the first two terms of Eq.~\eqref{eq:Ja} and up to smaller corrections, we recognize the Jacobian $\mat{K_\ast}$ of the corresponding model with stochastic switching only, because $\vec{k'}=\vec{k}+\mat{R}\cdot\vec{B_0}+O(\varepsilon)$. We use these expansions to sample coexistence equilibria of random systems with $\varepsilon\ll 1$ directly, and to determine their stability numerically.

\subsection{Permanent Coexistence in Eqs.~(\ref{eq:fullmodel})}
To determine whether coexistence is permanent where a coexistence equilibrium is unstable, we perturb the system away from that unstable equilibrium, and evolve it numerically using the stiff solver \texttt{ode15s} of \textsc{Matlab} (The MathWorks, Inc.). During the numerical solution, we repeatedly test for convergence to a stable equilibrium or a stable limit cycle. In particular, limit cycles and their stability are determined using the shooting method described in Ref.~\cite{parker89}.

There is one particular numerical difficulty associated with this: the numerical integration generally fails, even at stringent tolerances, if the dynamics approach an attractor intersecting one of the planes $B_n=0$ or $P_n=0$. Some species abundances become arbitrarily small (while remaining nonzero at finite times) on such a trajectory. Since such small abundances lack biological meaning, we simply remove these species from the system. In more detail, we declare species $n$ to go extinct at time $t_n=\operatorname{argmin}{\{\max{\{B_n(t),P_n(t)\}}<\epsilon\}}$, where $\epsilon\ll 1$ is fixed, and integrate the system constituted by the remaining species for $t>t_n$. We choose $\epsilon=10^{-6}$ and find empirically that this reduces considerably the proportion of systems for which the numerical integration would otherwise fail.

\subsection{Averaged Model without Phenotypic Variation}
In the same way as we have associated to Eqs.~\eqref{eq:fullmodel} an averaged model with stochastic switching only [Eqs.~\eqref{eq:fullmodels}], we can also associate to Eqs.~\eqref{eq:fullmodel} a model without phenotypic variation,
\begin{align}
\vec{\dot{B}'}=\vec{B'}\left(\vec{b'}-\mat{C'}\cdot\vec{B'}\right),\label{eq:modelav}
\end{align}
with a unique coexistence equilibrium $\vec{B'_\ast}=\mat{C'}^{-1}\cdot\vec{b'}$. As we have noted in Ref.~\cite{haas20}, this equilibrium is consistent with an equilibrium $\mathcalbf{E}=(\vec{B_\ast},\vec{P_\ast})$ of Eqs.~\eqref{eq:fullmodel} or \eqref{eq:fullmodels} if and only if the population sizes, births, and competition at equilibrium are equal, i.e.
\begin{subequations}\label{eq:cc}
\begin{align}
\vec{B'_\ast}&=\vec{B_\ast}+\vec{P_\ast}, \qquad\vec{b'B'_\ast}=\vec{b B_\ast}+\varepsilon\vec{pP_\ast}\\
\vec{B'_\ast}\mat{C'}\vec{B'_\ast}&=\vec{B_\ast}\mat{C}\vec{B_\ast}+\varepsilon\left(\vec{B_\ast}\mat{D}\vec{P_\ast}+\vec{P_\ast}\mat{E}\vec{B_\ast}+\vec{P_\ast}\mat{F}\vec{B_\ast}\right).
\end{align}
\end{subequations}
These conditions uniquely determine the effective parameters $\vec{b'}$ and $\mat{C'}$ of the averaged model and its equilibrium $\vec{B_\ast'}$. Again, the corresponding reduced model would inherit its birth rates $\vec{b}$ and competition parameters $\mat{C}$ from Eqs.~\eqref{eq:fullmodel}. Stability differences between this reduced model and Eqs.~\eqref{eq:modelav} need not therefore result from the phenotypic substructure, but could also stem from these parameter differences, i.e. evolution of the B phenotypes. We emphasize that this averaging does not imply that the dynamics of the sum $\vec{B}+\vec{P}$ resulting from Eqs.~\eqref{eq:fullmodel} are of the averaged form~\eqref{eq:modelav}, and so the dynamics of Eqs.~\eqref{eq:fullmodel} and Eqs.~\eqref{eq:modelav} are in general different away from $\mathcalbf{E}$.

\section{\uppercase{Analysis of Simplified Models}}\label{appB}
In this Appendix, we derive analytical results for three simplified models to establish some of the features seen numerically in the stability diagrams in \textwholefigref{fig6} and hence \textwholefigref{fig5}.

The reason for introducing these simplified models is that the full two-species model~\eqref{eq:model} does not easily lend itself to analytical progress. Indeed, on computing a Gr\"obner basis~\cite{groebner} for the steady-state version of Eqs.~\eqref{eq:model} using \textsc{Mathematica} (Wolfram, Inc.), we find that even just computing the coexistence equilibria of Eqs.~\eqref{eq:model} requires solving a quartic equation. To enable some analytical progress, we therefore introduce a simplified version of Eqs.~\eqref{eq:model},\begin{subequations}\label{eq:model2}
\begin{align}
\dot{B}&=B(1-\alpha A-B)-\beta AB-\gamma B+\delta P,\\
\dot{P}&=\beta A B+\gamma B-\delta P,\\
\dot{A}&=A(\zeta-\eta A-\vartheta B-\iota P).
\end{align}
\end{subequations}
Compared to the full system~\eqref{eq:model}, all but one of the competition terms involving the persisters have been removed in this system. We stress that Eqs.~\eqref{eq:model2} are not the asymptotic limit of Eqs.~\eqref{eq:model} for slowly growing and weakly competing persisters. Again, Eqs.~\eqref{eq:model2} are thus not an ecological model, but a mathematical model: including this one persister competition term, while leaving the system amenable to analytical progress, introduces nontrivial behavior.

We will also consider two other simplified models that are obtained similarly:
\begin{subequations}\label{eq:model2b}
\begin{align}
\dot{B}&=B(1-\alpha A-B)-\beta AB-\gamma B+\delta P,\\
\dot{P}&=\mu P+\beta A B+\gamma B-\delta P,\\
\dot{A}&=A(\zeta-\eta A-\vartheta B),
\end{align}
\end{subequations}
and
\begin{subequations}\label{eq:model2c}
\begin{align}
\dot{B}&=B(1-\alpha A-B-\kappa P)-\beta AB-\gamma B+\delta P,\\
\dot{P}&=\beta A B+\gamma B-\delta P,\\
\dot{A}&=A(\zeta-\eta A-\vartheta B).
\end{align}
\end{subequations}
The three simplified models \eqref{eq:model2}, \eqref{eq:model2b}, \eqref{eq:model2c} thus correspond to allowing exactly one of $\iota,\mu,\kappa$ to be nonzero. The equilibria of the analogous simplified models corresponding to the remaining logistic parameters involving persisters, viz. $\xi,\varpi,\varsigma$, are determined by equations that are at least cubic, and hence do not allow much analytical progress.

In what follows, we derive exact stability, feasibility, and permanence results for each of the models \eqref{eq:model2}, \eqref{eq:model2b}, \eqref{eq:model2c}.

\subsection{Stability and permanence of coexistence in Eqs.~(\ref{eq:model2})}
The simplified model \eqref{eq:model2} has five steady states: three trivial steady states,
\begin{subequations}
\begin{align}
&\mathcalbf{O}=(0,0,0),&&\mathcalbf{A}=(0,0,\zeta/\eta),&&\mathcalbf{B}=(1,\gamma/\delta,0),\label{eq:AB}
\end{align}
and, if $\beta\iota\not=0$, two coexistence equilibria,
\begin{align}
&\mathcalbf{C}_+=(b_+,p_+,a_+),&&\mathcalbf{C}_-=(b_-,p_-,a_-),\label{eq:CC}
\end{align}
\end{subequations}
where \begin{subequations}\label{eq:coex}
\begin{align}
a_\pm&=\dfrac{1}{2\alpha\beta\iota}\left(-Y\pm\sqrt{Y^2+4\alpha\beta\delta\iota V}\right),\label{eq:coex2a}\\
b_\pm&=\dfrac{1}{2\beta\iota}\left(X\mp\sqrt{X^2-4\beta\delta\iota U}\right),\\
p_\pm&=\dfrac{b_\pm}{\delta}\bigl(\beta a_\pm+\gamma\bigr),\label{eq:ppm}
\end{align}
\end{subequations}
wherein
\begin{subequations}\label{eq:UVXY}
\begin{align}
U&=\alpha\zeta-\eta,&V&=\vartheta-\zeta+\dfrac{\gamma\iota}{\delta},\label{eq:UV}\\
X&=\delta(U+\alpha V)+\iota\beta, &Y&=\delta(U+\alpha V)-\iota\beta,
\end{align}
\end{subequations}
so that $X^2-4\beta\delta\iota U=Y^2+4\alpha\beta\delta\iota V$ and $X=Y+2\beta\iota$. In particular, $X\geqslant Y$. If $\beta=0$ or $\iota=0$, then there is but a single coexistence state $\mathcalbf{C}=(b,p,a)$, with
\begin{align}
&b=\dfrac{U}{W}, &&a=\dfrac{V}{W}, &&p=\dfrac{b}{\delta}\bigl(\beta a+\gamma\bigr),\label{eq:coex2}
\end{align}
wherein $W=U+\alpha V$. From Eqs.~\eqref{eq:ppm} and \eqref{eq:coex2}, it is immediate that $p_\pm,p>0$ if $b_\pm,b>0$ and $a_\pm,a>0$, and so we need not consider $p_\pm,p$ to determine feasibility.

\begin{table}[t]
\caption{Feasibility of the coexistence equilibria $\mathcalbf{C}_\pm$ of Eqs.~\eqref{eq:model2}: discussion of the sixteen possible sign combinations of $U,V,X,Y$, defined in Eqs.~\eqref{eq:UVXY}. For some combinations, the resulting signs of $a_\pm$ or $b_\pm$, defined in Eqs.~\eqref{eq:coex}, are given, and the symbol $\mathbb{C}$ is used for some combinations to indicate that the resulting values of $a_\pm$ or $b_\pm$ may have nonzero imaginary parts. Some sign combinations, marked~\# in the final column, are inconsistent with definitions~\eqref{eq:UVXY}; for other sign combinations, this column gives the corresponding feasibility results.} \label{tab1}
\begin{ruledtabular}
\begin{tabular}{CCCCCCl}
U&V&X&Y&a_\pm&b_\pm&\\
\hline
+&+&+&+&\pm&+&only $\mathcalbf{C}_+$ is feasible\\
+&+&+&-&\pm&+&only $\mathcalbf{C}_+$ is feasible\\
+&+&-&+&&&\# ($X<0,Y>0\Rightarrow X<Y$)\\
+&-&+&+&-/\mathbb{C}&&$\mathcalbf{C}_\pm$ are not feasible\\
-&+&+&+&\pm&\mp&$\mathcalbf{C}_\pm$ are not feasible\\
+&+&-&-&&&\# ($U,V>0\Rightarrow X>0$)\\
+&-&+&-&+/\mathbb{C}&+/\mathbb{C}&$\mathcalbf{C}_\pm$ can both be feasible\\
-&+&+&-&\pm&\mp&$\mathcalbf{C}_\pm$ are not feasible\\
+&-&-&+&&&\# ($X<0,Y>0\Rightarrow X<Y$)\\
-&+&-&+&&&\# ($X<0,Y>0\Rightarrow X<Y$)\\
-&-&+&+&&&\# ($U,V<0\Rightarrow Y<0$)\\
+&-&-&-&&-/\mathbb{C}&$\mathcalbf{C}_\pm$ are not feasible\\
-&+&-&-&\pm&\mp&$\mathcalbf{C}_\pm$ are not feasible\\
-&-&+&-&+&\mp&only $\mathcalbf{C}_-$ is feasible\\
-&-&-&+&&&\# ($X<0,Y>0\Rightarrow X<Y$)\\
-&-&-&-&+&\mp&only $\mathcalbf{C}_-$ is feasible\\
\end{tabular}
\end{ruledtabular}
\end{table}

\subsubsection{Feasibility of the coexistence equilibria} 
We now ask whether the coexistence equilibria are feasible. In Table~\ref{tab1}, we analyze the possible sign combinations of the variables $U,V,X,Y$ defined in Eqs.~\eqref{eq:UVXY} and that appear in the coordinates of the equilibria in Eqs.~\eqref{eq:coex}. This shows that only $\mathcalbf{C}_+$ is feasible if $U,V>0$, while only $\mathcalbf{C}_-$ is feasible if $U,V<0$. Neither coexistence state is feasible if $U<0$, $V>0$, but it is possible for both coexistence states to be feasible if $U>0$, $V<0$ provided that $X>0$, $Y<0$ and that $\mathcalbf{C}_\pm$ are real (Table~\ref{tab1}). These conditions reduce to
\begin{align}
\beta\iota>\delta|U+\alpha V|\quad\text{and}\quad\bigl[\delta(U+\alpha V)+\beta\iota\bigr]^2>4\beta\delta\iota U.\label{eq:CCcond}
\end{align}
The second, quadratic condition implies that
\begin{align}
\beta\iota<\delta\left(\sqrt{U}-\sqrt{-\alpha V}\right)^2\quad\text{or}\quad\beta\iota>\delta\left(\sqrt{U}+\sqrt{-\alpha V}\right)^2. 
\end{align}
Since $(x-y)^2\leq\left|x^2-y^2\right|<(x+y)^2$ for all $x,y>0$, the first possibility is not consistent with the first condition in Eqs.~\eqref{eq:CCcond}, while this condition holds in the second case. It follows that both coexistence states are feasible if and only if
\begin{align}
U>0,\quad V<0,\quad\text{and}\quad\beta>\beta_\ast=\dfrac{\delta}{\iota}\left(\sqrt{U}+\sqrt{-\alpha V}\right)^2.\label{eq:coex3}
\end{align}
In particular, $U>0$ and $V<0$ requires
\begin{align}
\zeta>\max{\left\{\vartheta+\dfrac{\gamma\iota}{\delta},\dfrac{\eta}{\alpha}\right\}}. 
\end{align}
Moreover, letting $W=U+\alpha V$ again, the conditions $X>0$, $Y<0$ imply that $\beta_\ast\iota>\delta|W|$. Combining these results yields 
\begin{align}
\beta_\ast>\beta_\ast^{\min}=\left|\dfrac{\delta}{\iota}(\eta-\alpha\vartheta)-\alpha\gamma\right|.\label{eq:betamin}
\end{align}
This additional region of feasibility does not arise if $\beta=0$ or $\iota=0$. Indeed, it is immediate from Eqs.~\eqref{eq:coex2} that coexistence is feasible in that case if and only if $U,V,W=U+\alpha V$ all have the same sign, and hence if and only if $U,V>0$ or $U,V<0$.

\subsubsection{Stability of the coexistence equilibria}
We now turn to the question of stability of the coexistence equilibria. Before discussing the general case $\beta\iota\not=0$, we discuss two special cases with $\beta\iota=0$. The Jacobian evaluated at a coexistence equilibrium $(b,p,a)$ is
\begin{align}
\left(\begin{array}{ccc}
-b-\beta a-\gamma&\delta&-(\alpha+\beta)b\\
\beta a + \gamma&-\delta&\beta b\\
-\vartheta a&-\iota a&-\eta a
\end{array}\right). \label{eq:Jcoex}
\end{align}

\paragraph*{Stability of the coexistence equilibria if $\beta=0$.} In the absence of responsive switching, i.e. if $\beta=0$, and from Eq.~\eqref{eq:Jcoex}, the characteristic polynomial of the Jacobian at $\mathcalbf{C}=(b,p,a)$, defined by Eqs.~\eqref{eq:coex2}, is
\begin{align}
P(\lambda) &= W^2\lambda^3+c_2\lambda^2+c_1\lambda-\delta UVW,\label{eq:cpoly}
\end{align}
wherein
\begin{subequations}
\begin{align}
c_1&=\delta UW+(\gamma+\delta)\eta VW-UVw,\\
c_2&=\bigl[U+\eta V+(\gamma+\delta)W\bigr]W ,
\end{align}
\end{subequations}
with $w=W-\alpha\gamma\iota/\delta$. The Routh--Hurwitz conditions~\cite{murray} imply that $\mathcalbf{C}$ is stable only if $\delta UVW<0$. Recalling that $\mathcalbf{C}$ is feasible if and only if $U,V,W=U+\alpha V$ all have the same sign, it follows that $\mathcalbf{C}$ is stable only if $U,V<0$. Moreover, if $U,V<0$ and hence $W<0$, then $c_2>0$ and $c_1c_2>-\delta UVW^3$; the second inequality is easily checked by direct multiplication, noting that $w<W<0$. The Routh--Hurwitz conditions then imply that $\mathcalbf{C}$ is stable if and only if $U,V<0$.

\paragraph*{Stability of the coexistence equilibria if $\iota=0$.} In the case $\iota=0$, in which the competition dynamics do not involve $P$ directly, we find that the characteristic polynomial of the Jacobian at $\mathcalbf{C}=(b,p,a)$, defined by Eqs.~\eqref{eq:coex2}, still has the form in Eq.~(\ref{eq:cpoly}), with modified coefficients 
\begin{align}
\hat{c}_1&=c_1+\beta V(\eta V-\vartheta U),&\hat{c}_2&=c_2+\beta VW.\label{eq:polycoeff}
\end{align}
Similarly to the case $\beta=0$ discussed above, $\mathcalbf{C}$ is stable if and only if $U,V<0$ and $\hat{c}_1\hat{c}_2>-\delta UVW^3$ by the Routh--Hurwitz conditions~\cite{murray}. Noting that $U,V,W$ are independent of $\beta$, the latter condition can be written as a quadratic in~$\beta$, ${d_2\beta^2+d_1\beta+d_0>0}$. Since $d_0=c_1c_2+\delta UVW^3>0$ for $U,V<0$, this holds for small enough $\beta$. Moreover, if $U,V<0$ and $\eta V-\vartheta U<0$, all the terms in the definitions~\eqref{eq:polycoeff} are positive, so $d_1,d_2>0$ and hence $\hat{c}_1\hat{c}_2>-\delta UVW^3$. If $\eta V>\vartheta U$ however, we find $d_2=V^2W(\eta V-\vartheta U)<0$ and hence $d_2\beta^2+d_1\beta+d_0<0$ for sufficiently large $\beta$. Now $U,V<0\Longleftrightarrow\vartheta<\zeta<\eta/\alpha$, while $\eta V>\vartheta U$ if and only if $\zeta<2\eta\vartheta/(\eta+\alpha\vartheta)\equiv\zeta_\ast$, with $\vartheta<\zeta_\ast<\eta/\alpha$ since ${\alpha\vartheta-\eta=W<0}$. In particular, stable coexistence requires $\vartheta<\eta/\alpha$. If, additionally, $\zeta_\ast<\zeta<\eta/\alpha$, coexistence is stable for all $\beta$, but if $\vartheta<\zeta<\zeta_\ast$, coexistence is only stable for small enough $\beta$.

The condition $d_2<0$ is, if $U,V<0$ and hence $W<0$, equivalent with $\eta a<\vartheta b$. This says that destabilization at large $\beta$ requires the death rate of competitors at steady state due to inter-species competition to exceed that due to intra-species competition (Table~\ref{tab0}).
\paragraph*{Stability of the coexistence equilibria if $\beta\iota\neq0$.} Next, we discuss the stability of $\mathcalbf{C}_\pm=(b_\pm,p_\pm,a_\pm)$, defined in Eqs.~\eqref{eq:coex} for the case $\beta\iota\neq 0$. From Eq.~\eqref{eq:Jcoex}, the characteristic polynomial is 
\begin{align}
P_\pm(\lambda)=\lambda^3+c_2^\pm\lambda^2+c_1^\pm\lambda+c_0^\pm, 
\end{align}
where, in particular and using $W=U+\alpha V$,
\begin{align}
c_0^\pm=a_\pm b_\pm\bigl[-\delta W+\beta\iota(b_\pm-\alpha a_\pm)\bigr]=\mp a_\pm b_\pm\upDelta, \label{eq:c0}
\end{align}
wherein $\upDelta^2=X^2-4\beta\delta\iota U=Y^2+4\alpha\beta\delta\iota V$ and $\upDelta>0$. Hence $c_0^\pm\lessgtr0$, and the Routh--Hurwitz conditions~\cite{murray} imply in particular that $\mathcalbf{C}_+$ is unstable. Further, on comparing to our earlier feasibility results, stable coexistence is only possible at $\mathcalbf{C}_-$ and if $U,V<0$ or conditions~\eqref{eq:coex3} are satisfied. 

Moreover, $c_2^-=b_-+(\beta+\eta)a_-+\gamma+\delta>0$ if $\mathcalbf{C}_-$ is feasible, and hence, by the Routh--Hurwitz conditions, stability is, assuming feasibility, equivalent with $c_1^-c_2^->c_0^-$. We are left to analyze this necessary and sufficient condition.

We begin by noting that a region of instability must arise at large $\beta$ provided that $\iota<\vartheta$. Indeed, using \textsc{Mathematica} to simplify complicated expressions, we find that, for $\beta\gg1$,
\begin{align}
&c_0^-\sim -\delta V,\;c_1^-\sim\delta\left[1+V\left(\dfrac{\vartheta}{\iota}-1\right)\right],\;c_2^-\sim1+\gamma+\delta-\dfrac{V\delta}{\iota},
\end{align}
and hence $c_1^-c_2^--c_0^-\sim v_0+v_1V+v_2V^2$, with, in particular, $v_0=\delta(1+\gamma+\delta)>0$ and $v_2=-(\vartheta-\iota)\smash{(\delta/\iota)^2}$. Now, from the persister scalings~\eqref{eq:scalings}, we expect $\iota<\vartheta$, so that $v_2<0$. Since $v_0>0$, it follows that $\mathcalbf{C}_-$ is unstable at large $\beta$ if and only if $|V|$ is large enough. This condition is different from the one that we obtained above when discussing $\iota=0$, for which we showed that instability must occur at large $\beta$ if $|V|$ is small enough. This emphasizes that the limit $\iota=0$ is singular. 

This asymptotic condition for instability at large $\beta$ is independent of $\alpha,\eta$, and thus of $U$. Hence instability may, but need not occur in the region $U,V<0$. If such a region of instability exists then, because $\mathcalbf{A},\mathcalbf{B}$ are unstable if $U,V<0$, all steady states are in fact unstable in this region. This discussion also shows that instability must occur at large $\beta$ under the conditions described by Eqs.~\eqref{eq:coex3}. 

Coexistence is stable, however, for small $\beta$ if $U,V<0$ and hence $W<0$. Indeed, a straightforward calculation shows that $\mathcalbf{C}_-\rightarrow\mathcalbf{C}$ as $\beta\rightarrow 0$ at fixed $\iota\not=0$, with $\mathcalbf{C}$ defined as in Eqs.~\eqref{eq:coex2}. Since $\mathcalbf{C}$ is stable for $U,V<0$, so is $\mathcalbf{C}_-$ for sufficiently small $\beta$ by continuity.

Moreover, $\mathcalbf{C}_-$ is stable for $V<0$ and $|V|$ sufficiently small. Indeed, notice that $\upDelta=|\beta\iota-\delta U|+O(V)$. Now, if $U<0$, $\beta\iota-\delta U>0$. If $U>0$, feasibility requires $\beta\iota>\delta U+O(V)$ from Eqs.~\eqref{eq:coex3}, and so $\upDelta=\beta\iota-\delta U+O(V)$ in either case. Direct computation then yields $c_1^-c_2^--c_0^-=v_0+O(V)>0$, so $\mathcalbf{C}_-$ is stable, as claimed.

The question whether $\mathcalbf{C}_-$ is stable more generally under the conditions in Eqs.~\eqref{eq:coex3} requires somewhat more effort. First, we discuss the limit in which $\beta,\zeta\gg 1$, considering all other parameters to be $O(1)$ quantities. Moreover, we assume that $r=\beta\iota/\alpha\delta\zeta=O(1)$. With these scalings, $U>0$ and $V<0$, so the feasibility conditions~\eqref{eq:coex3} reduce to $r>4$. We then find
\begin{subequations}
\begin{align}
&c_0^-\sim\delta\zeta\sqrt{1-4r^{-1}},\quad  c_2^-\sim\dfrac{\beta}{2\alpha}\left(1-\sqrt{1-4r^{-1}}\right),\\
&c_1^-\sim\dfrac{\beta\eta}{2\alpha^2}\left(1-\sqrt{1-4r^{-1}}\right)-\dfrac{\delta\zeta}{\alpha\iota}\bigl[\alpha(\vartheta-\iota)+\eta\bigr],
\end{align}
\end{subequations}
and hence $c_1^-c_2^--c_0^->0$ for sufficiently large $\beta$ if and only if $c_1^->0$, which is if and only if
\begin{subequations}
\begin{align}
r\left(1-\sqrt{1-4r^{-1}}\right)>2s,\quad\text{where }s=1+\alpha(\vartheta-\iota)/\eta,
\end{align}
or, equivalently, if and only if
\begin{align}
4<r<r_\ast\equiv\dfrac{s^2}{s-1}\quad\text{and}\quad1<s<2. \label{eq:rast}
\end{align}
\end{subequations}
Assuming that $\vartheta>\iota$ as discussed above, $s>1$. We conclude that there exists a region of parameter space in which coexistence is stable for $\beta,\zeta\gg 1$ under the condition in Eqs.~\eqref{eq:coex3} if and only if $\alpha(\vartheta-\iota)<\eta$. We also note that $\alpha(\vartheta-\iota)<\eta$ is implied by the condition $W<0$. 

Finally, we consider stability near the feasibility boundary $\beta=\beta_\ast$ defined in Eqs.~\eqref{eq:coex3}. By continuity, stability near this boundary follows from stability at the boundary. Now, by definition $\upDelta=0$ and hence, from Eq.~\eqref{eq:c0}, $c_0^-=0$ at this boundary, so stability there is equivalent with $c_1^->0$ since $c_2^->0$. Using $\upDelta=0$, direct calculation shows that
\begin{subequations}
\begin{align}
(2\alpha\beta\iota)^2c_1^-&=[\eta-\alpha(\vartheta-\iota)]\beta(\beta\iota-\delta W)(\beta\iota+\delta W+2\delta\eta)\nonumber\\
&\qquad+\alpha(\eta-\alpha\vartheta)\left[(\beta\iota)^2-(\delta W)^2\right]\nonumber\\
&\qquad+2\alpha^2\delta\beta\iota(\beta\iota+\delta W)\label{eq:c1a} \\
&=[\eta-\alpha(\vartheta-\iota)](\beta\iota-\delta W)[\beta(\beta\iota+\delta W+2\delta\eta)\nonumber\\
&\hspace{42mm}+\alpha(\beta\iota+\delta W)]\nonumber\\
&\qquad+\alpha^2(\beta\iota+\delta W)[(2\delta-\iota)\beta\iota+\delta\iota W]. \label{eq:c1b}
\end{align}
The discussion around Eqs.~\eqref{eq:coex3} implies that ${\beta\iota>|\delta W|}$. Hence $c_1^->0$ if $\eta>\alpha\vartheta$ from Eq.~\eqref{eq:c1a}. If $\iota<\delta$, then $(2\delta-\iota)\beta\iota+\delta\iota W>\iota(\beta\iota+\delta W)>0$, so Eq.~\eqref{eq:c1b} shows that $c_1^->0$ continues to hold if ${\alpha(\vartheta-\iota)<\eta<\alpha\vartheta}$ provided that $\iota<\delta$. We expect this to be true from the persister scalings~\eqref{eq:scalings}. If $\iota>\delta$ however, parameter values such that $c_1^-<0$ can be found numerically (not shown). All of this shows that coexistence is stable if $\eta>\alpha(\vartheta-\iota)$, provided that $\iota<\delta$. Moreover, rearranging Eqs.~\eqref{eq:c1a} or \eqref{eq:c1b} yields
\begin{align}
(2\alpha\beta\iota)^2c_1^-&=\iota^2[\eta-\alpha(\vartheta-\iota)]\beta^3+d\beta^2+\beta\delta W\{\alpha\delta\iota\nonumber\\
&\qquad\qquad-[\eta-\alpha(\vartheta-\iota)](\alpha\delta\vartheta+\delta\eta+\alpha\gamma\iota)\}\nonumber\\
&\qquad+\alpha\delta^2W^2(\alpha\vartheta-\eta),\label{eq:c1c}
\end{align}
\end{subequations}
in which the value of the coefficient $d$ is of no consequence. Hence, if $\eta<\alpha(\vartheta-\iota)$, then $c_1^-<0$ and coexistence is unstable for sufficiently large $\beta$; we obtained the same result above. Here, we note additionally that, if $\eta<\alpha(\vartheta-\iota)$, then $W>0$, and so the linear and constant terms of the cubic~\eqref{eq:c1c} are positive, while its cubic term is negative, so it has exactly one positive real root. Since $\mathcalbf{C}_-$ is stable for $|V|$ sufficiently small, this root of $c_{\smash{1}}^-$ corresponds to a point on the feasibility boundary $\upDelta=0$. Thus the boundary of the unstable region intersects the feasibility boundary.

\subsubsection{Coexistence beyond coexistence equilibria}
We now extend these results on the coexistence of the two species described by Eqs.~\eqref{eq:model2} beyond coexistence at steady state. We begin by analyzing the stability of the trivial equilibria. We go on to discuss the alternative outcomes of extinction of one species and permanent coexistence~\cite{butler86} of the two species by proving that the two species coexist permanently if all trivial steady states are unstable.
\paragraph*{Stability of the trivial steady states.} First, we determine the stability of the trivial steady states $\mathcalbf{O},\mathcalbf{A},\mathcalbf{B}$ defined in Eqs.~\eqref{eq:AB}, and which are feasible for all parameter values. The Jacobian of Eqs.~\eqref{eq:model2} evaluated at $\mathcalbf{O}$ is 
\begin{subequations}
\begin{align}
\left(\begin{array}{ccc}
1-\gamma&\delta&0\\
\gamma&-\delta&0\\
0&0&\zeta
\end{array}\right).\label{eq:JOa} 
\end{align}
In particular, this Jacobian has an eigenvalue $\zeta>0$, so $\mathcalbf{O}$ is always unstable, with small perturbations expelled from the plane $A=0$.

The stability of the other trivial steady states depends on $U,V$. The Jacobian at $\mathcalbf{A}$ is
\begin{align}
\left(\begin{array}{ccc}
-(\gamma+\beta\zeta/\eta)-U/\eta&\delta&0\\
\gamma+\beta\zeta/\eta&-\delta&0\\
&&-\zeta
\end{array}\right), \label{eq:JAa}
\end{align}
in which the entries left blank clearly do not affect stability. Similarly, the Jacobian at $\mathcalbf{B}$ is
\begin{align}
\left(\begin{array}{ccc}
-1-\gamma&\delta&\\
\gamma&-\delta&\\
0&0&-V
\end{array}\right).\label{eq:JBa}
\end{align}
\end{subequations}
Direct computation of eigenvalues shows that $\mathcalbf{A}$ is stable if and only if $U>0$, while $\mathcalbf{B}$ is stable if and only if $V>0$. In more detail, $\mathcalbf{B}$ is an attractor in the plane $A=0$, but expels orbits out of that plane if and only if $V<0$.

It follows that, if $U>0$ or $V>0$, then there exist (feasible) initial conditions with which Eqs.~\eqref{eq:model2} converge to $\mathcalbf{A}$ or $\mathcalbf{B}$, and hence lead to extinction of one species, and so permanent coexistence of the two species is not possible in general.

\paragraph*{Permanent Coexistence.} The above shows that permanent coexistence irrespective of the initial conditions is only possible if $U,V<0$ and hence $\mathcalbf{A}$ and $\mathcalbf{B}$ are unstable, assumed henceforth. 

From the conditions obtained in Appendix~\ref{appA}, the dynamics of Eqs.~\eqref{eq:model2} are bounded. Hence extinction of one species requires the dynamics of Eqs.~\eqref{eq:model2} to converge to a (stable) limit set $\mathcalbf{L}$ that intersects the boundary $BPA=0$. We claim such a limit set $\mathcalbf{L}$ cannot exist if $\mathcalbf{A}$ and $\mathcalbf{B}$ are unstable.

\begin{figure}[t]
\includegraphics{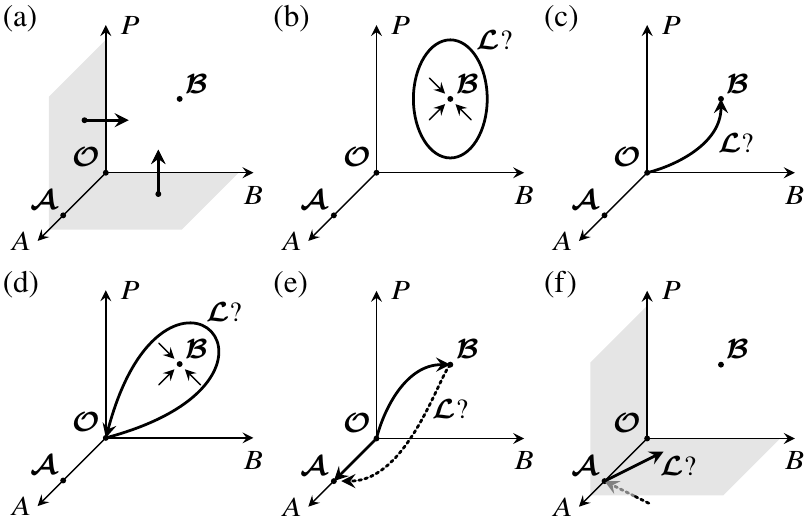}
\caption{Permanent coexistence of the two species described by Eqs.~\eqref{eq:model2} if $\mathcalbf{A},\mathcalbf{B}$ are unstable. (a) Orbits of Eqs.~\eqref{eq:model2} are expelled from $B=0$ and $P=0$. (b)~Non-existence of a (stable) limit cycle $\mathcalbf{L}$ in $\Pi=\{A=0;B,P>0\}$. (c)~Non-existence of a connection $\mathcalbf{L}$ containing $\mathcalbf{B}$ in $\Pi$. (d)~Non-existence of a homoclinic connection of $\mathcalbf{O}$ in $\Pi$. (e)~Non-existence of a connection containing $\mathcalbf{O}\rightarrow\mathcalbf{B}$. (f)~Non-existence of a connection containing $\mathcalbf{A}$, because one of the connections of $\mathcalbf{A}$ is not feasible.}\label{fig10}
\end{figure}

To prove~\cite{*[{The approach of proving permanence by ``chasing'' possible limit sets in this way has also been used for general three-species Lotka--Volterra systems without phenotypic variation: See, e.g., }] [{ and }] freedman85,*[{}][{. }]butler88,*[{A more general approach for establishing permanence involves identifying an appropriate ``averaged Lyapunov function'' [see, e.g., }][{], but the corresponding general results (\emph{vide ibid.}) cannot directly be applied to the systems with phenotypic switching considered here, because, for the two-species model~\eqref{eq:model} and its simplified versions~\eqref{eq:model2}, \eqref{eq:model2b}, \eqref{eq:model2c}, $\dot{B}/B$ and $\dot{P}/P$ diverge at $B=0$ and $P=0$, respectively.}]hofbauer87} this claim, we begin by noting that, from Eqs.~\eqref{eq:model2}, $B=0\Longrightarrow\dot{B}>0$ and $P=0\Longrightarrow\dot{P}>0$ unless $B=P=0$; also, if $A=0$, then $\dot{A}=0$. Hence $\mathcalbf{L}\cap\{BPA=0\}$ lies in the union of the plane ${\Pi=\{A=0;B,P>0\}}$ and the ray $\{B=P=0;A\geq0\}$~\figref{fig10}{a}.

Next, we observe that the dynamics of Eqs.~\eqref{eq:model2} do not allow $\mathcalbf{L}\subset\Pi$: if this were the case, the Poincaré--Bendixson theorem~\cite{glendinning} would imply that $\mathcalbf{L}$ is (1) a fixed point, (2) a limit cycle, or (3) a connection of equilibria. However, (1) is not possible because $\mathcalbf{O}$ and $\mathcalbf{B}$ are both unstable, the latter by assumption; (2) is not possible because a limit cycle would necessarily contain the only interior equilibrium, $\mathcalbf{B}$, which is impossible because the latter is, as we have noted below Eq.~\eqref{eq:JBa}, stable in the plane $A=0$~\figref{fig10}{b}; (3)~is not possible, because this connection would either contain the point $\mathcalbf{B}$~\figref{fig10}{c}, or be a homoclinic connection of $\mathcalbf{O}$ circling $\mathcalbf{B}$~\figref{fig10}{d}, both of which are impossible because $\mathcalbf{B}$ is stable in the plane $A=0$. That no limit cycle exists can also be established (less geometrically) using Dulac's criterion~\cite{glendinning}. 

Extending these arguments, $\mathcalbf{L}$ cannot in fact intersect $\Pi$, for if it did, then it would contain a connection ${\mathcalbf{O}\rightarrow\mathcalbf{B}}$, which is impossible because, as noted above, the directions transverse to $\Pi$ are unstable for both $\mathcalbf{O}$ and $\mathcalbf{B}$~\figref{fig10}{e}.

Hence $\mathcalbf{L}$ must intersect $\{B=P=0;A>0\}$, so must contain $\mathcalbf{A}$ and and two of its connections. Since it cannot contain the connection $\mathcalbf{O}\rightarrow\mathcalbf{A}$ by the above, it must contain the two other connections of $\mathcalbf{A}$. However, direct computation of the eigenvectors of the corresponding Jacobian in Eq.~\eqref{eq:JAa} shows that one of these is not feasible~\figref{fig10}{f}. This is the final contradiction showing that $\mathcalbf{L}$ cannot intersect $BPA=0$.

This argument shows that both species coexist permanently if $\mathcalbf{A},\mathcalbf{B}$ are unstable.

\subsubsection{Stability diagrams of Eqs.~(\ref{eq:model2})}
The exact results derived above yield the stability diagrams shown in \textwholefigref{fig11} for $\iota>0$; we will not discuss the singular case $\iota=0$. They reproduce some of the features of the numerical stability diagrams in Figs.~\figrefp{fig6}{a),(b}. 

\begin{figure}[t]
\includegraphics{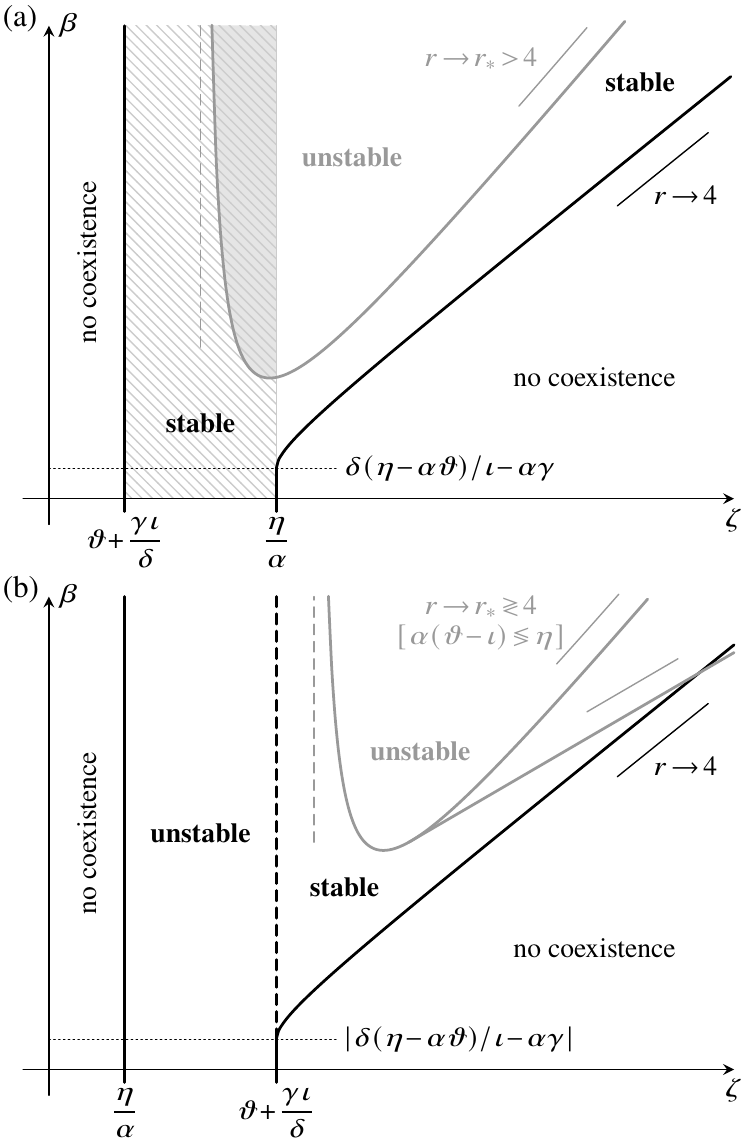}
\caption{Feasibility and stability of the coexistence states $\mathcalbf{C}_\pm$ of the simplified model~\eqref{eq:model2} in the $(\zeta,\beta)$ diagram for $\iota>0$, in the cases (a)~\mbox{$\eta/\alpha>\vartheta+\gamma\iota/\delta$} and (b) $\eta/\alpha<\vartheta+\gamma\iota/\delta$, assuming scalings~\eqref{eq:scalings}. Coexistence is feasible in the region bounded by the thick solid black lines. Both $\mathcalbf{C}_+$ and $\mathcalbf{C}_-$ are unstable in the region marked ``unstable'', but $\mathcalbf{C}_-$ is stable (and $\mathcalbf{C}_+$ is unstable) in the region marked ``stable''. The exact boundary of the region of instability that arises at sufficiently large $\beta$ (grey lines) must be computed numerically; a region of stability at large $\zeta,\beta$ must exist in case (a), but only exists in case~(b) if $\alpha(\vartheta-\iota)<\eta$. The boundary of this region asymptotes to the straight line $\beta\iota/\alpha\delta\zeta\equiv r=r_\ast$, where $r_\ast$ is defined in Eq.~\eqref{eq:rast} and depends only on $\alpha(\vartheta-\iota)/\eta$. In the hatched region of parameter space, the trivial steady states $\mathcalbf{A}$ and $\mathcalbf{B}$ are unstable, and the two species coexist permanently. This is also the region of stable steady-state coexistence for an average of Eqs.~\eqref{eq:model2} with respect to $\mathcalbf{C}_-$ without phenotypic variation and with stochastic switching only (Appendix~\ref{appC}). In the shaded region of parameter space (which need not exist), all steady states of Eqs.~\eqref{eq:model2} are unstable.}\label{fig11}
\end{figure}

Figure~\ref{fig11} shows how the combination of responsive switching and persister-competitor interactions (${\beta\iota>0}$) leads to new behavior compared to case in which these effects are absent ($\beta\iota=0$): There are additional regions of feasibility and stability at large enough rates of responsive switching $\beta>\beta_\ast^{\min}$, with $\beta_\ast^{\min}$ given by Eq.~\eqref{eq:betamin}. Given the persister scalings~\eqref{eq:scalings}, it is important to note that $\beta_\ast^{\min}=O(\varepsilon)$ is possible even if $\iota=O(\varepsilon)$, provided that $\eta-\alpha\vartheta\lesssim O\bigl(\varepsilon^2\bigr)$. This condition expresses the requirement that the intra-species competitions of phenotypes B and A be sufficiently close to their inter-species competitions (Table~\ref{tab0}).

To understand the stabilization of coexistence by responsive switching observed numerically [Figs.~\figrefp{fig6}{a),(b}, insets], we compare Eqs.~\eqref{eq:model} to their averages without phenotypic variation and with stochastic switching only. The calculations in Appendix~\ref{appC} show that, if $\beta=0$, there is a one-to-one correspondence, both in terms of parameters and in terms of feasibility and stability, between Eqs.~\eqref{eq:model2} and this averaged model without phenotypic variation. In other words, stochastic switching on its own does not affect stability. However, this correspondence breaks down if $\beta>0$ (Appendix~\ref{appC}): Responsive switching stabilizes coexistence in a region in which the competitor growth rate $\zeta$ is sufficiently large. 

The conditions for stability of the trivial steady states $\mathcalbf{A}$ and $\mathcalbf{B}$ derived above are independent of the rate of responsive switching $\beta$. Responsive switching does not therefore help in driving the competitors to extinction, but the above shows that it makes stable steady-state coexistence possible in $\zeta>\max{\{\vartheta+\gamma\iota/\delta,\eta/\alpha\}}$ where extinction of bacteria and persisters is the only possible steady state at $\beta=0$. All of these observations show how the combination of responsive switching and persister-competitor interactions ($\beta\iota>0$) favors coexistence.

The shaded region of parameter space in \textfigref{fig11}{a}, in which all steady states of Eqs.~\eqref{eq:model} are unstable, stresses the importance of non-steady attractors: Since $\mathcalbf{A}$ and $\mathcalbf{B}$ are thus unstable, coexistence is permanent, but is not at steady state, since $\mathcalbf{C}_-,\mathcalbf{C}_+$ are unstable, too. In particular, the mathematical observation that responsive switching destabilizes the coexistence equilibrium in this region of parameter space does not contradict the ecological picture of responsive switching stabilizing coexistence that we have painted above: It simply implies that coexistence cannot be at steady state in this case, and hence that responsive switching induces oscillatory population dynamics. 

Classifying all attractors (i.e. not only the stable equilibria) of Eqs.~\eqref{eq:model2} and (even for stable steady states) the initial conditions that lead to them is beyond the scope of this paper. In particular, our result that responsive switching is stabilizing means that, for some initial conditions, Eqs.~\eqref{eq:model2} must converge to stable steady-state coexistence, while these same initial conditions must lead to extinction of one species in the absence of phenotypic variation, because the two-species model without phenotypic variation has no non-steady-state attractors~\cite{murray}. We have no analytical proof of the non-existence of such attractors for the averaged model with stochastic switching only, so these initial conditions could lead to permanent coexistence (albeit not at steady state) in the averaged model with stochastic switching only. These analytical results cannot therefore exclude that it might be any phenotypic switching, rather than responsive switching specifically, that makes coexistence permanent. That it is indeed responsive switching that stabilizes coexistence must be shown numerically, as we have done in Figs.~\ref{fig6} and~\ref{fig7}.

\subsection{Stability and permanence of coexistence in Eqs.~(\ref{eq:model2b})}
The simplified model~\eqref{eq:model2b} has three trivial steady states similar to those of Eqs.~\eqref{eq:model2} defined in Eqs.~\eqref{eq:AB}. They are 
\begin{subequations}
\begin{align}
&\mathcalbf{O}=(0,0,0),&&\mathcalbf{A}=(0,0,\zeta/\eta), &&\mathcalbf{B}=I\left(1,\gamma/(\delta-\mu),0\right),\label{eq:tss2b}
\end{align}
wherein $I=1+\gamma\mu/(\delta-\mu)$. Clearly, $\mathcalbf{O}$ and $\mathcalbf{A}$ are feasible, but, letting $\tau=\delta-\mu$, $\mathcalbf{B}$ is feasible if and only if $\tau>0$. Moreover, model~\eqref{eq:model2b} has a single coexistence equilibrium,
\begin{align}
\mathcalbf{C}=(b,p,a), 
\end{align}
\end{subequations}
with \begin{align}
b=\dfrac{U}{W},&& a=\dfrac{V}{W},&& p=\dfrac{b}{\tau}(\beta a+\gamma),\label{eq:coex2b}
\end{align}
wherein
\begin{subequations}\label{eq:UVW2b}
\begin{align}
U&=\alpha\zeta-\eta-\dfrac{\mu}{\tau}\bigl(\gamma\eta+\beta\zeta\bigr),\quad V=\vartheta\left(1+\dfrac{\mu\gamma}{\tau}\right)-\zeta,\label{eq:UV2b}\\
W&=\alpha\vartheta-\eta-\dfrac{\mu\beta\vartheta}{\tau}=\dfrac{\vartheta U+\eta V}{\zeta}.&\label{eq:W}
\end{align}
\end{subequations}
If $a,b>0$, then $p>0$ if and only if $\tau>0$, assumed henceforth. The coexistence state $\mathcalbf{C}$ is then feasible if and only if $U,V,W$ have the same sign, which is, from Eq.~\eqref{eq:W}, if and only if $U,V$ have the same sign. We also note that the results of Appendix~\ref{appA} show that $\tau>0$ is a sufficient condition for the dynamics of model~\eqref{eq:model2b} to be bounded.

\subsubsection{Stability of the coexistence equilibrium}
We now analyze the stability of the coexistence equilibrium. We will assume that $\mu>0$; the case $\mu=0$ is equivalent to the case $\iota=0$ for Eqs.~\eqref{eq:model2} analyzed in the first part of this Appendix. We will first discuss the case $\beta=0$ before analyzing $\beta\not=0$. The Jacobian evaluated at $\mathcalbf{C}$ is
\begin{align}
\left(\begin{array}{ccc}
-b-(\beta a+\gamma)\delta/\tau&\delta&-(\alpha+\beta)b\\
\beta a+\gamma&-\tau&\beta b\\
-\vartheta a&0&-\eta a
\end{array}\right). \label{eq:JJ}
\end{align}
\paragraph*{Stability of coexistence if $\beta=0$.} If $\beta=0$, i.e. in the absence of responsive switching, the Jacobian~\eqref{eq:JJ} has characteristic polynomial
\begin{align}
P(\lambda)= W^2\lambda^3+c_2\lambda^2+c_1\lambda-\tau UVW,\label{eq:charpoly2}
\end{align}
wherein 
\begin{subequations}\label{eq:c1c20}
\begin{align}
c_1&=\bigl[\tau U+\eta V(\gamma\delta/\tau+\tau)-UV\bigr]W,\\
c_2&=\bigl[U+\eta V+(\gamma\delta/\tau+\tau)W\bigr]W.
\end{align}
\end{subequations}
The Routh--Hurwitz conditions~\cite{murray} imply that coexistence is stable only if $\tau UVW<0$, i.e. only if $U,V<0$ using the feasibility conditions. Conversely, if $U,V<0$ and hence $W<0$, then $c_2>0$ and $c_1c_2>-\tau UVW^3$, of which the second inequality is easily checked by direct multiplication. The Routh--Hurwitz conditions thus imply that coexistence is stable if and only if $U,V<0$.
\paragraph*{Stability of coexistence if $\beta\not=0$.} If $\beta\not=0$, the characteristic polynomial of the Jacobian~\eqref{eq:JJ} still has the form in Eq.~\eqref{eq:charpoly2}, with modified coefficients
\begin{align}
\hat{c}_1&=c_1+\beta V(\eta V-\vartheta U)\delta/\tau,&\hat{c}_2&=c_2+\beta\delta VW/\tau.\label{eq:c1c2}
\end{align}
As in the above analysis of the case $\beta=0$, the Routh--Hurwitz conditions~\cite{murray} imply that coexistence is feasible and stable if and only if $U,V<0$ and $\hat{c}_1\hat{c}_2>-\tau UVW^3$. 

The latter condition is not implied by $U,V<0$, although coexistence is stable if $|U|$ or $|V|$ is sufficiently small away from the singular point $U=V=0$. We prove this claim by expanding the final Routh--Hurwitz condition in $U$ and $V$, assuming all other parameters to be $O(1)$ quantities. Using \textsc{Mathematica} to handle complicated expressions, we obtain 
\begin{subequations}\label{eq:c1c2hat}
\begin{align}
\hat{c}_1\hat{c}_2+\tau UVW^3&=\dfrac{V^4\eta^2u(u+\zeta\eta\tau)}{\zeta^3\tau^2}+O(U),\label{eq:c1c21}\\
&=\dfrac{U^4\vartheta^2}{\zeta^3}\left(\gamma\delta\vartheta+\zeta\tau+\vartheta\tau^2\right)+O(V)\label{eq:c1c22},
\end{align}
wherein $u=\beta\delta\zeta+\gamma\delta\eta+\eta\tau^2$, which proves our claim. A region of instability must however arise for $\beta,\zeta\gg 1$. This follows from the expansion
\begin{align}
&\hat{c}_1\hat{c}_2+\tau UVW^3=-\left[\dfrac{\vartheta^2\mu^2}{\tau^3}\left(\delta+\mu\right)\right]\zeta^3\beta^4+O\left(\zeta^2\beta^4,\zeta^3\beta^3\right).
\end{align}
\end{subequations}
Within this region of instability, $U,V<0$, and so $\mathcalbf{A},\mathcalbf{B}$ are unstable there, too, whence all steady states of Eqs.~\eqref{eq:model2b} are unstable in that region.

Moreover, Eqs.~\eqref{eq:c1c2} show that $\hat{c}_1\rightarrow c_1$ and ${\hat{c}_2\rightarrow c_2}$ as $\beta\rightarrow 0$. Since ${c_1c_2+\tau UVW^3>0}$, it follows that, for sufficiently small $\beta$, $\hat{c}_1\hat{c}_2+\tau UVW^3>0$, too. In other words, for small $\beta$, $\mathcalbf{C}$ is stable if and only if $U,V<0$.

If $\vartheta<\eta/\alpha$, then $U,V<0$ is possible if $\beta=0$, but, if $\vartheta>\eta/\alpha$, this requires $\zeta>\vartheta(1+\mu\gamma/\tau)$ and
\begin{align}
\beta>\beta_\ast=\dfrac{\tau}{\mu}\alpha-\left(\gamma+\dfrac{\tau}{\mu}\right)\dfrac{\eta}{\zeta}, 
\end{align}
using Eqs.~\eqref{eq:UV2b}. In particular, this implies the lower bound
\begin{align}
\beta_\ast>\beta_\ast^{\min}=\dfrac{\delta-\mu}{\mu\vartheta}(\alpha\vartheta-\eta). \label{eq:betaminb}
\end{align}
We are left to discuss the stability of $\mathcalbf{C}$ near the singular point $U=V=0$ for $\vartheta>\eta/\alpha$. This corresponds, in the $(\zeta,\beta)$ plane, to $\zeta=\zeta_\ast^{\min}\equiv\vartheta(1+\mu\gamma/\tau)$ and $\beta=\beta_\ast^{\min}>0$ for $\vartheta>\eta/\alpha$. Near this point, we write
\begin{align}
&\dfrac{\zeta}{\zeta_\ast^{\min}}=1+\bar{\zeta},&&\dfrac{\beta}{\beta_\ast^{\min}}=1+\bar{\beta},
\end{align}
with $\bar{\zeta},\bar\beta>0$. Inserting these definitions into Eqs.~\eqref{eq:UV2b} shows that the domain $U,V<0$ in which $\mathcalbf{C}$ is feasible corresponds, at leading order, to $\bar\beta>\eta\bar\zeta/w$, with ${w=\alpha\vartheta-\eta>0}$. Similarly, from Eqs.~\eqref{eq:UVW2b}, \eqref{eq:c1c20}, and \eqref{eq:c1c2}, the stability boundary $\hat{c}_1\hat{c}_2+\tau UVW^3=0$ corresponds, again at leading order, to the straight lines $\bar\beta=0$, $\bar{\beta}=g_0\bar\zeta$, where
\begin{align}
g_0=\dfrac{\eta}{w}\dfrac{\gamma\mu+\tau}{(\gamma\mu+\tau)+(\gamma\delta+\tau^2)}\left(1-\vartheta-\dfrac{\delta w}{\mu\eta}\right),
\end{align}
and additionally, if the roots $g_\pm$ of the quadratic
\begin{align}
&\bigl(\mu\tau^2w\bigr)g^2-\bigl[\mu\tau^2\eta+w\delta \vartheta(\gamma\mu+\tau)-\mu\eta\vartheta\bigl(\delta\gamma+\tau^2\bigr)\bigr]g\nonumber\\
&\qquad+2\delta\eta\vartheta(\gamma\mu+\tau)=0 
\end{align}
are real, to the straight lines $\bar\beta=g_+\bar\zeta$, $\bar\beta=g_-\bar\zeta$. Clearly, $\bar\beta=0$ is outside $\bar\beta>\eta\bar\zeta/w$. It is also easy to see that $g_0<\eta/w$, whence so is $\bar{\beta}=g_0\bar\zeta$. A region of instability near the singular point $U=V=0$ can therefore arise only if $g_\pm$ are real, and $g_+>\eta/w$ or $g_->\eta/w$. Now $g_\pm$ are real if and only if
\begin{align}
&D\equiv\bigl[\mu\tau^2\eta+w\delta \vartheta(\gamma\mu+\tau)-\mu\eta\vartheta\bigl(\delta\gamma+\tau^2\bigr)\bigr]^2\nonumber\\
&\qquad- 8\mu \tau^2w\delta\eta\vartheta(\gamma\mu+\tau)\geq 0,\label{eq:D2b}
\end{align}
while $g_\pm>\eta/w$ if and only if, additionally,
\begin{subequations}
\begin{align}
w\delta\vartheta(\gamma\mu+\tau)-\mu\eta\vartheta\bigl(\delta\gamma+\tau^2\bigr)-\mu\tau^2\eta\geq \mp\sqrt{D}. \label{eq:ineq1}
\end{align}
Direct computation shows that
\begin{align}
&\bigl[w\delta\vartheta(\gamma\mu+\tau)-\mu\eta\vartheta\bigl(\delta\gamma+\tau^2\bigr)-\mu\tau^2\eta\bigr]^2-D \nonumber\\
&\qquad=4\eta\vartheta\mu\tau^2\bigl[w\delta(\gamma\mu+\tau)+\eta\mu\bigl(\delta\gamma+\tau^2\bigr)\bigr]\geq 0,
\end{align}
and so inequality~\eqref{eq:ineq1} holds if and only if
\begin{align}
w\delta\vartheta(\gamma\mu+\tau)\geq\mu\eta\bigl[\vartheta\bigl(\delta\gamma+\tau^2\bigr)+\tau^2\bigr].\label{eq:ineq2}
\end{align}
\end{subequations}
This implies in particular that $g_+>\eta/w\Longleftrightarrow g_->\eta/w$. We could have obtained this result directly: if only one of $\bar{\beta}=g_+\bar\zeta$, $\bar\beta=g_-\bar\zeta$ intersected $\bar\beta>\eta\bar\zeta/w$, $\mathcalbf{C}$ would be unstable at one of $U=0$, $V=0$ for $|V|$ or $|U|$ sufficiently small, but this would contradict Eq.~\eqref{eq:c1c21} or Eq.~\eqref{eq:c1c22}. Equations \eqref{eq:D2b} and \eqref{eq:ineq2} give the necessary and sufficient conditions for a region of instability to arise near the singular point $U=V=0$.

\subsubsection{Stability of the trivial steady states and permanence}
As in the analysis of Eqs.~\eqref{eq:model2} in the first part of this Appendix, we discuss the stability of the trivial steady states to obtain conditions for permanent coexistence. Again, $\mathcalbf{O}$ is clearly unstable, because the Jacobian of Eqs.~\eqref{eq:model2b} at this equilibrium is
\begin{subequations}
\begin{align}
\left(\begin{array}{ccc}
1-\gamma&\delta&0\\
\gamma&-\tau&0\\
0&0&\zeta
\end{array}\right), 
\end{align}
with an unstable eigenvalues $\zeta>0$. The Jacobian at $\mathcalbf{A}$ is
\begin{align}
\left(\begin{array}{ccc}
-U/\eta-(\gamma+\beta\zeta/\eta)\delta/\tau&\delta&0\\
\gamma+\beta\zeta/\eta&-\tau&0\\
&&-\zeta
\end{array}\right). \label{eq:JA}
\end{align}
Again, entries that do not affect stability have been left blank. Similarly, the Jacobian evaluated at $\mathcalbf{B}$ is
\begin{align}
\left(\begin{array}{ccc}
-1-\gamma(\delta+\mu)/\tau&\delta&\\
\gamma&-\tau&\\
0&0&-V
\end{array}\right). 
\end{align}
\end{subequations}
On computing the eigenvalues of these Jacobians and since we assume that $\tau>0$, we conclude, again, that $\mathcalbf{A}$ is stable if and only if $U>0$, while $\mathcalbf{B}$ is stable if and only if $V>0$. We note from Eqs.~\eqref{eq:UV2b} that $U>0$ is not possible and hence that $\mathcalbf{A}$ is unstable if $\beta>\alpha(\delta/\mu-1)$.

The geometric properties of these Jacobians are identical to those of the corresponding equilibria of Eqs.~\eqref{eq:model2} analyzed in the first part of this Appendix; this follows from direct computation of their eigenvectors. Hence the argument deployed there to establish permanence of coexistence carries over to model~\eqref{eq:model2b}: if $\mathcalbf{A}$ and $\mathcalbf{B}$ are both unstable, i.e. if $U,V<0$, then the two species coexist permanently. 

\subsubsection{Stability diagrams of Eqs.~(\ref{eq:model2b})}
These results yield the stability diagrams drawn in \textwholefigref{fig12} for $\mu>0$; again, we will not discuss the singular case $\mu=0$. These diagrams confirm some of the features present in the numerical stability diagrams in Figs.~\figrefp{fig6}{c),(d}, too.

\begin{figure}[b]
\includegraphics{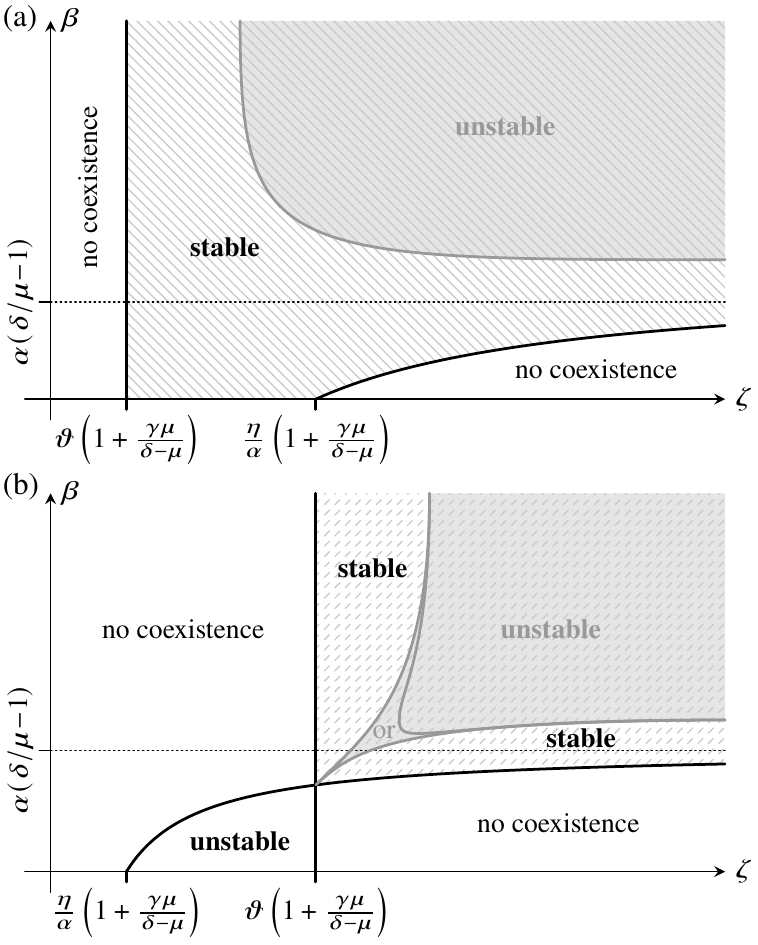}
\caption{Feasibility and stability of the coexistence state $\mathcalbf{C}$ of the simplified model \eqref{eq:model2b} in the $(\zeta,\beta)$ diagram for $\mu>0$, in the cases (a) $\eta/\alpha>\vartheta$ and (b) $\eta/\alpha<\vartheta$. Coexistence is feasible in the region bounded by the thick solid black lines, and is stable or unstable in the regions marked ``stable'' or ``unstable'', respectively. A region of instability (grey lines) arises at sufficiently large $\beta,\zeta$. The exact boundary of this region must be computed numerically. Equations~\eqref{eq:D2b} and \eqref{eq:ineq2} express the necessary and sufficient conditions for the two different behaviors (``or'') that are possible  in panel (b). In the hatched region of parameter space, the trivial steady states $\mathcalbf{A}$ and $\mathcalbf{B}$ are unstable, and the two species coexist permanently. The solidly hatched region in panel (a) is also the region of stable steady-state coexistence for an average of Eqs.~\eqref{eq:model2b} without phenotypic variation and with stochastic switching only (Appendix~\ref{appC}); the dashed hatching in panel~(b) signifies that steady-state coexistence cannot be stable in the averaged model for $\eta/\alpha<\vartheta$ (Appendix~\ref{appC}). In the shaded regions of parameter space, all steady states of Eqs.~\eqref{eq:model2b} are unstable.}\label{fig12}
\end{figure}

The combination of responsive switching and persister growth ($\beta\mu>0$) leads to new behavior compared to $\beta\mu=0$ for $\beta>\beta_\ast^{\min}$, with $\beta_\ast^{\min}$ now given by Eq.~\eqref{eq:betaminb}. Again, $\beta_\ast^{\min}=O(\varepsilon)$ is possible even if $\mu=O(\varepsilon)$, provided that $\alpha\vartheta-\eta\lesssim O\bigl(\varepsilon^2\bigr)$. This is precisely the condition that we discussed when analyzing model~\eqref{eq:model2}.

To assess the effect of responsive switching on coexistence in Eqs.~\eqref{eq:model2b}, we consider again the corresponding averaged models without phenotypic variation and with stochastic switching only. The calculations in Appendix~\ref{appC} determine the range of stability of these averaged models \wholefigref{fig12}. Again, they show that stochastic switching on its own does not affect the stability of the system. Moreover, these results emphasize the importance of establishing correspondences to averaged models when comparing the stability properties of different models: the range of competitor growth rates $\zeta$ for which coexistence is stable increases with the rate of responsive switching $\beta$ initially~\figref{fig12}{a}, but the parameters of the averaged models vary correspondingly, and so responsive switching is neither stabilizing nor destabilizing for small $\beta$. However, large levels of responsive switching destabilize $\mathcalbf{C}$ if $\vartheta<\eta/\alpha$~\figref{fig12}{a}. This destabilization of the coexistence equilibrium does not mean, however, that responsive switching destabilizes coexistence: as shown above, in the region of large $\beta,\zeta$ where $\mathcalbf{C}$ is feasible, but unstable, the trivial steady states $\mathcalbf{A}$ and $\mathcalbf{B}$ are unstable, too, so the two species still coexist permanently. The fact that all steady states of Eqs.~\eqref{eq:model2b} are unstable there simply means that the two species do not coexist at steady state. Similarly, \textfigref{fig12}{b} shows that the combination of responsive switching and persister growth is stabilizing if $\vartheta>\eta/\alpha$ and the competitors grow sufficiently fast. Again, for large $\beta,\zeta$, the two species cannot coexist at steady state, but our calculations imply that they do coexist permanently. This emphasizes the importance of non-steady-state attractors for coexistence. As already discussed for model~\eqref{eq:model2}, this stabilization argument cannot exclude the possibility of permanent, non-steady-state coexistence in the model with stochastic switching only (which is not possible in the averaged model without phenotypic variation); this question must be addressed numerically, as we have done in the discussion of Figs.~\ref{fig6} and~\ref{fig7}.

The condition for stability of $\mathcalbf{B}$, which we have computed above, is independent of the rate of responsive switching, $\beta$. Again, responsive switching cannot therefore drive the competitors to extinction. Moreover, we have shown that $\mathcalbf{A}$, in which the bacteria and persisters are extinct, is stable only if ${\beta<\alpha(\delta/\mu-1)}$. Not only does responsive switching thus destabilize $\mathcalbf{A}$ and hence extinction of bacteria and persisters. It also makes permanent coexistence (albeit not necessarily at steady state) possible at large enough competitor growth rates ${\zeta>I\max{\{\eta/\alpha,\vartheta\}}}$~\wholefigref{fig12}, where extinction of bacteria and persisters is the only possible steady state in the absence of responsive switching, i.e. at $\beta=0$. All of this supports the idea that the combination of responsive switching and persister growth ($\beta\mu>0$) favors coexistence.

\subsection{Stability and permanence of coexistence in Eqs.~(\ref{eq:model2c})}
Model~\eqref{eq:model2c} has three trivial steady states and two coexistence equilibria. They are
\begin{subequations} 
\begin{align}
&\mathcalbf{O}=(0,0,0),&&\mathcalbf{A}=(0,0,\zeta/\eta),&&\mathcalbf{B}=J^{-1}(1,\gamma/\delta,0), \label{eq:tss2c}
\end{align}
with $J=1+\gamma\kappa/\delta$, and
\begin{align}
&\mathcalbf{C}_+=(b_+,p_+,a_+), &&\mathcalbf{C}_-=(b_-,p_-,a_-), \label{eq:css2c}
\end{align}
\end{subequations}
where
\begin{subequations}\label{eq:coexc}
\begin{align}
a_\pm&=\dfrac{1}{2\beta\eta\kappa}\left(Y\mp\sqrt{Y^2-4\beta\delta\eta\kappa V}\right),\label{eq:coex2ca}
\end{align}
\begin{align}
b_\pm&=\dfrac{1}{2\beta\vartheta\kappa}\left(-X\pm\sqrt{X^2+4\beta\delta\vartheta\kappa U}\right),\\
p_\pm&=\dfrac{b_\pm}{\delta}\bigl(\beta a_\pm+\gamma\bigr),\label{eq:ppmc}
\end{align}
\end{subequations}
wherein
\begin{subequations}\label{eq:UVXYc}
\begin{align}
U&=\alpha\zeta-\eta,&V&=\vartheta-\left(1+\dfrac{\gamma\kappa}{\delta}\right)\zeta,\label{eq:UV2c}\\
X&=\dfrac{\delta}{\zeta}(\vartheta U+\eta V)-\beta\kappa\zeta, &Y&=\dfrac{\delta}{\zeta}(\vartheta U+\eta V)+\beta\kappa\zeta,\label{eq:XY2c}
\end{align}
\end{subequations}
so that $X^2+4\beta\delta\vartheta\kappa U=Y^2-4\beta\delta\eta\kappa V$ and $Y=X+2\beta\kappa\zeta$. In particular, $X\leqslant Y$. Similarly to the analysis of model~\eqref{eq:model2} in the first part of this Appendix, if $\beta=0$ or $\kappa=0$, then there is but a single coexistence state ${\mathcalbf{C}=(b,p,a)}$, where
\begin{align}
&b=\dfrac{U}{W}, &&a=\dfrac{V}{W}, &&p=\dfrac{b}{\delta}\bigl(\beta a+\gamma\bigr),\label{eq:coex2c}
\end{align}
with $W=(\vartheta U+\eta V)/\zeta$. From Eqs.~\eqref{eq:ppmc} and \eqref{eq:coex2c}, it is again immediate that $p_\pm,p>0$ if $b_\pm,b>0$ and $a_\pm,a>0$, and so, again, we need not consider $p_\pm,p$ to determine feasibility.

\begin{table}[b]
\caption{Feasibility of the coexistence equilibria $\mathcalbf{C}_\pm$ of Eqs.~\eqref{eq:model2c}: discussion of the sixteen possible sign combinations of $U,V,X,Y$, defined in Eqs.~\eqref{eq:UVXYc}. For some combinations, the resulting signs of $a_\pm$ or $b_\pm$, defined in Eqs.~\eqref{eq:coexc}, are given, and the symbol $\mathbb{C}$ is used for some combinations to indicate that the resulting values of $a_\pm$ or $b_\pm$ may have nonzero imaginary parts. Some sign combinations, marked \# in the final column, are inconsistent with definitions~\eqref{eq:UVXYc}; for other sign combinations, this column gives the corresponding feasibility results.} \label{tab2}
\begin{ruledtabular}
\begin{tabular}{CCCCCCl}
U&V&X&Y&a_\pm&b_\pm&\\
\hline
+&+&+&+&+&\pm&only $\mathcalbf{C}_+$ is feasible\\
+&+&+&-&&&\# ($X>0,Y<0\Rightarrow X>Y$)\\
+&+&-&+&+&\pm&only $\mathcalbf{C}_+$ is feasible\\
+&-&+&+&\mp&\pm&$\mathcalbf{C}_\pm$ are not feasible\\
-&+&+&+&&-/\mathbb{C}&$\mathcalbf{C}_\pm$ are not feasible\\
+&+&-&-&&&\# ($U,V>0\Rightarrow Y>0$)\\
+&-&+&-&&&\# ($X>0,Y<0\Rightarrow X>Y$)\\
-&+&+&-&&&\# ($X>0,Y<0\Rightarrow X>Y$)\\
+&-&-&+&\mp&\pm&$\mathcalbf{C}_\pm$ are not feasible\\
-&+&-&+&+/\mathbb{C}&+/\mathbb{C}&$\mathcalbf{C}_\pm$ can both be feasible\\
-&-&+&+&&&\# ($U,V<0\Rightarrow X<0$)\\
+&-&-&-&\mp&\pm&$\mathcalbf{C}_\pm$ are not feasible\\
-&+&-&-&-/\mathbb{C}&&$\mathcalbf{C}_\pm$ are not feasible\\
-&-&+&-&&&\# ($U,V<0\Rightarrow X<0$)\\
-&-&-&+&\mp&+&only $\mathcalbf{C}_-$ is feasible\\
-&-&-&-&\mp&+&only $\mathcalbf{C}_-$ is feasible\\
\end{tabular}
\end{ruledtabular}
\end{table}

\subsubsection{Feasibility of the coexistence equilibria}
We now have to ask whether the coexistence equilibria $\mathcalbf{C}_\pm$ and $\mathcalbf{C}$ are feasible. The calculations are similar to those for model~\eqref{eq:model2} in the first part of this Appendix. In Table~\ref{tab2}, we analyze the possible sign combinations of the variables $U,V,X,Y$, defined in Eqs.~\eqref{eq:UVXYc}, that appear in the coordinates of the equilibria in Eqs.~\eqref{eq:coexc}. We infer that only $\mathcalbf{C}_+$ is feasible if $U,V>0$, while only $\mathcalbf{C}_-$ is feasible if $U,V<0$. Neither coexistence state is feasible if $U>0$, $V<0$, but both coexistence states are feasible if $U<0$, $V>0$ provided that $X<0$, $Y>0$ and that $\mathcalbf{C}_\pm$ are real (Table~\ref{tab2}). Similarly to the analysis of Eqs.~\eqref{eq:model2}, we conclude that both coexistence states are feasible if and only if
\begin{align}
U<0,\quad V>0,\quad\text{and}\quad\beta>\beta_\ast=\dfrac{\delta}{\kappa\zeta^2}\left(\sqrt{-\vartheta U}+\sqrt{\eta V}\right)^2.\label{eq:coex3c}
\end{align}
In particular, $U<0$ and $V>0$ requires
\begin{align}
\zeta<\min{\left\{\vartheta\left(1+\dfrac{\gamma\kappa}{\delta}\right)^{-1},\dfrac{\eta}{\alpha}\right\}}. \label{eq:coex4c}
\end{align}
Moreover, letting $W=(\vartheta U+\eta V)/\zeta$ again, the conditions $X<0$, $Y>0$ imply that $\beta_\ast>\delta|W|/\kappa\zeta$. These results also yield the lower bound \begin{align}
\beta_\ast>\beta_\ast^{\min}=\dfrac{\left|\delta(\eta-\alpha\vartheta)/\kappa-\gamma\eta\right|}{\min{\left\{\eta/\alpha,\vartheta/(1+\gamma\kappa/\delta)\right\}}^2}.\label{eq:betaminc}
\end{align}
This additional region of feasibility does not arise if $\beta=0$ or $\kappa=0$. Indeed, similarly to the analysis of model~\eqref{eq:model2}, Eqs.~\eqref{eq:coex2c} show that coexistence is feasible in that case if and only if $U$, $V$, $W=(\vartheta U+\eta V)/\zeta$ all have the same sign, and hence if and only if $U,V>0$ or $U,V<0$.

\subsubsection{Stability of the coexistence equilibria}
Next, we analyze the stability of the coexistence equilibria. We can assume that $\kappa>0$, since the case $\kappa=0$ is equivalent to the case $\iota=0$ for Eqs.~\eqref{eq:model2} discussed in the first part of this Appendix. In now familiar fashion, we discuss the cases $\beta=0$ and $\beta\not=0$ separately, but note that, in both cases, the Jacobian evaluated at a coexistence equilibrium $(b,p,a)$ is
\begin{align}
\left(\begin{array}{ccc}
-b-(\beta a+\gamma)&\delta-\kappa b&-(\alpha+\beta)b\\
\beta a+\gamma&-\delta&\beta b\\
-\vartheta a&0&-\eta a
\end{array}\right).\label{eq:JJc}
\end{align} 
\paragraph*{Stability of coexistence if $\beta=0$.} If $\beta=0$, the characteristic polynomial of Eq.~\eqref{eq:JJc} evaluated at $\mathcalbf{C}=(b,p,a)$ defined in Eqs.~\eqref{eq:coex2c} is
\begin{align}
P(\lambda)= W^2\lambda^3+c_2\lambda^2+c_1\lambda-\delta UVW,
\end{align}
where
\begin{subequations}
\begin{align}
c_1&=(\delta+\gamma\kappa) UW+VW(\gamma+\delta)\eta-UV(W+\gamma\kappa\eta/\delta),\\
c_2&=W\left[U+\eta V+(\gamma+\delta)W\right]. 
\end{align}
\end{subequations}
In particular, the Routh--Hurwitz conditions~\cite{murray} imply that $\mathcalbf{C}$ is stable only if $\delta UVW<0$. Since $\mathcalbf{C}$ is feasible if and only if $U,V,W$ have the same sign, a necessary condition for stability is $U,V<0$. Now, using $\zeta W=\vartheta U+\eta V$,
\begin{align}
c_1c_2+\delta UVW^3&=W\left[U+\eta V+(\gamma+\delta)W\right]\bigl\{U^2\vartheta(\delta+\gamma\kappa)/\zeta\nonumber\\
&\qquad +V^2(\gamma+\delta)\eta^2/\zeta+\eta UV[(\gamma+\delta)\vartheta/\zeta\nonumber\\
&\qquad\qquad +(\delta+\gamma\kappa)/\zeta-\gamma\kappa/\delta]\bigr\}\nonumber\\
&\quad-UVW^2(U+\eta V+\gamma W).
\end{align}
In particular, a sufficient condition for $c_1c_2+\delta UVW^3>0$ under $U,V,W<0$ and hence for $\mathcalbf{C}$ to be stable by the Routh--Hurwitz conditions is $\delta[\delta+\vartheta(\gamma+\delta)]+\gamma\kappa(\delta-\zeta)>0$. If $\delta>\zeta$, this holds true; if $\delta<\zeta$, it holds assuming the persister scalings~\eqref{eq:scalings}, but $c_1c_2+\delta UVW^3<0$ is possible and hence instability can occur if $\delta<\zeta$ and the persister scalings are not satisfied (not shown).
\paragraph*{Stability of coexistence if $\beta\not=0$.} Finally, we discuss the stability of the two coexistence states $\mathcalbf{C}_\pm=(b_\pm,p_\pm,a_\pm)$, defined in Eqs.~\eqref{eq:coexc} for $\beta\kappa\neq 0$. The Jacobian in Eq.~\eqref{eq:JJc} has characteristic polynomial
\begin{align} 
P_\pm(\lambda)=\lambda^3+c_2^\pm\lambda^2+c_1^\pm\lambda+c_0^\pm, 
\end{align}
where, in particular and using $W=(\vartheta U+\eta V)/\zeta$,
\begin{align}
c_0^\pm=a_\pm b_\pm\bigl[-\delta W+\beta\kappa(\eta a_\pm-\vartheta b_\pm)\bigr]=\mp a_\pm b_\pm\upDelta, 
\end{align}
wherein $\upDelta^2=X^2+4\beta\delta\vartheta\kappa U=Y^2-4\beta\delta\eta\kappa V$ and $\upDelta>0$. Hence $c_0^\pm\lessgtr0$, and the Routh--Hurwitz conditions~\cite{murray} imply in particular that $\mathcalbf{C}_+$ is unstable. Further, on comparing to our earlier feasibility results, stable coexistence is only possible at $\mathcalbf{C}_-$ and if $U,V<0$ or conditions~\eqref{eq:coex3c} are satisfied. 

Moreover, $c_2^-=b_-+(\beta+\eta)a_-+\gamma+\delta>0$ if $\mathcalbf{C}_-$ is feasible, and hence, by the Routh--Hurwitz conditions, if $\mathcalbf{C}_-$ is feasible, it is stable if and only if $c_1^-c_2^->c_0^-$. Although, as in the analysis of Eqs.~\eqref{eq:model2} and~\eqref{eq:model2b} in the previous parts of this Appendix, we have no complete characterization of the region of parameter space in which this condition holds, we can show that $\mathcalbf{C}_-$ is stable in certain limits.

First, we notice that $\mathcalbf{C}_-$ is stable for sufficiently large $\beta$. This follows from the expansion\begin{subequations}
\begin{align}
c_1^-c_2^--c_0^- =\dfrac{\zeta^3}{\eta^2}\beta^2+O(\beta),\label{eq:c1c2c0c}
\end{align}
obtained using \textsc{Mathematica} to assist with manipulating complicated algebraic expressions.

Next, $\upDelta=\smash{\left|\beta\zeta\kappa-V\delta\eta/\zeta\right|}+O(U)$ from its definition. If ${V<0}$, then $\beta\zeta^2\kappa>V\delta\eta$ is clearly true. This inequality also holds if ${V>0}$ and $\mathcalbf{C}_-$ is feasible, from Eqs.~\eqref{eq:coex3c}. Hence ${\upDelta=(\beta\zeta\kappa-V\delta\eta/\zeta)+O(U)}$, and we find
\begin{align}
c_1^-c_2^--c_0^- = \dfrac{\zeta}{\eta^2}u(u+\zeta\eta)+O(U)>0,\label{eq:c1c2c0c2}
\end{align}
\end{subequations}
where $u=\beta\zeta+(\gamma+\delta)\eta$. Hence $\mathcalbf{C}_-$ is stable for small enough $|U|$ if it is feasible, i.e. if $U,V<0$ or conditions~\eqref{eq:coex3c} hold.

Moreover, as $\beta\rightarrow 0$ at $\kappa\not=0$, $\mathcalbf{C}_-\rightarrow\mathcalbf{C}$, so, by continuity, $\mathcalbf{C}_-$ is stable for small enough $\beta$, since $\mathcalbf{C}$ is, at least under the conditions discussed above and in particular if the persister scalings~\eqref{eq:scalings} are satisfied.

Finally, we discuss the stability of $\mathcalbf{C}_-$ on, and hence by continuity near, the feasibility boundary $\beta=\beta_\ast$ defined in Eqs.~\eqref{eq:coex3c}. On this boundary, $\upDelta=0$ and hence $c_0^-=0$. Since $c_2^->0$, stability is equivalent with $c_1^->0$ there, as in the analysis of Eqs.~\eqref{eq:model2}. Direct computation yields
\begin{subequations}
\begin{align}
(2\beta\kappa)^2\eta\vartheta c_1^-&=\beta\kappa(\beta\zeta\kappa-\delta W)\left[\beta\zeta\kappa+\delta W+2\eta(\delta+\gamma\kappa)\right]\nonumber\\
&\quad+(\beta\zeta\kappa+\delta W)\left[(\alpha\vartheta-\eta)(2\beta\delta\vartheta-\beta\zeta\kappa+\delta W)\right.\nonumber\\
&\hspace{27mm}+\left.2\beta\delta\kappa\eta\vartheta\right].\label{eq:c12c}
\end{align}
From Eqs.~\eqref{eq:coex3c}, feasibility requires $V>0$, and so, from definition~\eqref{eq:UV2c}, $\vartheta>\zeta$. Moreover, we expect, from the persister scalings~\eqref{eq:scalings}, that $\delta>\kappa$. Thus $2\beta\delta\vartheta-\beta\zeta\kappa>\beta\zeta\kappa$. Since ${\beta\zeta\kappa>\delta|W|}$, Eq.~\eqref{eq:c12c} then shows that $c_1^->0$ provided that $\alpha\vartheta>\eta$. Conversely, if $\alpha\vartheta<\eta$ or $\kappa<\delta$, then $c_1^-<0$ is possible (not shown), but we are not aware of simple conditions that ensure the presence or absence of instability in that case. However, after inserting the explicit expressions for $\beta=\beta_\ast$ and $W$ in terms of the other model parameters into Eq.~\eqref{eq:c12c} and expanding using \textsc{Mathematica},
\begin{align}
c_1^-=\dfrac{64\delta^3\eta^3\vartheta^3}{\zeta^4}+O\bigl(\zeta^{-3}\bigr), 
\end{align}
and hence $\mathcalbf{C}_-$ is stable on the feasibility boundary $\beta=\beta_\ast$ for sufficiently small $\zeta$. We stress that this does not follow from Eq.~\eqref{eq:c1c2c0c}, since the ``sufficiently large'' there depends on the other model parameters and, in particular, on $\zeta$. Again expressing $\beta=\beta_\ast$ and $W$ in terms of the other model parameters, we also find
\begin{align}
c_1^-&=\dfrac{4\delta\eta}{\vartheta}(\delta+\gamma\kappa)^2W^2&&\text{if }\zeta=\vartheta\left(1+\dfrac{\gamma\kappa}{\delta}\right)^{-1},\\
&=\dfrac{4\alpha\delta^3\vartheta}{\eta\kappa}(\alpha\vartheta-\eta+\eta\kappa)W^2&&\text{if }\zeta=\dfrac{\eta}{\alpha}.
\end{align}
\end{subequations}
If $\eta/\alpha<\vartheta/(1+\gamma\kappa/\delta)<\vartheta$, then $\alpha\vartheta-\eta>0$. This therefore shows that $c_1^->0$ at $\zeta=\min{\{\eta/\alpha,\vartheta/(1+\gamma\kappa/\delta)\}}$, and hence, on referring to Eq.~\eqref{eq:coex4c}, that $\mathcalbf{C}_-$ is stable at the ``endpoint'' of the feasibility boundary under discussion. For $\zeta=\eta/\alpha$, this result confirms a particular case of the conclusion that we drew from Eq.~\eqref{eq:c1c2c0c2}.

\subsubsection{Stability of the trivial steady states and permanence}
Finally, we analyze the stability of the trivial steady states to obtain conditions for permanent coexistence. As usual, $\mathcalbf{O}$ is always unstable, with the Jacobian of Eqs.~\eqref{eq:model2c} at this fixed point being
\begin{subequations}\label{eq:Jc}
\begin{align}
\left(\begin{array}{ccc}
1-\gamma&\delta&0\\
\gamma&-\delta&0\\
0&0&\zeta
\end{array}\right),
\end{align}
with an unstable eigenvalue $\zeta>0$. Again omitting entries that do not affect the stability, the Jacobian at $\mathcalbf{A}$ is
\begin{align}
\left(\begin{array}{ccc}
-U/\eta-(\gamma+\beta\zeta/\eta)&\delta&0\\
\gamma+\beta\zeta/\eta&-\delta&0\\
&&-\zeta
\end{array}\right).
\end{align}
The Jacobians at $\mathcalbf{O}$ and $\mathcalbf{A}$ are thus actually equal to the Jacobians of the corresponding steady states for model~\eqref{eq:model2} given by Eqs.~\eqref{eq:JOa} and \eqref{eq:JAa}. Moreover, the Jacobian evaluated at $\mathcalbf{B}$ is 
\begin{align}
\left(\begin{array}{ccc}
-\gamma-J^{-1}&\delta-\kappa J^{-1}&\\
\gamma&-\delta&\\
0&0&-VJ^{-1}\!\!
\end{array}\right),  \label{eq:JBc}
\end{align}
\end{subequations}
with $J=1+\gamma\kappa/\delta$ again. On referring to the results for Eqs.~\eqref{eq:model2} in the first part of this Appendix, we find that $\mathcalbf{A}$ is stable if and only if $U>0$. Moreover, on computing the eigenvalues of the matrix in Eq.~\eqref{eq:JBc}, we obtain again that $\mathcalbf{B}$ is stable if and only if $V>0$.

What is more, the eigenvectors of Jacobians in Eqs.~\eqref{eq:Jc} are easily seen to have the same geometric properties as those of the corresponding Jacobians of model~\eqref{eq:model2} analyzed in the first part of this Appendix. We can therefore conclude, as we have done there, that the two species coexist permanently, irrespective of the initial conditions, if $\mathcalbf{A}$ and $\mathcalbf{B}$ are unstable, i.e. if $U,V<0$.

\subsubsection{Stability diagrams of Eqs.~(\ref{eq:model2c})}
We assemble all of these analytical results into the stability diagrams shown in \textwholefigref{fig13} for $\kappa>0$. Again, we will not discuss the singular case $\kappa=0$. These exact diagrams establish some of the features that we have already seen in the numerical stability diagrams in Figs.~\figrefp{fig6}{e),(f}.

\begin{figure}[t]
\includegraphics{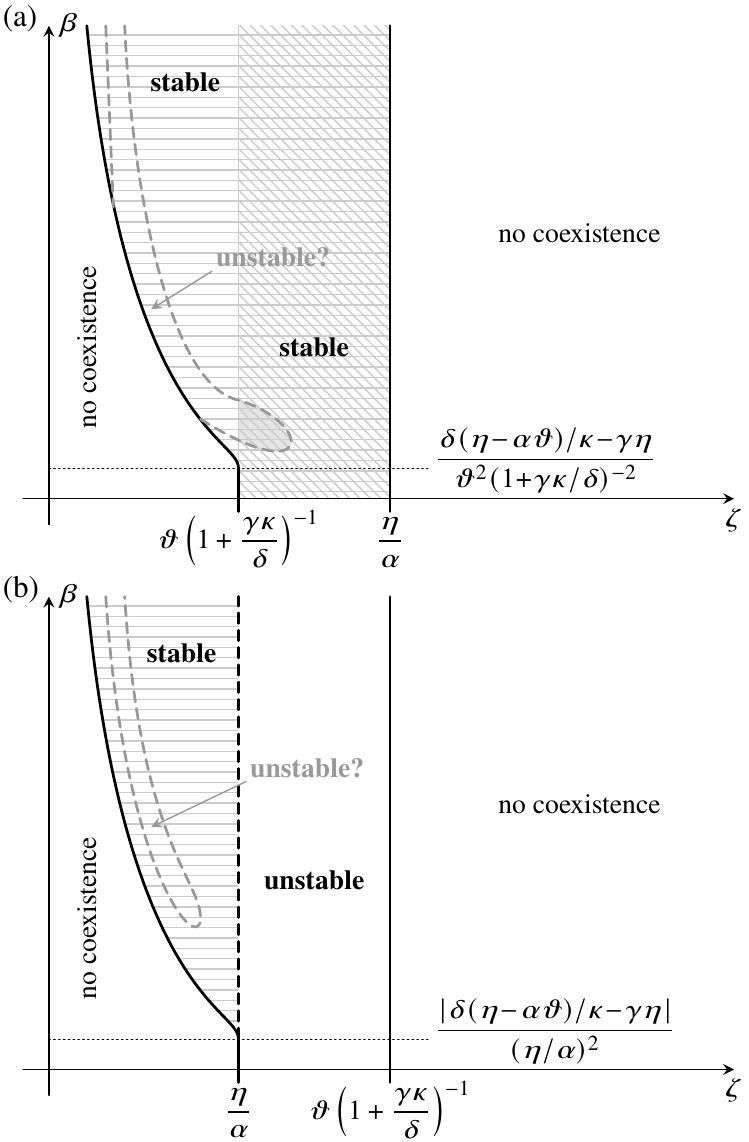}
\caption{Feasibility and stability of the coexistence states $\mathcalbf{C}_\pm$ of the simplified model~\eqref{eq:model2c} in the $(\zeta,\beta)$ diagram for $\kappa>0$, in the cases (a)~\mbox{$\eta/\alpha>\vartheta/(1+\kappa\gamma/\delta)$} and (b) $\eta/\alpha<\vartheta/(1+\kappa\gamma/\delta)$, assuming that the persister scalings~\eqref{eq:scalings} hold. Coexistence is feasible in the region bounded by the thick solid black lines. Both $\mathcalbf{C}_+$ and $\mathcalbf{C}_-$ are unstable in the region marked ``unstable'', but $\mathcalbf{C}_-$ is stable (and $\mathcalbf{C}_+$ is unstable) in the region marked ``stable''. Whether $\mathcalbf{C}_-$ is unstable in a subregion of the ``stable'' region must be determined numerically. In the diagonally hatched region of parameter space in panel (a), the trivial steady states $\mathcalbf{A}$ and $\mathcalbf{B}$ are unstable, and the two species coexist permanently. If the ``unstable'' region intersects this region, all steady states of Eqs.~\eqref{eq:model2c} are unstable within this intersection. The horizontally hatched region is the region of stable steady-state coexistence for averaged models of Eqs.~\eqref{eq:model2c} without phenotypic variation and with stochastic switching only; coexistence may be stable in the average with respect to $\mathcalbf{C}_-$ and in that with respect to $\mathcalbf{C}_+$ (Appendix~\ref{appC}).}\label{fig13}
\end{figure}

Once again, the combination of responsive switching and bacteria-persister interactions ($\beta\kappa>0$) leads to expanded regions of feasibility and stability, compared to $\beta\kappa=0$, if $\beta>\beta_\ast^{\min}$, with $\beta_\ast^{\min}$ now given by Eq.~\eqref{eq:betaminc}. As for models~\eqref{eq:model2} and~\eqref{eq:model2b}, $\beta_\ast^{\min}=O(\varepsilon)$ is possible even if $\kappa=O(\varepsilon)$ provided that $\eta-\alpha\vartheta\lesssim O\bigl(\varepsilon^2\bigr)$. This is the now familiar requirement that the intra-species competitions of bacteria and competitors be sufficiently close to their inter-species competitions (Table~\ref{tab0}).

We conclude our discussion by comparing Eqs.~\eqref{eq:model2c} to their averages without phenotypic variation and with stochastic switching only. As shown in Appendix~\ref{appC}, the conditions for stability of coexistence if $\beta=0$, i.e. for stochastic switching only, is precisely that of the averaged model without phenotypic variation. This equivalence assumes that ${\delta[\delta+\vartheta(\gamma+\delta)]+\gamma\kappa(\delta-\zeta)>0}$, expected to hold given the persister scalings~\eqref{eq:scalings}. Thus stochastic switching on its own has again no effect on stability. The comparison for $\beta>0$, shown in \textwholefigref{fig13}, stresses again the importance of careful model averaging: Although the region of parameter space in which coexistence is stable widens with increasing $\beta$, the corresponding variation of the model parameter means that coexistence is stable in the averaged model if and only if it is feasible. Thus, if regions of instability as sketched in~\textwholefigref{fig13} exist, steady-state coexistence is destabilized there by the combination of responsive switching and competition between bacteria and persisters ($\beta\kappa>0$). Again, this argument does not preclude non-steady-state coexistence there, unless there is a region, as shown in \textfigref{fig13}{a} in which all steady states are unstable and hence coexistence is permanent there. 

Anyway, numerical experiments (not shown) indicate that, while such regions of instability do exist, they are rare for parameter values consistent with the persister scalings~\eqref{eq:scalings}. We have no analytical understanding of this observation, however. Given that signatures of the possible regions of instability for model~\eqref{eq:model2c} sketched in~\textwholefigref{fig13} are absent from the numerical stablity diagrams in Figs.~\figrefp{fig6}{e),(f}, too, this suggests that the destabilizing effect of the combination of responsive switching and bacteria-persister competition ($\beta\kappa>0$) is weaker than the stabilizing effect of responsive switching with persister-competitor interactions ($\beta\iota>0$) and persister growth ($\beta\mu>0$) revealed respectively by the analyses of models~\eqref{eq:model2} and~\eqref{eq:model2b} and also seen in the numerical results in Figs.~\figrefp{fig6}{a)--(c}.

\section{\uppercase{Coexistence without Phenotypic Variation}}\label{appC}
In this Appendix, we briefly rederive the classical results for the stability of two-species Lotka--Volterra competition models~\cite{murray}. We then obtain the stability conditions for averaged models of this Lotka--Volterra form and that correspond to the simplified models \eqref{eq:model2}, \eqref{eq:model2b}, \eqref{eq:model2c}. We combine these results with the stability calculations in Appendix~\ref{appB} to compare the stability of coexistence in these simplified models and in their averages without phenotypic variation and with stochastic switching only.  
\subsection{Coexistence in a two-species Lotka--Volterra model without phenotypic variation}
Coexistence in a two-species Lotka--Volterra competition model without phenotypic variation is a classical problem, discussed, for example, in \citeref{murray}. With the aim in mind of comparing this model to the models with phenotypic variation considered in the main text, it will be useful to rederive the results briefly. We consider the competition between species $A',B'$ described by the differential equations
\begin{align}
&\dot{B}'=B'\bigl(\omega'-\alpha' A'-\chi'B'\bigr),&& \dot{A}'=A'\bigl(\zeta'-\eta' A'-\vartheta' B'\bigr),\label{eq:lv}
\end{align}
wherein $\alpha',\zeta',\eta',\vartheta',\chi',\omega'\geqslant 0$ are parameters. As in our derivation of Eqs.~\eqref{eq:model} and \eqref{eq:model2}, we could have scaled time and $A',B'$ to set some parameters equal to $1$, but we have not done so in order to be able to relate this model to a model with phenotypic variation in the next subsection~\cite{Note3}. Equations~\eqref{eq:lv} have a single coexistence state $\mathcalbf{C}'=(b',a')$, where $b'=u'/w'$ and $a'=v'/w'$, with
\begin{subequations}
\begin{align}
u'&=\alpha'\zeta'-\eta'\omega',\quad v'=\vartheta'\omega'-\zeta'\chi',\\
w'&=\alpha'\vartheta'-\eta'\chi'=\dfrac{\chi'u'+\alpha'v'}{\omega'}.
\end{align}
\end{subequations}
Hence the coexistence state is feasible if and only if $u',v'$ have the same sign. The Jacobian matrix evaluated at this steady state is
\begin{align}
\left(\begin{array}{cc}
-\chi'b'&-\alpha' b'\\-\vartheta' a'&-\eta' a'       
\end{array}\right).
\end{align}
Since $\mathrm{tr}=-\chi'b-\eta' a'<0$, classical stability results~\cite{murray} imply that coexistence is stable if and only if $0<\det=-u'v'/w'$. Hence coexistence state is feasible and stable if and only if $u',v'<0$, or equivalently, if and only if
\begin{align}
\dfrac{\vartheta'\omega'}{\chi'}<\zeta'<\dfrac{\eta'\omega'}{\alpha'}.\label{eq:lvstab}
\end{align}
Equations~\eqref{eq:lv} have two additional nonzero steady states, namely $\mathcalbf{A}'=(0,\zeta'/\eta')$ and $\mathcalbf{B}'=(\omega'/\chi',0)$, at which the Jacobian matrix evaluates to
\begin{align}
\left(\begin{array}{cc}
-u'/\eta'&0\\-\vartheta'\zeta'/\eta'&-\zeta'       
\end{array}\right)\quad\text{and}\quad\left(\begin{array}{cc}
-\omega'&-\alpha'\omega'/\chi'\\0&-v'/\chi'
\end{array}\right), 
\end{align}
respectively. Hence $\mathcalbf{A}'$ and $\mathcalbf{B}'$ are stable if $u'>0$ and $v'>0$, respectively. Comparing these parameter ranges for stability, it follows that either exactly one of $\mathcalbf{A}',\mathcalbf{B'},\mathcalbf{C}'$ is stable, or $\mathcalbf{A}',\mathcalbf{B}'$ are both stable. From arbitrary initial conditions, Eqs.~\eqref{eq:lv} converge to the stable steady state if is unique; if $\mathcalbf{A'},\mathcalbf{B}'$ are both stable, a separatrix through $\mathcalbf{C}'$ separates initial conditions converging to $\mathcalbf{A}'$ from those converging to $\mathcalbf{B}'$~\cite{murray}.

\subsection{Averages without phenotypic variation of the two-species model~(\ref{eq:model}) and of the simplified models~(\ref{eq:model2}), (\ref{eq:model2b}), (\ref{eq:model2c})}
We seek to describe the populations $B,P,A$ that evolve according to Eqs.~\eqref{eq:model} by an averaged model without phenotypic variation and two populations $B'$, which corresponds to $B$ and $P$, and $A'$, which corresponds to $A$~\figref{fig1}{b}. We have introduced this averaging in Ref.~\cite{haas20}, and we have motivated it again here, in Section~\ref{sec:model}.
\subsubsection{Derivation of the averaged model}
A coexistence equilibrium $\mathcalbf{C}=(b,p,a)$ of Eqs.~\eqref{eq:model} is consistent with the equilibrium $\mathcalbf{C}'=(b',a')$ of Eqs.~\eqref{eq:lv} if and only if the populations, the births, and the competition are equal at equilibrium, i.e. if and only if
\begin{subequations}\label{eq:cca}
\begin{align}
b'&=b+p ,&a'&=a,\\
\omega'b'&=b+\mu p,&\zeta'a'&=\zeta a,\\
\alpha'a'b'&=a(\alpha b+\xi p),&\vartheta'a'b'&=a(\vartheta b+\iota p),\\
\chi'b'^2&=b^2+(\kappa+\varpi)bp+\varsigma p^2,&\eta'a'^2&=\eta a^2.
\end{align}
\end{subequations}
These conditions are to Eqs.~\eqref{eq:model} what conditions~\eqref{eq:cc} are to Eqs.~\eqref{eq:fullmodel}. We let $q=p/b$, so that, on introducing $Q=(1+q)^{-1}$, they reduce to
\begin{subequations}
\begin{align}
\alpha'&=Q(\alpha+\xi q),&\zeta'&=\zeta,&\omega'&=Q(1+\mu q),\\
\vartheta'&=Q(\vartheta+\iota q),&\eta'&=\eta,&\chi'&=Q^2\left[1+(\kappa+\varpi)q+\varsigma q^2\right].
\end{align}
\end{subequations}
Hence Eq.~\eqref{eq:lvstab}, the stability condition for the averaged model, becomes
\begin{align}
\dfrac{\vartheta+\iota q}{1+(\kappa+\varpi)q+\varsigma q^2}<\dfrac{\zeta}{1+\mu q}<\dfrac{\eta}{\alpha+\xi q}. \label{eq:sint}
\end{align}
\subsubsection{Equivalence of the simplified models~(\ref{eq:model2}), (\ref{eq:model2b}), (\ref{eq:model2c}) with $\beta=0$ \\ to a model without phenotypic variation}
If $\xi=\varpi=\varsigma=0$ and $\beta=0$, then $q=\gamma/\tau$, where $\tau=\delta-\mu$. Feasibility requires $\tau>0$. In the cases $\mu=\kappa=0$, $\iota=\kappa=0$, $\mu=\iota=0$, the interval in Eq.~\eqref{eq:sint} thus reduces to
\begin{align}
\vartheta+\dfrac{\gamma\iota}{\delta}<\zeta<\dfrac{\eta}{\alpha},\;\vartheta<\dfrac{\zeta}{1+\mu\gamma/\tau}<\dfrac{\eta}{\alpha},\;\dfrac{\vartheta}{1+\kappa\gamma/\delta}<\zeta<\dfrac{\eta}{\alpha},\label{eq:equiv} 
\end{align}
respectively, which, from the calculations in Appendix~\ref{appB}, are precisely the stability conditions for Eqs.~\eqref{eq:model2}, \eqref{eq:model2b}, and~\eqref{eq:model2c}, respectively, with $\beta=0$. This establishes the one-to-one correspondence between these simplified models and the corresponding averaged models claimed in the main text. We note that, for model~\eqref{eq:model2c}, this requires an additional condition on the model parameters discussed in the analysis of that model; this condition follows from the persister scalings~\eqref{eq:scalings}.

\subsubsection{Stability of the averages of the simplified models~(\ref{eq:model2}), (\ref{eq:model2b}), (\ref{eq:model2c}) \\ for $\beta\not=0$}
If now $\xi=\varpi=\varsigma=0$, but $\beta\not=0$, then $q=(\gamma+\beta a)/\tau$, where, again, $\tau=\delta-\mu>0$. We discuss the three simplified models~\eqref{eq:model2}, \eqref{eq:model2b}, \eqref{eq:model2c} severally, simplifying the interval~\eqref{eq:sint} using the explicit expressions for $a$ derived in Appendix~\ref{appB}.

For model~\eqref{eq:model2}, $\mu=\kappa=0$ and $a=a_\pm$, defined in Eq.~\eqref{eq:coex2a} and corresponding to the equilibria $\mathcalbf{C}_\pm$. The stability conditions~\eqref{eq:sint} of the averaged model become
\begin{align}
&U<0,\;V+\dfrac{\iota\beta}{\delta}a_\pm<0,&&\mbox{with }U=\alpha\zeta-\eta,\;V=\vartheta-\zeta+\dfrac{\gamma\iota}{\delta}, 
\end{align}
as in definitions~\eqref{eq:UV}. Since $a_\pm>0$ by feasibility, a necessary condition for stability is $U,V<0$, but this is inconsistent with feasibility of $\mathcalbf{C}_+$ (Appendix~\ref{appB}), which is therefore unstable in the averaged model. Now, using the expression for $a_-$ in Eq.~\eqref{eq:coex2a}, we find 
\begin{align}
V+\dfrac{\iota\beta}{\delta}a_-&=\dfrac{1}{2\alpha\delta}\left(Z-\sqrt{Z^2+\alpha\delta^2UV}\right),
\end{align}
where $Z=\delta(\alpha V-U)+\iota\beta$, and infer that the second condition above is satisfied if $U,V<0$. Since this condition is necessary for stability, $\mathcalbf{C}_-$ is stable in the average of model~\eqref{eq:model2} if and only if $U,V<0$. 

For model~\eqref{eq:model2b}, $\iota=\kappa=0$ and $a$ is given in Eqs.~\eqref{eq:coex2b} and corresponds to the single coexistence equilibrium $\mathcalbf{C}$. The interval~\eqref{eq:sint} yields the inequalities
\begin{align}
V\left(1+\dfrac{\mu\vartheta\beta}{\tau W}\right)<0,\quad U\left(1+\dfrac{\mu\vartheta\beta}{\tau W}\right)<0,\label{eq:sint2b}
\end{align}
wherein 
\begin{align}
U&=\alpha\zeta-\eta-\dfrac{\mu}{\tau}\bigl(\gamma\eta+\beta\zeta\bigr),&V=\vartheta\left(1+\dfrac{\mu\gamma}{\tau}\right)-\zeta, 
\end{align}
are as in Eqs.~\eqref{eq:UV2b}, and $W=(\theta U+\eta V)/\zeta$. Feasibility of $\mathcalbf{C}$ requires $U,V$, and hence $W$ to be of the same sign (Appendix~\ref{appB}). Clearly, $U,V,W>0$ is not compatible with inequalities~\eqref{eq:sint2b}. If $U,V,W<0$, then they require $1+\mu\beta\vartheta/\tau W<0$, which reduces to $\vartheta<\eta/\alpha$. Hence coexistence in the averaged model corresponding to Eqs.~\eqref{eq:model2b} is stable (and feasible) if and only if $\vartheta<\eta/\alpha$ and $U,V<0$.

Finally, for model~\eqref{eq:model2c}, $\iota=\mu=0$ and $a=a_\pm$, where $a_\pm$ are defined in Eq.~\eqref{eq:coex2ca} and correspond to the equilibria $\mathcalbf{C}_\pm$. The stability interval~\eqref{eq:sint} reduces to
\begin{align}
&U<0,\;V<\dfrac{\beta\kappa\zeta}{\delta}a_\pm,\;\text{with }U=\alpha\zeta-\eta,\;V=\vartheta-\left(1+\dfrac{\gamma\kappa}{\delta}\right)\zeta,\label{eq:sint2c}
\end{align}
as defined in Eqs.~\eqref{eq:UV2c}, too. Since $a_\pm>0$ by feasibility, the second condition is clearly satisfied if $V<0$. We may therefore suppose that $U<0$ and $V>0$. Now, on letting $Y=\delta(\vartheta U+\eta V)/\zeta+\beta\kappa\zeta$ as in Eqs.~\eqref{eq:XY2c} and using the explicit form of $a_\pm$ given in Eq.~\eqref{eq:coex2ca}, the second condition in Eqs.~\eqref{eq:sint2c} becomes
\begin{align}
\mp\sqrt{Y^2-4\beta\delta\eta\kappa V}>\dfrac{2\delta\eta}{\zeta}V-Y. \label{eq:sint2c2}
\end{align}
We observe that
\begin{align}
\dfrac{2\delta\eta}{\zeta}V-Y<0\quad\Longleftrightarrow\quad \beta>\dfrac{\delta}{\kappa\zeta^2}\left(\eta V-\vartheta U\right).\label{eq:intYY}
\end{align}
Since we assume that $U<0,V>0$, feasibility of $\mathcalbf{C}_\pm$ requires, from Eq.~\eqref{eq:coex3c},
\begin{align}
\beta>\dfrac{\delta}{\kappa\zeta^2}\left(\sqrt{\eta V}+\sqrt{-\vartheta U}\right)^2>\dfrac{\delta}{\kappa\zeta^2}\left(\eta V-\vartheta U\right), 
\end{align}
where the second inequality holds since $(x+y)^2>x^2+y^2$ for $x,y>0$. This shows that inequalities~\eqref{eq:intYY} hold true. Moreover,
\begin{align}
\left(Y^2-4\beta\delta\eta\kappa V\right)-\left(\dfrac{2\delta\eta}{\zeta}V-Y\right)^2=\dfrac{4UV\delta^2\eta\vartheta}{\zeta^2}<0, \label{eq:intYY2}
\end{align}
since $U<0$, $V>0$. It follows from inequalities~\eqref{eq:intYY} and \eqref{eq:intYY2} that condition~\eqref{eq:sint2c2} holds true. On comparing with the feasibility conditions derived in Appendix~\ref{appB}, we infer that $\mathcalbf{C}_-$ is stable in the averaged model if and only if it is feasible. Moreover, $\mathcalbf{C}_+$ is stable (and feasible) in its averaged model (which is in general different from that for $\mathcalbf{C}_+$) if and only if conditions~\eqref{eq:coex3c} hold. Appendix~\ref{appB} shows that $\mathcalbf{C}_+$ is an unstable equilibrium of Eqs.~\eqref{eq:model2c}. We therefore emphasize that, while responsive switching ($\beta>0$) thus destabilizes the coexistence equilibrium $\mathcalbf{C}_+$, coexistence in the unaveraged Eqs.~\eqref{eq:model2c} may still be stable at the other coexistence equilibrium $\mathcalbf{C}_-$.

These results also enable us to compare the simplified models~\eqref{eq:model2}, \eqref{eq:model2b}, \eqref{eq:model2c} with responsive switching to averaged models with stochastic switching only: The latter have an effective switching rate $\gamma'=\gamma+\beta a$. This follows similarly to the correspondence of Eqs.~\eqref{eq:fullmodel} and \eqref{eq:fullmodels} established in Sec.~\ref{sec:model} and implies that $q'=\gamma'/\tau=(\gamma+\beta a)/\tau$ as above, i.e. conditions~\eqref{eq:equiv}, with $q$ replaced by this $q'$, are precisely the conditions that we have just analyzed. This means that the averaged model with stochastic switching only is stable if and only if the averaged model without phenotypic variation is stable. We stress however that there is no reason to expect this ``complete'' correspondence to hold for the full two-species model~\eqref{eq:model}.

\section{\uppercase{Stability of the trivial steady states of Eqs.}~(\ref{eq:model}) \uppercase{and permanent coexistence}}\label{appD}
While we did not obtain any analytical results bearing on the stability (or indeed the feasibility) of the coexistence equilibria of Eqs.~\eqref{eq:model}, more meaningful progress can be made as far as the stability of the trivial steady states is concerned. These results extend our results for the simplified models by proving that coexistence is permanent for model~\eqref{eq:model} whenever these trivial steady states are all unstable.   

We begin by noting that the trivial steady state $\mathcalbf{O}=(0,0,0)$ of model~\eqref{eq:model} is always unstable. Next, Eqs.~\eqref{eq:model} have a trivial steady state $\mathcalbf{A}=(0,0,\zeta/\eta)$, which is feasible for all parameter values. The Jacobian of Eqs.~\eqref{eq:model} evaluated at $\mathcalbf{A}$ is
\begin{subequations}
\begin{align}
\left(\begin{array}{ccc}
1-(\alpha+\beta)\zeta/\eta-\gamma&\delta&0\\
\gamma+\beta\zeta/\eta&-\tau-\xi\zeta/\eta&0\\
&&-\zeta
\end{array}\right),
\end{align}
wherein, again, $\tau=\delta-\mu$ and entries left blank clearly do not affect stability; we shall assume that $\tau>0$, consistently with the persister scalings~\eqref{eq:scalings}. This Jacobian has one eigenvalue $-\zeta<0$, which results from the trivial connection $\mathcalbf{O}\rightarrow\mathcalbf{A}$. Classical stability results~\cite{murray} imply that $\mathcalbf{A}$ is stable if and only if the sub-Jacobian
\begin{align}
\left(\begin{array}{cc}
1-(\alpha+\beta)\zeta/\eta-\gamma&\delta\\
\gamma+\beta\zeta/\eta&-\tau-\xi\zeta/\eta\\
\end{array}\right),\label{eq:sJF}
\end{align}
\end{subequations}
has $\text{tr}<0$ and $\det>0$. These conditions are at most quadratic in $\zeta$ and can therefore be solved to show that 
\begin{align}
&\text{tr}<0\Longleftrightarrow \zeta>\zeta_0, && \det>0\Longleftrightarrow \zeta<\zeta_-\text{ or }\zeta>\zeta_+,\label{eq:detD}
\end{align}
where 
\begin{subequations}
\begin{align}
\zeta_0&=\dfrac{(1-\gamma-\tau)\eta}{\alpha+\beta+\xi},\\
\zeta_\pm&=\dfrac{\eta}{2(\alpha+\beta)\xi}\left(-z\pm\sqrt{z^2+4(\alpha+\beta)\xi(\tau+\gamma\mu)}\right),
\end{align}
\end{subequations}
with $z=\alpha\tau-\beta\mu-(1-\gamma)\xi$. In particular, $\zeta_-<0$. Also, if $\zeta=\zeta_0$, then $\det=-[1-\gamma-(\alpha+\beta)\zeta_0/\eta]^2-\delta(\gamma+\beta\zeta_0/\eta)<0$, so $\zeta_-<\zeta_0<\zeta_+$ using the second of Eqs.~\eqref{eq:detD}. Hence the necessary and sufficient condition for stability of $\mathcalbf{A}$ is $\zeta>\zeta_+$.

More importantly, if the sub-Jacobian in Eq.~\eqref{eq:sJF} is unstable, its eigenvectors are
\begin{align}
\left(-\phi\pm \sqrt{\phi^2+4\delta\eta(\beta\zeta+\gamma\eta)},2(\beta\zeta+\gamma\eta)\right), 
\end{align}
with $\phi=(\alpha+\beta)\zeta-(1+\tau-\gamma)\eta-\zeta\xi$, and so one of the eigendirections is not feasible. This shows that the geometric properties of $\mathcalbf{A}$ match those of the corresponding trivial steady state of Eqs.~\eqref{eq:model2} analyzed in Appendix~\ref{appB}.

If any more feasible trivial steady states exist, they are of the form $\mathcalbf{B}=(b,p,0)$, where $b,p>0$ satisfy the simultaneous equations
\begin{align}
&b(1-\gamma-b-\kappa p)+\delta p=p(-\tau-\varpi b-\varsigma p)+\gamma b=0. \label{eq:tssfull}
\end{align}
The Jacobian of Eqs.~\eqref{eq:model} evaluated at $\mathcalbf{B}$ is, on simplification using Eqs.~\eqref{eq:tssfull},
\begin{subequations}
\begin{align}
\left(\begin{array}{ccc}
-\delta q-b&\delta-\kappa b&\\
\gamma-\varpi p&-\gamma/q-\varsigma p&\\
0&0&\zeta-\vartheta b-\iota p
\end{array}\right),
\end{align}
where $q=p/b$ and, once again, entries left blank do not affect stability. This Jacobian has one eigenvalue $\zeta-\vartheta b-\iota p$, associated with perturbations out of the plane $A=0$, which may be of either sign. We note that the sub-Jacobian\begin{align}
\left(\begin{array}{cc}
-\delta q-b&\delta-\kappa b\\
\gamma-\varpi p&-\gamma/q-\varsigma p
\end{array}\right), 
\end{align}\end{subequations}has trace and determinant
\begin{subequations}
\begin{align}
\text{tr}&=-\delta q-b-\gamma/q-\varsigma p<0,\\
\det&=\delta p(\varsigma q+\varpi)+\gamma b(1/q+\kappa)+(\varsigma-\kappa\varpi)pb>0,
\end{align}
\end{subequations}
respectively. We have assumed, in the final line and consistently with the persister scalings~\eqref{eq:scalings}, that $\varsigma>\kappa\varpi$. [It is because of the need for this additional assumption in this argument that we have provided separate proofs of our permanence result for the simplified models~\eqref{eq:model2}, \eqref{eq:model2b}, \eqref{eq:model2c}, which do not require such an additional assumption.] It follows that the remaining eigenvalues at $\mathcalbf{B}$ have negative real parts, and that the stability of $\mathcalbf{B}$ is determined by the sign of $\zeta-\vartheta b-\iota p$; determining this sign requires solving Eqs.~\eqref{eq:tssfull}. For our purposes, it suffices to note that this implies that if $\mathcalbf{B}$ is unstable, then the direction transverse to $A=0$ is unstable. Hence the geometric properties of $\mathcalbf{B}$ match those of the corresponding steady state of Eqs.~\eqref{eq:model2} in Appendix~\ref{appB}.

Compared to the discussion in Appendix~\ref{appB}, there remains however one more case to be discussed before we can conclude that coexistence in Eqs.~\eqref{eq:model} is permanent if all trivial steady states are unstable by the results of Appendix~\ref{appB}. Indeed, there could be multiple steady states of the form $\mathcalbf{B}=(b,p,0)$, and so, using the notation introduced in Appendix~\ref{appB}, the limit set $\mathcalbf{L}$ could also be a connection of several such states. This is however impossible because each of them is stable in the plane $A=0$. We can therefore conclude that, also in the full model~\eqref{eq:model}, coexistence is permanent if all trivial steady states are unstable.

\bibliography{per}
\end{document}